\documentclass[a4paper,twocolumn,10pt,accepted=2023-10-23]{quantumarticle}
\pdfoutput=1

\usepackage[utf8]{inputenc}
\usepackage[english]{babel}
\usepackage[T1]{fontenc}

\usepackage{amsmath, amsthm,,amssymb,mathtools}
\usepackage{amsmath}
\usepackage{graphicx}
\usepackage{dcolumn}
\usepackage{dsfont}
\usepackage{bm}
\usepackage{color}
\usepackage[usenames,dvipsnames]{xcolor}

\usepackage{IEEEtrantools}

\usepackage{changes}
\usepackage{verbatim}

\usepackage{geometry}
\usepackage[numbers,sort&compress]{natbib}

\DeclareGraphicsExtensions{.png,.pdf,.eps}
\graphicspath{ {./figs/} }
\renewcommand{\figurename}{Fig.}
\usepackage{booktabs}
\usepackage{array}

\definecolor{myurlcolor}{rgb}{0,0,0.7}
\definecolor{myrefcolor}{rgb}{0.8,0,0}
\usepackage[unicode=true,pdfusetitle, bookmarks=false,bookmarksnumbered=false, bookmarksopen=false, breaklinks=false,pdfborder={0 0 0},backref=false, colorlinks=true, linkcolor=myrefcolor,citecolor=myurlcolor,urlcolor=myurlcolor]{hyperref}

\newcommand{\eref}[1]{(\ref{#1})}
\newcommand{\eqnref}[1]{Eq.~(\ref{#1})}

\newcommand{\figref}[1]{Fig.~\ref{#1}}
\newcommand{\tabref}[1]{Tab.~\ref{#1}}
\newcommand{\secref}[1]{Sec.~\ref{#1}}
\newcommand{\appref}[1]{App.~\ref{#1}}

\newcommand{\citeref}[1]{Ref.~\cite{#1}}

\makeatletter
\newcommand\footnoteref[1]{\protected@xdef\@thefnmark{\ref{#1}}\@footnotemark}
\makeatother

\def \diracspacing {0.7pt}
\newcommand{\bra}[1]{\langle #1 \hspace{\diracspacing} |} % bra
\newcommand{\ket}[1]{| \hspace{\diracspacing} #1 \rangle} % ket
\newcommand{\ketbra}[2]{| \hspace{\diracspacing} #1 \rangle \langle #2 \hspace{\diracspacing} |} % ketbra with different vectors
\newcommand{\ketbraq}[1]{\ketbra{#1}{#1}} % ketbra with the same vector
\newcommand{\mrm}[1]{\mathrm{#1}}
\newcommand{\trm}[1]{\textrm{#1}}
\renewcommand{\t}[1]{\text{#1}}

\newcommand{\cQ}{\mathcal{Q}}
\newcommand{\cL}{\mathcal{L}}
\newcommand{\cNL}{\mathcal{N\!L}}
\newcommand{\cH}{\mathcal{H}}
\newcommand{\cB}{\mathcal{B}}

\newcommand{\cS}{\mathcal{S}}

\newcommand{\tr}[1]{\mrm{Tr}\!\left\{{#1}\right\}}

\newcommand{\UB}[1]{#1^\uparrow}

\newcommand{\DW}[1]{{#1}_{\text{DW}}}

\newcommand{\ad}[1]{{#1}_{\text{a.d.}}}

\newcommand{\eqdef}{\coloneqq}

\usepackage{xparse}
\NewDocumentCommand\roneway{g}{
  \IfNoValueTF{#1}
    {r_{\text{1-way}}}
    {r_{\text{1-way}\!,\text{#1}}}
}

\NewDocumentCommand\rtwoway{g}{
  \IfNoValueTF{#1}
    {r_{\text{2-way}}}
    {r_{\text{2-way}\!,\text{#1}}}
}

\newcommand{\rDW}{\DW{r}}

\newcommand{\neta}{\bar{\eta}}

\newcommand{\sub}[1]{\mrm{#1}}
\newcommand{\p}[1]{p_{#1}}

\newcommand{\pAB}{\p{AB}}
\newcommand{\pABobs}{\p{AB}^\trm{obs}} %observed
\newcommand{\pobs}{p^\trm{obs}} %observed

\newcommand{\pABE}{\p{ABE}}
\newcommand{\pA}{\p{A}}
\newcommand{\pAobs}{\p{A}^\trm{obs}}
\newcommand{\pB}{\p{B}}
\newcommand{\pBobs}{\p{B}^\trm{obs}}

\newcommand{\qAB}{Q_{AB}}

\newcommand{\QA}[1]{\mathsf{Q}^{\mrm{A}}_{#1}}
\newcommand{\QB}[1]{\mathsf{Q}^{\mrm{B}}_{#1}}
\newcommand{\PA}[1]{\mathsf{P}^{\mrm{A}}_{#1}}
\newcommand{\PB}[1]{\mathsf{P}^{\mrm{B}}_{#1}}

\newcommand{\rhoAB}{\rho_\sub{AB}}
\newcommand{\rhoABE}{\rho_\sub{ABE}}

\newcommand{\mA}{m_\sub{A}}
\newcommand{\mB}{m_\sub{B}}
\newcommand{\nA}{n_\sub{A}}
\newcommand{\nB}{n_\sub{B}}
\newcommand{\dA}{d_\sub{A}}
\newcommand{\dB}{d_\sub{B}}

\newcommand{\xk}{x^*}
\newcommand{\yk}{y^*}

\newcommand{\pvec}{\textbf{p}}
\newcommand{\qvec}{\textbf{q}}

\newcommand{\crit}[1]{{#1}_\mrm{crit}}
\newcommand{\critmin}[1]{\check{#1}_\mrm{crit}}

\newcommand{\pp}{\mathsf{p}}

\newcommand{\qq}{\mathsf{q}}

\begin{document}

\title{Upper bounds on key rates in device-independent quantum key distribution based on convex-combination attacks}

\author{Karol \L ukanowski}
\email{k.lukanowski@cent.uw.edu.pl}
\affiliation{Centre for Quantum Optical Technologies, Centre of New Technologies, University of Warsaw, Banacha 2c, 02-097 Warszawa, Poland}
\affiliation{Faculty of Physics, University of Warsaw, Pasteura 5, 02-093 Warszawa, Poland}
\author{Maria Balanz\'o-Juand\'o}
\affiliation{ICFO -- Institut de Ciencies Fotoniques, The Barcelona Institute of Science and Technology, 08860 Castelldefels, Spain}
\author{M\'at\'e Farkas}
\affiliation{Department of Mathematics, University of York, Heslington, York, YO10 5DD, United Kingdom}
\affiliation{ICFO -- Institut de Ciencies Fotoniques, The Barcelona Institute of Science and Technology, 08860 Castelldefels, Spain}
\author{Antonio Ac\'in}
\affiliation{ICFO -- Institut de Ciencies Fotoniques, The Barcelona Institute of Science and Technology, 08860 Castelldefels, Spain}
\affiliation{ICREA-Instituci\'o Catalana de Recerca i Estudis Avan\c{c}ats, Lluis Companys 23, 08010 Barcelona, Spain}
\author{Jan Ko\l ody\'nski}
\email{j.kolodynski@cent.uw.edu.pl}
\affiliation{Centre for Quantum Optical Technologies, Centre of New Technologies, University of Warsaw, Banacha 2c, 02-097 Warszawa, Poland}

\begin{abstract}
The device-independent framework constitutes the most pragmatic approach to quantum protocols that does not put any trust in their implementations. It requires all claims, about e.g.~security, to be made at the level of the final classical data in hands of the end-users. This imposes a great challenge for determining attainable key rates in \emph{device-independent quantum key distribution} (DIQKD), but also opens the door for consideration of eavesdropping attacks that stem from the possibility of a given data being just generated by a malicious third-party. In this work, we explore this path and present the \emph{convex-combination attack} as an efficient, easy-to-use technique for upper-bounding DIQKD key rates. It allows verifying the accuracy of lower bounds on key rates for state-of-the-art protocols, whether involving one-way or two-way communication. In particular, we demonstrate with its help that the currently predicted constraints on the robustness of DIQKD protocols to experimental imperfections, such as the finite visibility or detection efficiency, are already very close to the ultimate tolerable thresholds.
\end{abstract}

\maketitle

%%%%%%%%%%%%%%%%%%%%%%%%%%%%%%%%%%%%%%%%%%%%%%%%%%%%%%%%%%%%%%%%%%%%%%%%%%%%%%%%%%%%%%%%%%%%%
%%%%%%%%%%%%%%%%%%%%%%%%%%%%%%%%%%%%%%%%%%%%%%%%%%%%%%%%%%%%%%%%%%%%%%%%%%%%%%%%%%%%%%%%%%%%%
\section{Introduction}
\label{sec:intro}
%%%%%%%%%%%%%%%%%%%%%%%%%%%%%%%%%%%%%%%%%%%%%%%%%%%%%%%%%%%%%%%%%%%%%%%%%%%%%%%%%%%%%%%%%%%%%
%%%%%%%%%%%%%%%%%%%%%%%%%%%%%%%%%%%%%%%%%%%%%%%%%%%%%%%%%%%%%%%%%%%%%%%%%%%%%%%%%%%%%%%%%%%%%
%
Device-independent quantum key distribution (DIQKD) is the strongest form of quantum cryptographic protocols~\cite{ABG+07,PAB+09}. It does not require the honest users to make any assumptions about the inner workings of devices at their hands and, hence, opens doors to assuring security without putting any trust into the manufacturer providing a given key-distribution system. As long as the parties can assure the classical data they generate during the protocol does not leak out without control---an assumption at the foundation of any cryptographic protocol~\cite{Shannon1949}---then by revealing some of the data and verifying it to exhibit non-local correlations~\cite{BCP+14}, they can extract a cryptographic key whose security is guaranteed by the correctness of quantum theory~\cite{ABG+07,PAB+09}, or even just the no-signalling paradigm~\cite{BHK05,AGM06,AMP06}. As this makes DIQKD immune to all quantum implementation flaws, these cannot be exploited anymore to perform any hacking attack~\cite{ZFQ+08,xu_experimental_2010,LWW+10,GLL+11}.

\begin{figure}[t]
\centering
\includegraphics[width=\columnwidth]{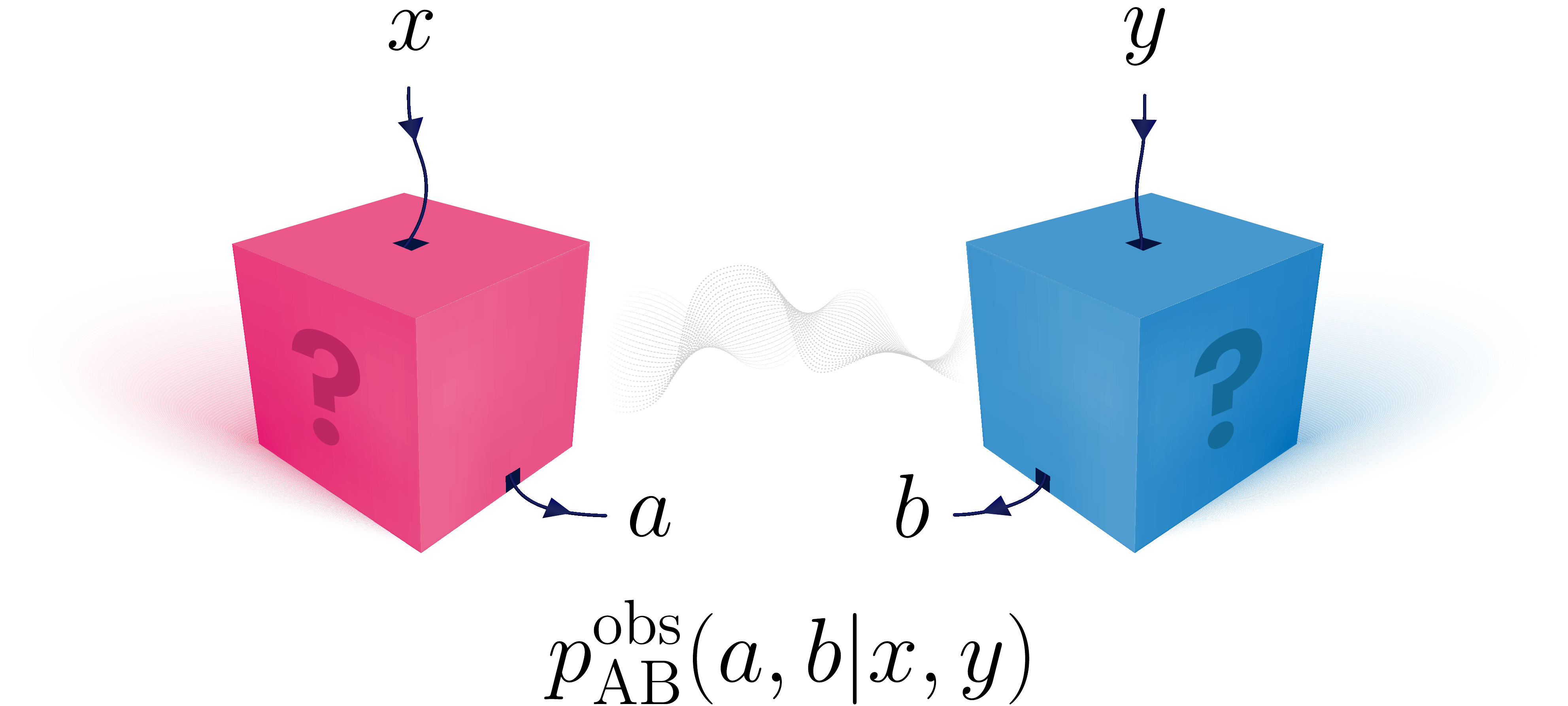}
\caption{\textbf{Device-independent view of a QKD protocol.} Each honest user, Alice (or Bob), ignores the inner workings of her (his) device and treats it as a ``black box'', here red (or blue), which in each round of the protocol takes the \emph{setting} $x$ ($y$) as an input and outputs the \emph{outcome} $a$ ($b$). As a result, by publicly revealing some of the generated data, the parties can verify to be sharing devices whose behaviour is described by a particular \emph{correlation} $\pABobs(a,b | x,y)$---the only property to be trusted when validating the security of the protocol.
}
\label{fig:diqkd_protocol}
\end{figure}

Quantum key distribution (QKD) protocols are based on the setting in which two distant honest parties, Alice and Bob, aim at sharing a cryptographic key, while assuring it to be unknown to any potential eavesdropper. In QKD, this can be achieved by distributing entangled quantum states between Alice and Bob in each round of the protocol, during which they then measure their corresponding part of the state~\cite{Scarani2009}. Within the DIQKD framework, however, the ``black-box'' approach depicted schematically in \figref{fig:diqkd_protocol} is pursued. From the perspective of the users, they are just provided with devices that allow to vary the type of measurement being implemented in each round, the measurement \emph{setting}, whose \emph{outcome} is then outputted by the device. Still, by revealing some of the results between each other, Alice and Bob can verify what is the probability distribution---the \textit{correlation}---describing the operation of their devices (boxes), i.e.~specifying the probabilities with which the outcomes occur for each of the chosen setting. Note that, for the sake of simplicity, we adopt here a terminology in terms of the observed correlation that is meaningful in a scenario consisting of independent and identically distributed (i.i.d.) realisations of the experiment. In a general security proof, one should consider the estimated frequencies of all the observed events for a given finite number of rounds.

Crucially, within the DI paradigm the users do not assume anything about the origin of the correlation, apart from the fact that it must be compliant with the laws of quantum mechanics. Nonetheless, if the observed correlation violates a Bell inequality~\cite{BCP+14}, Alice and Bob can estimate the information that any potential eavesdropper may have about their recorded outcomes---the \emph{raw data}---opening up the possibility for the parties to extract a secure key. Although establishing performance when only finite amount of data is available is important for real-life implementations~\cite{ADF+18,MDR+19,SGP+20}, the first step is always to verify whether the asymptotic key rate---which we refer to here as just the (DIQKD) \emph{key rate}---can be even positive, i.e.~the number of secret bits being finally shared by Alice and Bob divided by the number of protocol rounds employed (the size of the raw data), in the limit of the latter going to infinity.

The task of estimating key rates and proving the security of DIQKD protocols constitutes a great challenge, being a subject of intensive theoretical research. In the \emph{one-way} scenario, in which the parties can publicly communicate only in one direction when distilling the key from the raw data, say from Alice to Bob, the amount of secrecy in any of Alice's outcomes can be quantified by its corresponding von Neumann entropy conditioned on an eavesdropper's quantum side information, $H(A|\mrm{E})$. Thanks to recent developments, such a statement is now crucially true not only when considering \emph{collective attacks}~\cite{DW05,RGK05}, but also when allowing for the most powerful \emph{coherent attacks}~\cite{ADF+18,Arn20,Zhang2020}. Still, the challenge is to compute (or at least lower-bound) $H(A|\mrm{E})$ for a given non-local correlation being shared, in order to determine (lower-bound) the corresponding DIQKD key rate. The first approaches (c.f.~\cite{ABG+07,PAB+09}) have succeeded in providing lower bounds based on the violation of the Clauser--Horne--Shimony--Holt (CHSH) Bell inequality~\cite{CHSH69}, while the more recent works generalised these to include biased CHSH inequalities~\cite{AMP12,WAP20}, accounting also for \emph{noisy preprocessing} of the raw data~\cite{WAP20,HST+20,SBV+20}. Another valid approach is based on lower-bounding $H(A|\mrm{E})$ via the min-entropy~\cite{konig_operational_2009}, which apart from being again relatable to the violation of CHSH~\cite{Masanes2011}, can be accurately lower-bounded by resorting to numerical convex-programming methods~\cite{nieto-silleras_using_2014,bancal_more_2014,mattar_optimal_2015,kolodynski2020}, based on a convergent hierarchy of relaxations~\cite{navascues2007,navascues2008}. This has been lately done also for DIQKD protocols involving \emph{random postselection} of the raw data~\cite{xu_device-independent_2021}---a procedure performed jointly by the users for which, however, the security beyond i.i.d.~attacks~\cite{thinh2015} has not been proven so far. Moreover, a convergent hierarchy has been recently proposed for the conditional von Neumann entropy, $H(A|\mrm{E})$, itself~\cite{brown_device-independent_2021,brown_computing_2021}, see also~\cite{tan_computing_2020}.

In parallel, complementary methods of upper-bounding the DIQKD key rates have been proposed~\cite{KWW20,CFH20,AL20,Farkas2021}. These were put forward, however, with general aim in mind of dealing with \emph{all} potential DIQKD protocols that may involve even \emph{two-way} communication between the parties---e.g.~advantage distillation of the raw data~\cite{TLR20}---so that the upper bounds may then serve as ultimate benchmarks beyond which no DIQKD protocol can venture. In this work, we follow this path but focus instead on attacks that an eavesdropper may adapt for a \emph{particular} DIQKD scenario. As a result, the upper bounds on key rates we obtain account for the special features of the DIQKD protocol considered, e.g.:~whether it involves one-way or two-way communication, the type of preprocessing used, or the postselection stage.

In particular, our approach is to propose concrete strategies a malicious third-party can play when distributing and controlling the devices, so that the data in hands of the users is consistently recovered, but some information about it---which can be explicitly quantified---remains in possession of the eavesdropper, Eve. We consider \emph{individual attacks}~\cite{Scarani2009} that yield a tripartite (Alice, Bob and Eve) classical (i.i.d.) model that in case of one-way scenarios describes a broadcast (wiretap) channel~\cite{CK78}, while in the two-way case allows for unconstrained public discussion~\cite{Maurer93,AC93}. The key rate attained between Alice and Bob within such a model constitutes then an upper bound on the DIQKD key rate associated with their shared data. Moreover, as allowing for more powerful eavesdroppers (collective or coherent attacks) can only decrease the attainable key rate, such an upper bound remains valid beyond individual attacks. In particular, in case it vanishes, it is assured that no secure key can be distilled by the honest parties from a given data set.

In our work, we focus on a class of individual attacks---dubbed \emph{convex-combination} (CC) attacks---in which Eve randomly alternates between distributing either devices that yield stronger non-local correlations than the ones exhibited by the data being shared, or devices that yield classical (local) correlations with Eve possessing full knowledge about the output data of both Alice and Bob. We show that, while correctly reproducing the shared data on average, the CC attack can be optimised by means of linear programming to maximise the probability of the local correlation being shared. As a result, it provides a direct method of upper-bounding the key rates that turns out to be very effective in predicting the zero-key regions, in which no DIQKD is possible. Although this is not our motivation here, let us note that in the non-zero regions our technique could also be merged with the other methods~\cite{KWW20,CFH20,AL20} to determine the tightest overall upper bounds on the key rate~\cite{kaur_upper_2021}.

In contrast to our previous work~\cite{Farkas2021}, in which we have focused on the two-way scenario in order to find regimes in which the CC attack precludes any key to be extracted despite the shared correlation exhibiting non-locality, here we study the limits the attack imposes on imperfections within the correlations being shared---given the DIQKD protocol (incl.~any processing of the raw data) followed by the parties.

In particular, we use the CC attack to determine thresholds on experimentally motivated parameters:~\emph{visibility} and \emph{detection efficiency} (level of losses) beyond which DIQKD cannot be made possible both in the one-way and two-way scenarios---no matter how well one improves current techniques of lower-bounding the exact DIQKD key rates~\cite{tan_computing_2020,brown_device-independent_2021,brown_computing_2021,masini_simple_2021}. To our knowledge, at the time of preparing this manuscript, the best known values of tolerable detection efficiency above which fully secure one-way DIQKD becomes feasible read $\eta\ge80.00\%$~\cite{brown_device-independent_2021} and $\eta\ge80.26\%$~\cite{masini_simple_2021}, while the CC attack allows us to verify that without changing the structure of these protocols, the thresholds could at most be improved to $79.04\%$ and $79.15\%$, respectively. While noisy preprocessing of the shared data is crucial for the parties to reach the above tolerable efficiencies~\cite{brown_device-independent_2021,masini_simple_2021}, it simultaneously makes the CC attack more efficient, so that it provides very tight lower bounds. In general, the CC-based value diminishes to $75\%$ when optimising over all forms of (one-way~\cite{RGK05}) preprocessing potentially employed by the parties. On the other hand, when the eavesdropper can be assumed to perform at most collective attacks, one may further allow the parties to publicly perform random postselection before one-way communication. In such a scenario, the CC-based threshold decreases further to $\approx\!66.(6)\%$---the fundamental value imposed by non-locality of the shared correlation~\cite{E93}---what is consistent with the smallest up-to-date known tolerable efficiency of $\eta\ge68.5\%$ established with random postselection~\cite{xu_device-independent_2021}. Finally, by considering the recent protocol of~\citeref{Gonzales2021}, we demonstrate that the CC attack can be easily applied also to DIQKD schemes motivated by Bell violations with more than two measurement settings and outcomes.

Our results make us believe that the CC attack constitutes a useful tool that, not only allows to easily verify whether there is much room for improvement of the state-of-the-art estimates~\cite{WAP20,HST+20,SBV+20,Masanes2011,bancal_more_2014,nieto-silleras_using_2014,tan_computing_2020,brown_device-independent_2021,masini_simple_2021,xu_device-independent_2021,Gonzales2021} of the key rate for a given DIQKD protocol, but also can be very helpful in seeking ways to modify the protocol in order to improve its robustness to imperfections. Finally, it is also useful to benchmark the lower bounds on Eve's entropy obtained through the existing hierarchies and understand how much can be gained by increasing the level in the hierarchy.

The manuscript is structured as follows. In \secref{sec:DIQKD_ind_attacks}, we discuss the formulation of standard DIQKD protocols under individual attacks and, in particular, the upper bounds on one-way and two-way key rates such attacks yield. We then introduce the CC attack as a special case of an individual attack in \secref{sec:CC}, including its geometric formulation and optimisation via a linear program. In \secref{sec:noise}, we describe the noise models of finite visibility and detection efficiency that we will use to benchmark the robustness of current DIQKD schemes by means of the CC attack. In particular, in \secref{sec:appl_to_DIQKD}, we firstly apply the CC attack to both one-way and two-way protocols that rely on non-local correlations arising from maximally entangled states, discussing in detail the construction and its consequences. We then move onto one-way protocols involving partially entangled states in \secref{sec:DIQKD_part_ent}, which currently provide the state-of-the-art key rates and robustness to noise. In \secref{sec:beyond_two_set}, we further demonstrate that the CC attack can be straightforwardly applied also to scenarios in which the parties employ more than two measurement settings and outcomes. Finally, we conclude our findings in \secref{sec:conclusions}.

%%%%%%%%%%%%%%%%%%%%%%%%%%%%%%%%%%%%%%%%%%%%%%%%%%%%%%%%%%%%%%%%%%%%%%%%%%%%%%%%%%%%%%%%%%%%%
%%%%%%%%%%%%%%%%%%%%%%%%%%%%%%%%%%%%%%%%%%%%%%%%%%%%%%%%%%%%%%%%%%%%%%%%%%%%%%%%%%%%%%%%%%%%%
\section{DIQKD under individual attacks}
\label{sec:DIQKD_ind_attacks}
%%%%%%%%%%%%%%%%%%%%%%%%%%%%%%%%%%%%%%%%%%%%%%%%%%%%%%%%%%%%%%%%%%%%%%%%%%%%%%%%%%%%%%%%%%%%%
%%%%%%%%%%%%%%%%%%%%%%%%%%%%%%%%%%%%%%%%%%%%%%%%%%%%%%%%%%%%%%%%%%%%%%%%%%%%%%%%%%%%%%%%%%%%%

%%%%%%%%%%%%%%%%%%%%%%%%%%%%%%%%%%%%%%%%%%%%%%%%%%%%%%%%%%%%%%%%%%%%%%%%%%%%%%%%%%%%%%%%%%%%%
\subsection{Standard DIQKD protocols}
In a DIQKD protocol two parties, Alice and Bob, have access to a bipartite quantum state, $\rho_\mrm{AB}\in \cB({\cH_\sub{A} \otimes \cH_\sub{B}})$, defined on the tensor product of their corresponding Hilbert spaces. The protocol consists of several rounds, in each of which Alice and Bob choose a particular quantum measurement to measure their part of a fresh copy of $\rhoAB$. In particular, Alice chooses her measurement according to a random variable $X$, whose instance $x$ labels the (measurement) \emph{setting} selected out of $|X|=\mA$ possibilities. Similarly, Bob chooses his measurement according to $Y=y$ with $|Y|=\mB$. The (measurement) \emph{outcome} $a$ ($b$) recorded then by Alice (Bob) corresponds to an instance of the random variable $A$ ($B$) that we assume, without loss of generality, to take the same number of $|A|=\nA$ ($|B|=\nB$) values for any setting, $x$ ($y$), chosen.

According to quantum theory, each of the $\mA$ ($\mB$) measurements of Alice (Bob) is described by a positive-operator-valued measure $\{M^x_a\}_{a=1}^{\nA}$ ($\{N^y_b\}_{b=1}^{\nB}$), so that the \emph{correlation} shared by the parties generally reads
\begin{equation}\label{eq:correlation}
\pABobs(a,b | x,y) = \tr{\,\rhoAB \,(M^x_a \otimes N^y_b)},
\end{equation}
specifying the probability of obtaining the outcomes $a$ and $b$, given that the measurements $x$ and $y$ were selected. We say that $\pABobs$ in \eqnref{eq:correlation} is the \emph{observed} correlation within the $\mA\mB\nA\nB$\emph{-scenario}~\cite{CG04}.

After each protocol round, Alice and Bob store both the observed measurement outcomes, $a$ and $b$, as well as the measurement settings, $x$ and $y$, they employed. The records constitute then the \emph{raw data}, out of which Alice and Bob distil a secret key with help of public communication, so that at the end of the procedure they aim at holding identical strings that appear perfectly random to any third party. In this work, we focus on estimating the asymptotic \textit{key rate}, i.e.~the length of such secret strings divided by the overall number of protocol rounds, in the limit of the latter going to infinity.

We consider here \emph{standard} DIQKD protocols, i.e.~ones in which both parties announce publicly the measurement settings employed in each round~\cite{Farkas2021}. Although we primarily focus on protocols that further use the outcomes of a pre-agreed fixed pair of settings to extract the key, let us emphasise already that the CC attack, which is our main interest, can be applied to any scenario by following step-by-step every stage of a given protocol within the attack, e.g.~see~\cite{Farkas2021} for its application to the scheme of~\cite{SGP+20} involving multiple key-settings. In standard protocols, Alice and Bob firstly record strings of outcomes and settings over sufficiently many protocol rounds. Since individually they only have access to the marginal distributions, they publicly reveal part of their data in order to estimate the full correlation $\pABobs(a,b | x,y)$. This part of the dataset is then discarded. They also reveal the settings for the remaining dataset, and keep those outcomes that correspond to a pre-agreed \textit{key setting} pair, ($\xk,\yk$), distributed according to $\pABobs (a, b | \xk, \yk)$. If estimation shows that the error probability is low enough, they extract the final key from this dataset, using either two-way or one-way public communication schemes known as \textit{privacy amplification} and \textit{error correction}~\cite{CK78,AC93,Maurer93}---and abort the protocol otherwise.

%%%%%%%%%%%%%%%%%%%%%%%%%%%%%%%%%%%%%%%%%%%%%%%%%%%%%%%%%%%%%%%%%%%%%%%%%%%%%%%%%%%%%%%%%%%%%
\subsection{Individual attacks}
In this work, we consider \emph{individual attacks}~\cite{Scarani2009} of the eavesdropper, Eve, in which her register at the end of each protocol round corresponds to a random variable, $E$, being somehow correlated with the outcomes of Alice and Bob, determined by $A$ and $B$, respectively. As $E$ may take as many values as required, it may, for example, consist of doubles (ordered pairs), i.e.~$e=(\tilde{a},\tilde{b})$ where $|E|=\nA\nB$, and $\tilde{a}$ and $\tilde{b}$ stand for Eve's guesses of Alice's and Bob's outcomes, respectively. In such a case, the situation in which Eve knows perfectly both the outcomes corresponds simply to the (tripartite) correlation $\pABE(a,b,e\!=\!(\tilde{a},\tilde{b}))=\delta_{a\tilde{a}}\delta_{b\tilde{b}}/(\nA\nB)$ with $\delta_{\alpha\beta}$ denoting the Kronecker delta function. Note that, generalising naturally \eqnref{eq:correlation} to $\pABE$, such attacks ``force'' Eve to measure her part of now a tripartite state $\rhoABE$ in the same way at the end of each protocol round~\cite{Scarani2009}, and exclude the possibility of her possessing a quantum memory~\cite{PMLA14}.

As a result, each round of the protocol and, hence, the protocol on the whole, is then completely described by a tripartite correlation incorporating also the eavesdropper:
\begin{align}
& \pABE(a,b,e | x,y) \label{eq:p_ABE_ind_attack}\\
&\text{s.t.} \quad  \forall_{a,b,x,y}:\;
\sum_e \pABE(a,b,e | x,y)= \pABobs(a,b|x,y), \nonumber
\end{align}
where the above constraint assures that a given quantum correlation observed by Alice and Bob is indeed recovered on average, despite the presence of Eve.

In general, in order to consistently define the attack one should specify the form of the correlation \eref{eq:p_ABE_ind_attack}, in particular, its quantum origin, i.e.~the state being shared between all three parties and the measurements they perform~\cite{Scarani2009}. However, for our purposes we consider individual attacks in which the strategy of Eve is to simply distribute different boxes---bipartite correlations shared by Alice and Bob (known to her and labelled by $\lambda$)---in each protocol round, so that \eqnref{eq:p_ABE_ind_attack} takes the
%a special 
form:

\begin{align}
& \pABE(a,b,e | x,y) \nonumber \\ & \qquad = \sum_\lambda q(\lambda)\,p(e|\lambda)\,\pAB(a,b | x,y,\lambda) \label{eq:class_ind_attacks}\\
&\text{s.t.} \quad  \forall_{x,y}:\;
\sum_\lambda q(\lambda)\,\pAB(a,b | x,y,\lambda) = \pABobs(a,b|x,y). \nonumber
\end{align}

Such an attack is then specified by the probabilities, $q(\lambda)$, of Eve distributing each bipartite correlation $\pAB(a,b | x,y,\lambda)$, each of which must be decomposable as in \eqnref{eq:correlation} to be consistent with quantum theory, and her knowledge $p(e|\lambda)$ about the outcomes of Alice and Bob for each of these correlations.

Note that this individual attack can be implemented by Eve via sharing the same tripartite state in each measurement round and measuring her part of the state, producing the outcome $e$. Importantly, this can be done such that $e$ preserves all information about $\lambda$ that can later be used by Eve to post-process the variable $e$. This will be important in standard DIQKD protocols, in which Alice and Bob at some point reveal their measurement settings for each round, and this information---together with $\lambda$---can be used by Eve to improve her guess on Alice's and Bob's outcomes. The knowledge of $\lambda$ practically means that Eve always knows which term in the convex decomposition of $\pABobs$ in \eqnref{eq:class_ind_attacks} is used, whereas Alice and Bob have access only to the average distribution $\pABobs$. For an explicit construction of the tripartite state and the measurements of Alice, Bob and Eve, see \appref{sec:ind_attack_explicit}.

%%%%%%%%%%%%%%%%%%%%%%%%%%%%%%%%%%%%%%%%%%%%%%%%%%%%%%%%%%%%%%%%%%%%%%%%%%%%%%%%%%%%%%%%%%%%%
\subsection{Upper bounds on one-way key rates}
Formally, the key rate for all QKD and, hence, also DIQKD protocols assisted by one-way communication (say from Alice to Bob) includes a maximisation over all \emph{preprocessing} maps (performed then by Alice), i.e.~\cite{RGK05}:
\begin{equation}
\roneway(A\to B) \eqdef    \max\limits_{p_{A'|A},\;p_{M|A'}} \roneway(A\to B|A',M),
\label{eq:oneway_rate}
\end{equation}
where $A$ and $B$ are the outcome variables of Alice and Bob when they both select the key settings, $\xk$ and $\yk$, respectively. The mapping $A\rightarrow A'$ described by the stochastic map $p_{A'|A}$ is applied by Alice on her outcome, followed by $A'\rightarrow M$ (described by $p_{M|A'}$), whose output is then sent to Bob over a public channel\footnote{Without loss of generality, the ranges of $M$ and $A'$ can be set to be finite \cite{CK78,AC93}.}. In contrast, all the operations performed by Bob on his outcomes can be ignored, as this would lead to an underestimation of the rate due to an overestimation of the fraction of bits required to perform the error correction~\cite{ML11}. 

For any given preprocessing strategy, however, the one-way key rate may be generally lower-bounded by the so-called \emph{Devetak-Winter} (DW) rate~\cite{DW05}, which is valid for all \emph{collective attacks} (more powerful than individual~\cite{Scarani2009}) that Eve may perform, i.e.~\cite{RGK05}:
\begin{align}
&\roneway(A\to B|A',M) \;\ge\; \nonumber \\
& \qquad \rDW \eqdef  H(A'|\mrm{E},M)-H(A'|B,M),
\label{eq:DW_rate}
\end{align}
where $H(A'|\mrm{E},M)$ is the von Neumann entropy conditioned on the information possessed by the most general \emph{quantum} eavesdropper---denoted here by a roman letter $\mrm{E}$ to explicitly distinguish quantum side-information from random variables signified throughout the text by italic characters---while $H(A'|B, M)$ is the conditional (Shannon) entropy between Alice's and Bob's outcomes for the key settings, both conditioned also on the classical data $M$ revealed by Alice during the preprocessing stage. 

The DW rate \eref{eq:DW_rate} can be intuitively understood as the difference between the contributions attributed to privacy amplification (PA) and error correction (EC). In particular, the PA-term, $H(A'|\mrm{E},M)$, represents the fraction of bits that are at least available to Alice after she compresses her bit-string sufficiently to ensure that it is no longer correlated anyhow with any eavesdropper. The EC-term, $H(A'|B,M)$, denotes instead the fraction of bits that she must still publicly communicate to Bob for him to correct his bit-string to be perfectly matching the one of hers. However, note that the latter is fully determined by the correlation $\pABobs(a,b|\xk,\yk)$ shared by Alice and Bob (and the chosen preprocessing) and, hence, is actually unaffected by the presence of any eavesdropper. 

Strikingly, within the DIQKD framework the inequality in \eqnref{eq:DW_rate} has been shown via the entropy accumulation theorem (EAT)~\cite{ADF+18} to hold for the most powerful quantum eavesdroppers, i.e.~all \emph{coherent attacks}~\cite{Scarani2009}, as long the data announced publicly in a given round is independent from the device outputs generated in preceding rounds~\cite[Section~6.1]{primaatmaja22}. This is true, in particular, for a large family of DIQKD protocols where the publicly disclosed data is restricted to (random) device inputs, whereas the preprocessing performed by Alice is limited to some stochastic mapping $A \to A'$, in which case the DW rate \eref{eq:DW_rate} just reads $\rDW = H(A'|\mrm{E})-H(A'|B)$~\cite{ADF+18,Arn20,Zhang2020}. This applies, for instance, to scenarios when $A$ and $B$ constitute dichotomous variables, while \emph{noisy preprocessing} of the raw data is included~\cite{TSB+20}. In particular, Alice applies then a symmetric bit-flip map%
\footnote{With some probability $0<\pp<1$ of flipping `0' onto `1', and symmetrically `1' onto `0'.}
$A\to A'$ to introduce extra randomness (errors) and make her outputs less correlated with Eve by an amount larger than the one required for them to be corrected during the EC stage, so that the DW rate goes up overall~\cite{RGK05}---as recently demonstrated also within the context of DIQKD~\cite{WAP20,HST+20,SBV+20}.

%Strikingly, within the DIQKD framework the inequality in \eqnref{eq:DW_rate} has been shown to hold for the most powerful quantum eavesdroppers, i.e.~all \emph{coherent attacks}~\cite{Scarani2009}, when no preprocessing is performed by Alice, i.e.~in the absence of $A\to A'\to M$, in which case the DW rate \eref{eq:DW_rate} just reads $\rDW = H(A|\mrm{E})-H(A|B)$~\cite{ADF+18,Arn20,Zhang2020}. Moreover, it has been proven to hold in scenarios when $A$ and $B$ constitute dichotomous variables, while \emph{noisy preprocessing} of the raw data is included~\cite{TSB+20}. In particular, Alice applies then a symmetric bit-flip map%
%\footnote{With some probability $0<\pp<1$ of flipping `0' onto `1', and symmetrically `1' onto `0'.}
%$A\to A'$ to introduce extra randomness (errors) and make her outputs less correlated with Eve by an amount larger than the one required for them to be corrected during the EC stage, so that the DW rate goes up overall~\cite{RGK05}---as recently demonstrated also within the context of DIQKD~\cite{WAP20,HST+20,SBV+20}.

On the other hand, by considering any particular \emph{individual attack} and fixing the preprocessing strategy, we may construct an upper bound on the one-way rate (DIQKD or not) that is valid for all one-way protocols employing this strategy~\cite{CK78,AC93}:
\begin{align}
&\roneway(A\to B|A',M) \;\le\; \nonumber \\
& \qquad H(A'|E,M) - H(A'|B,M),
\label{eq:oneway_UB_ind_attack}
\end{align}
which, in contrast to \eqnref{eq:DW_rate}, assumes a \emph{classical} eavesdropper, i.e.~\eqnref{eq:oneway_UB_ind_attack} is completely determined by the tripartite distribution \eref{eq:p_ABE_ind_attack} for the key settings, $\pABE(a,b,e|\xk,\yk)$, and, in particular, its marginals $\p{AE}$ and $\pAB=\pABobs$ specifying the PA- and EC-terms, respectively. Note that this upper bound remains valid also when stronger attacks are considered, as these may only decrease the rate. Furthermore, by maximising the upper bound \eref{eq:oneway_UB_ind_attack} (typically by numerical heuristic methods) over all preprocessing strategies, i.e.~maps $p_{A'|A}$ and $p_{M|A'}$, we obtain an upper bound on the key rate that is universally valid for one-way protocols \eqref{eq:oneway_rate}.

%%%%%%%%%%%%%%%%%%%%%%%%%%%%%%%%%%%%%%%%%%%%%%%%%%%%%%%%%%%%%%%%%%%%%%%%%%%%%%%%%%%%%%%%%%%%%
\subsection{Upper bounds on two-way key rates}
\label{sec:UBs-twoway}
When it comes to two-way protocols within the DIQKD framework, i.e.~the setting in which Alice and Bob are allowed to perform unconstrained public discussion~\cite{AC93,Maurer93}, lower bounds on the corresponding two-way key rates have been established only when constraining the power of Eve to collective attacks and the communication between Alice and Bob to the so-called \emph{advantage distillation} protocol~\cite{TLR20}. On the other hand, universal upper bounds on the two-way DIQKD rates have been recently proposed~\cite{KWW20,CFH20,AL20,Farkas2021} that base on, e.g., measures of reduced entanglement~\cite{CFH20} or the CHSH-inequality violation~\cite{AL20}.

Here, following the approach based on individual attacks described above and our previous work~\cite{Farkas2021}, we consider upper bounds on the two-way DIQKD key rate constructed with help of intrinsic information~\cite{MW97,CEH+07}---originally employed when considering non-signalling eavesdroppers~\cite{AGM06,AMP06}. In particular, for any individual attack of Eve described by the tripartite correlation \eqref{eq:p_ABE_ind_attack}, the following upper bound on the two-way key rate generally holds~\cite{MW97,CEH+07}: 
\begin{equation}
\rtwoway(A\leftrightarrow B) \;\le\;  I(A \! : \! B \! \downarrow \! E), 
\label{eq:r_twoway_UB_ind_attack}
\end{equation}
where the \emph{intrinsic information},
\begin{equation}
I(A \! : \! B  \! \downarrow \! E) \eqdef \min_{ p_{F|E} }  I(A\!:\!B | F),
\label{eq:intrinsic_info}
\end{equation}
is defined as the conditional mutual information evaluated on the tripartite correlation  
\begin{align}
& \p{ABF}(a,b,f|\xk,\yk)= \nonumber \\
& \qquad \sum_e p_{F|E}(f|e)\,\p{ABE}(a,b,e|\xk,\yk),
\end{align}
which is further minimised over all potential mappings $E\to F$ that Eve can perform on her variable $E$\footnote{Without loss of generality, $F$ can be taken to have the same number of outcomes as $E$~\cite{CEH+07}.}.

In general, the computation of \eqnref{eq:intrinsic_info} may require heuristic methods, as the minimisations over mappings $E\to F$ constitutes a non-convex optimisation problem. However, any map $p_{F|E}$ provides a valid upper bound on two-way rate, since ${I(A\!:\!B \! \downarrow \! E)} \le {I(A\!:\!B|F)}$. Moreover, let us note that the conditional mutual information ${I(A\!:\!B|F)}$ and, hence, the intrinsic information \eref{eq:intrinsic_info} is a monotonic decreasing function under stochastic maps applied on either $A$ or $B$. Thus, as the two-way rate ${\rtwoway(A\!\leftrightarrow \!B)}$ by definition involves maximisation over all stochastic maps that the parties may apply on their bits (supplemented by any two-way communication), the r.h.s.~in \eqnref{eq:r_twoway_UB_ind_attack} correctly incorporates already such a maximisation in its form and is thus a universal upper bound---in stark contrast to the upper bound \eref{eq:oneway_UB_ind_attack} on the one-way rate, which similarly to the DW rate \eref{eq:DW_rate} may increase under preprocessing. In fact, by applying any transformations $A\to A'$ and $B\to B'$ on the r.h.s.~of \eqnref{eq:r_twoway_UB_ind_attack}, we obtain a valid upper bound on the two-way rate in protocols with fixed preprocessing that may only be smaller then the universal bound, i.e., 
\begin{equation}
\rtwoway(A' \! \leftrightarrow \! B') \leq I(A'\!:\!B'\!\downarrow \!E) \leq I(A\!:\!B\!\downarrow \!E),
\end{equation}
where the preprocessed $A'$ and $B'$ are now the random variables initially available to the parties.

%%%%%%%%%%%%%%%%%%%%%%%%%%%%%%%%%%%%%%%%%%%%%%%%%%%%%%%%%%%%%%%%%%%%%%%%%%%%%%%%%%%%%%%%%%%%%
%%%%%%%%%%%%%%%%%%%%%%%%%%%%%%%%%%%%%%%%%%%%%%%%%%%%%%%%%%%%%%%%%%%%%%%%%%%%%%%%%%%%%%%%%%%%%
\section{The convex-combination attack}
\label{sec:CC}
%%%%%%%%%%%%%%%%%%%%%%%%%%%%%%%%%%%%%%%%%%%%%%%%%%%%%%%%%%%%%%%%%%%%%%%%%%%%%%%%%%%%%%%%%%%%%
%%%%%%%%%%%%%%%%%%%%%%%%%%%%%%%%%%%%%%%%%%%%%%%%%%%%%%%%%%%%%%%%%%%%%%%%%%%%%%%%%%%%%%%%%%%%%
%
We consider a subclass of individual attacks taking the form \eref{eq:class_ind_attacks}, in particular, the \emph{convex-combination (CC) attacks} introduced by us in~\cite{Farkas2021}, being inspired by the considerations of~\cite{AMP06,AGM06} in which, however, Eve is allowed to possess even stronger than quantum, but still non-signalling, correlations with the raw data.

In short, within the CC attack, Eve mimics the `observed' non-local correlation (pair of boxes) being shared between Alice and Bob, $\pABobs(a,b|x,y)$, by distributing interchangeably `local' (exhibiting a local-hidden-variable model~\cite{BCP+14}) and `non-local' correlations, in such a way that on average the `observed' correlation is recovered and the attack proceeds unnoticed by the parties. Here we are interested in protocols involving two parties, but such a strategy may be analogously generalised to scenarios in which more parties are involved~\cite{winczewski_limitations_2020}.

An (overpessimistic%
\footnote{If Eve possesses any information about the outcomes also in the `non-local' rounds, this may only improve the CC attack further---yield even tighter upper bounds on the key rates.}%
) assumption is then made, restricting Eve to possess no knowledge about the outcomes of the honest parties whenever she distributes any `non-local' correlation. This contrasts strongly the case of distributing `local' correlations, for each of which Eve can be shown to possess full knowledge about all the outcomes. Motivated by this difference, it is further assumed within the CC attack that it is best for Eve to maximise the overall probability of using local boxes. As a result, once the `non-local' boxes to be used by Eve are specified, the optimal `local' correlation to be distributed most frequently by her can always be found by means of linear programming.

In what follows, we first provide a geometrical interpretation of the CC attack, in order then to describe its optimisation in terms of a linear program, which we subsequently employ in \secref{sec:appl_to_DIQKD} to find the tightest upper bounds on the DIQKD key rates that the CC attack can provide.

%%%%%%%%%%%%%%%%%%%%%%%%%%%%%%%%%%%%%%%%%%%%%%%%%%%%%%%%%%%%%%%%%%%%%%%%%%%%%%%%%%%%%%%%%%%%%
\subsection{Geometric formulation of the CC attack}
As stated above, the CC attack constitutes an example of the individual attack described by \eqnref{eq:class_ind_attacks}. In particular, in its simplest form, Eve distributes either a local or a non-local correlation, denoted by $\pAB^\cL$ or $\pAB^\cNL$, respectively, such that the tripartite correlation \eref{eq:p_ABE_ind_attack} reads:
\begin{align}
\pABE(a,b,e|x,y) &  =  q^\cL \, \pAB^\cL(a,b | x,y)\,  \delta_{e,(a,b)} \nonumber\\
&\qquad + q^\cNL \, \pAB^\cNL(a,b| x, y)\, \delta_{e,?},
\label{eq:CC_tripart}
\end{align}
which corresponds to setting $\lambda=\{0,1\}$ in \eqnref{eq:class_ind_attacks} to a binary variable, whose outcome heralds that either a local or a non-local correlation is distributed by Eve, with probabilities $p(\lambda\!=\!0)=q^\cL$ and $p(\lambda\!=\!1)=q^\cNL\!=\!1\!-\!q^\cL$, respectively. Moreover, Eve knows and controls which boxes are being used in each protocol round, so whenever $\pAB^\cL$ is distributed she has perfect knowledge and $p(e|\lambda=0)=\delta_{e,(a,b)}$ in \eqnref{eq:class_ind_attacks}, i.e.~her outcome is perfectly correlated with the outcomes of Alice and Bob; while in case $\pAB^\cNL$ is used $p(e|\lambda=1)=\delta_{e,?}$ in \eqnref{eq:class_ind_attacks}, i.e.~she registers a special extra outcome ``$?$'' giving her no knowledge about the outcomes of the honest parties. Note that for simplicity, we collected all the `local' terms in $\lambda = 0$. In practice, every local correlation can be decomposed as a convex combination of deterministic correlations. The $\lambda = 0$ term contains this convex decomposition, and as stated earlier, Eve knows exactly which term in the convex decomposition is being used. Hence, once the inputs of Alice and Bob are announced, Eve knows their outcomes exactly, which explains the $p(e|\lambda) = \delta_{e,(a,b)}$ term for each $\lambda = 0$ case.

Recall from \eqnref{eq:class_ind_attacks} that for such an individual attack to be valid the actual correlation observed by the parties, $\pABobs(a,b|x,y)$, must be recovered on average. Evaluating the relevant marginal of \eqnref{eq:CC_tripart}, this corresponds to the following constraint:
\begin{equation}
q^\cL \, \pAB^\cL(a,b | x,y) + q^\cNL \, \pAB^\cNL(a,b| x, y)=\pABobs(a,b|x,y).
\label{eq:CC_constr}
\end{equation}
Therefore, ``reversing'' the above construction, any \emph{convex combination}---hence, the name of the attack---of a local and a non-local quantum correlation satisfying \eqnref{eq:CC_constr} can be used to construct a valid CC attack defined by the tripartite classical correlation \eref{eq:CC_tripart}. Still, the best choice of the convex decomposition \eref{eq:CC_constr} may strongly depend on the setting in which the CC attack is applied.

\begin{figure}[t!]%[h]
\centering
\includegraphics[width=\columnwidth]{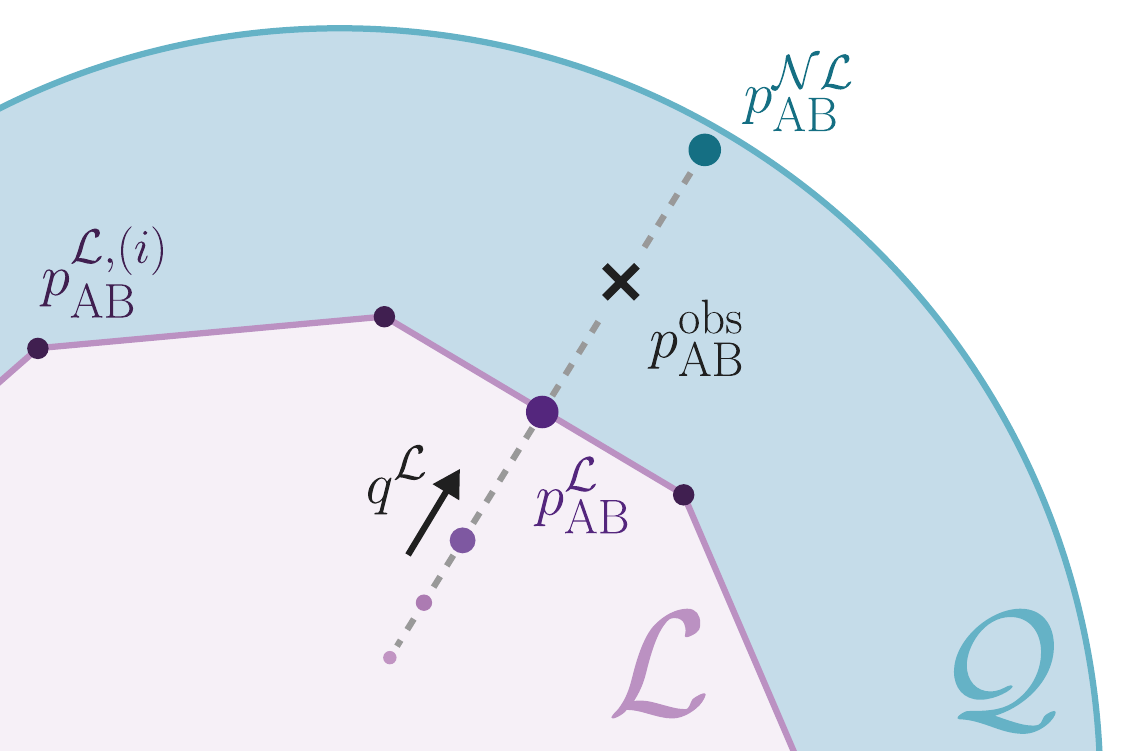}
\caption{\textbf{Geometric formulation of the CC attack.} The correlation $\pABobs(a,b|x,y)$ observed by Alice and Bob resides in the quantum set $\mathcal{Q}$ of the correlation space. In the CC attack, Eve decomposes $\pABobs$ into a local correlation $\pAB^\cL$, contained within the local set $\cL$, and a nonlocal, quantum correlation $\pAB^\cNL$, contained within $\mathcal{Q}$ but outside of $\cL$. She performs the decomposition under the constraint that the convex combination of $\pAB^\cL$ and $\pAB^\cNL$, with weights $q^\cL$ and $1-q^\cL$ respectively, reproduces $\pABobs$ on average. At the same time she strives to maximise the local weight $q^\cL$, which, after fixing $\pAB^\cNL$, corresponds to moving $\pAB^\cL$ in the correlation space along the line connecting $\pAB^\cNL$ and $\pABobs$ in the direction of $\pABobs$. This finally results in $\pAB^\cL$ lying at the boundary of $\cL$. Moreover, because $\cL$ is a convex polytope, $\pAB^\cL$ can be further decomposed into a convex combination of the vertices of $\cL$, corresponding to deterministic correlations $\pAB^{\cL,(i)}$. This allows Eve to possess perfect knowledge of the outcomes whenever distributing a local correlation to Alice and Bob, as she may then equivalently distribute deterministic strategies with predetermined outputs.}
\label{fig:CC_geom_pic}
\end{figure}

As depicted geometrically in \figref{fig:CC_geom_pic}, for our purposes of constructing upper bound DIQKD key rates, we assume that it is best for Eve to maximise the probability with which she distributes the local correlation, i.e.~$q^\cL$ in \eqnref{eq:CC_constr} that we refer to as the \emph{local weight}. A particular correlation being shared, $\pABobs$, constitutes a point within the probability space that we mark in \figref{fig:CC_geom_pic}. Then, maximising $q^\cL$ corresponds to finding two other points collinear with it:~$p^\cL$ contained within the local set $\cL$, and $\pAB^\cNL$ outside of $\cL$ but within the quantum set $\cQ$;~such that the ratio of distances of $\pAB^\cL$ to $\pABobs$ and $\pAB^\cNL$ to $\pABobs$ is minimised, see \figref{fig:CC_geom_pic}.

In the above argumentation, we have stated that Eve perfectly knows all the outcomes whenever she distributes a local correlation to Alice and Bob. This follows from the fact that, because the local set $\cL$ forms a convex polytope in the probability space~\cite{BCP+14}, any local correlation $\pAB^\cL$ can always be decomposed into the extremal points of the polytope, i.e.~$\pAB^\cL=\sum_i\mu_i\, \pAB^{\cL,(i)}$ where it is actually the extremal local correlations $\pAB^{\cL,(i)}$ that are distributed by Eve in every protocol round, each with probability $\mu_i q^\cL$. Now, as any such $\pAB^{\cL,(i)}$ corresponds to a deterministic strategy with predetermined outputs~\cite{BCP+14}, by tracking which extremal local correlation she uses in every round, Eve is able to perfectly infer the outcomes of both Alice and Bob---for whom it still appears that $\pAB^\cL$ is being shared (on average).

Moreover, as displayed in \figref{fig:CC_geom_pic} within the geometric construction, the maximisation of the local weight, $q^\cL$, leads to an optimal local correlation $\pAB^\cL$ lying at the border of the local polytope $\cL$. As a consequence, the optimal $\pAB^\cL$ must always belong to one of the facets of $\cL$. This means that not only one may perform such maximisation by solving a linear program, as we now show, but also the facet at which the optimal $\pAB^\cL$ lies can be unambiguously determined. Although identifying facets of the local set may be hard in general~\cite{Avis_2005}, once the Bell inequality associated with a particular facet is identified, one can in principle determine the corresponding expression for the local weight analytically, and hence provide an analytic solution to the problem.

In what follows, we succeed in doing so when the observed correlation $\pABobs$ of Alice and Bob arises from a maximally entangled state being shared by them, so that the non-local correlation $\pAB^\cNL$ used in the attack by Eve corresponds to Tsirelson boxes~\cite{tsirelson1980}, i.e.~the non-local correlation violating maximally the CHSH inequality~\cite{BCP+14}. Moreover, when analysing the robustness of DIQKD to finite detection efficiency (losses), which requires the use of partially entangled states by the honest users~\cite{E93}, we also obtain analytic results in the limit of the shared state approaching its product form---a feature typically required to reach the highest robustness to losses~\cite{E93}.

%%%%%%%%%%%%%%%%%%%%%%%%%%%%%%%%%%%%%%%%%%%%%%%%%%%%%%%%%%%%%%%%%%%%%%%%%%%%%%%%%%%%%%%%%%%%%
\subsection{Optimisation of the CC attack via a linear program}
For brevity, we drop within this section the subscript $p \equiv \pAB$, as all the probabilities refer here to bipartite correlations shared by the honest users---which Eve distributes within the CC attack---unless specified otherwise.

The extremal correlations defining the local polytope, $\cL$ in \figref{fig:CC_geom_pic}, correspond to deterministic strategies of assigning particular outcomes, $a$ and $b$, for each combination of measurement settings, $x$ and $y$~\cite{BCP+14}. Hence, for a given $\mA\mB\nA\nB$-scenario considered, there exists $\nA^{\mA}\nB^{\mB}$ such extremal points. We shall label by the vector $\pvec^{\cL}=\left( p_i^\cL\right)_{i}$ the set of all such extremal local correlations, and by $\qvec^{\cL}=\left( q_{i}^{\cL}\right)_{i}$ the vector of the corresponding probabilities that Eve assigns to each of them within the CC attack. On the other hand, we assume the average non-local correlation that she distributes to be a mixture of pre-chosen non-local quantum correlations forming a vector $\pvec^{\cNL}=\left( p^\cNL_j\right)_{j}$, each of which is distributed by Eve with the corresponding probability from the vector $\qvec^{\cNL}=\left( q_{j}^{\cNL}\right)_{j}$. Finally, let us recall that for the attack to succeed Eve must reproduce on average the true correlation observed by Alice and Bob, i.e.~$\pobs \equiv \pABobs(a,b|x,y)$ of \eqnref{eq:correlation}, for \emph{all} the measurement settings $x$ and $y$.

In order to optimise the CC attack, Eve seeks a probability vector $\qvec=\qvec^{\cL}\oplus\qvec^{\cNL}$ such that the local correlations are distributed as frequently as possible. This corresponds to solving the following linear program~\cite{BV04}, which maximises the overall probability of sending any local boxes:
\begin{align}
\label{eq:linear_program}
\qvec_{\text{CC}}\!\left[\pvec^{\cNL}, \pobs \right] = \; 
& \underset{\qvec}{\text{argmax}} \;  \sum_i q_i^\cL  \\
& \text{s.t.}\quad  \qvec^{\cL}\cdot\pvec^{\cL}+\qvec^{\cNL}\cdot\pvec^{\cNL}=\pobs, \nonumber\\
& \qquad \sum_i q_i^\cL + \sum_j q_j^\cNL = 1, \nonumber\\
& \qquad\; \forall_{i,j}\!:\quad 0\leq q_{i}^{\cL}, q_{j}^{\cNL} \leq1, \nonumber
\end{align}
where the first constraint is just the generalisation of \eqnref{eq:CC_constr} enforcing Eve to distribute on average the observed correlation, while the other constraints ensure $\qvec$ to constitute a valid probability vector. Note that the set of extremal local correlations, $\pvec^{\cL}$, is not an input to the linear program. Rather, it is a predetermined collection defined by the considered scenario and remains fixed for all programs computed within that scenario, for different choices of $\pvec^{\cNL}$ and $\pobs$.

The above construction requires to specify multiple local correlations, $\pvec^{\cL}$ (extremal points of the local polytope), and for generality we have also allowed for multiple non-local boxes, $\pvec^{\cNL}$. However, by defining now the effective local correlation as the average $p^\cL\eqdef \qvec^{\cL}\cdot\pvec^{\cL}$, and similarly $p^\cNL\eqdef \qvec^{\cNL}\cdot\pvec^{\cNL}$ for the non-local case, we always recover the binary setting described in the previous section and \figref{fig:CC_geom_pic}. In particular, the resulting CC attack is completely specified by the tripartite correlation \eref{eq:CC_tripart}, where now the local weight reads $q^{\cL}\eqdef \sum_{i} q_{i}^{\cL}$ (and similarly $q^{\cNL}\eqdef \sum_{i} q_{i}^{\cNL}$ as the non-local weight), while one must substitute the solution of \eqnref{eq:linear_program}, $\qvec_{\text{CC}}=\qvec_{\text{CC}}^{\cL}\oplus\qvec_{\text{CC}}^{\cNL}$, for the corresponding local and non-local probability vectors.

%%%%%%%%%%%%%%%%%%%%%%%%%%%%%%%%%%%%%%%%%%%%%%%%%%%%%%%%%%%%%%%%%%%%%%%%%%%%%%%%%%%%%%%%%%%%%
%%%%%%%%%%%%%%%%%%%%%%%%%%%%%%%%%%%%%%%%%%%%%%%%%%%%%%%%%%%%%%%%%%%%%%%%%%%%%%%%%%%%%%%%%%%%%
\section{Robustness of non-local correlations}
\label{sec:noise}
%%%%%%%%%%%%%%%%%%%%%%%%%%%%%%%%%%%%%%%%%%%%%%%%%%%%%%%%%%%%%%%%%%%%%%%%%%%%%%%%%%%%%%%%%%%%%
%%%%%%%%%%%%%%%%%%%%%%%%%%%%%%%%%%%%%%%%%%%%%%%%%%%%%%%%%%%%%%%%%%%%%%%%%%%%%%%%%%%%%%%%%%%%%
%
It should be clear from the previous section and the geometric picture that for the CC attack to be applicable the observed correlation, $\pABobs$ in \figref{fig:CC_geom_pic}, cannot lie at the border of the quantum set $\mathcal{Q}$, in which case $q^\cL$ is necessarily zero. This, however, never happens in real-life implementations due to the inevitable noise perturbing the desired correlation and forcing it to be decomposable in the form of a convex combination depicted in \figref{fig:CC_geom_pic}. The two models applicable to experimental realisations~\cite{MDR+19,kolodynski2020,zapatero_long-distance_2019}, commonly used to verify robustness of DIQKD protocols~\cite{ABG+07,PAB+09,HST+20,TLR20,WAP20,SBV+20}, are the scenarios of finite visibility and finite detection efficiency that we summarise below.   

%%%%%%%%%%%%%%%%%%%%%%%%%%%%%%%%%%%%%%%%%%%%%%%%%%%%%%%%%%%%%%%%%%%%%%%%%%%%%%%%%%%%%%%%%%%%%
\subsection{Finite visibility}
Although within the DI framework we are restricted to perform the analysis at the level of correlations, the noise models associated with particular implementations are typically defined assuming certain form of quantum states and measurements employed. The finite \textit{visibility}, in particular, is associated with the probability $V \in [0,1]$ with which Alice and Bob succeed in sharing the intended bipartite state $\rhoAB$, while with probability $1-V$ it is the maximally mixed state that is rather distributed. As a result, the actual state they share becomes
\begin{equation}
\rhoAB(V) := V \, \rhoAB + \frac{1 - V}{\dA\dB}\,\openone_{\dA\dB},
\label{eq:rhoV}
\end{equation}
where $\dA=\dim\cH_\sub{A}$ and $\dB=\dim\cH_\sub{B}$.

However, given that Alice and Bob perform projective (von Neumann) measurements for which $\nA=\dA$ and $\nB=\dB$, we may then write their observed correlation \eref{eq:correlation} as
\begin{equation}
\pABobs(a,b | x,y) = V \, \qAB(a,b | x,y) + \frac{1 - V}{\nA \nB},
\label{eq:visibility}
\end{equation}
where by $\qAB$ we denote the ideal correlation shared at $V=1$. Hence, the finite visibility model is then equivalent to the uniform noise being admixed with all the $\nA\nB$ outcomes occurring with equal probability, so that at $V=0$ a uniformly random distribution of the outcomes is always observed by the parties, i.e.~independently of the measurement settings chosen.

%%%%%%%%%%%%%%%%%%%%%%%%%%%%%%%%%%%%%%%%%%%%%%%%%%%%%%%%%%%%%%%%%%%%%%%%%%%%%%%%%%%%%%%%%%%%%
\subsection{Finite detection efficiency}
The second model of finite detection efficiency is attributed to the problem of photonic losses in optical implementations of DIQKD~\cite{MDR+19,zapatero_long-distance_2019,kolodynski2020}. At the level of the shared correlation, this results in both Alice and Bob failing to detect %the quantum state 
any signal with probability $\neta \eqdef  1-\eta$, where $\eta \in [0,1]$ is  the \textit{detection efficiency} parameter. Such a non-detection event  constitutes then an additional outcome for any of the measurement used, a `\emph{no-click}', that we denote by $\varnothing$. For example, if the original outputs of the parties' devices are binary, $a, b \in \{0,1\}$, after the inclusion of non-detection events (which changes the scenario to have $n_A = n_B = 3$ outcomes) the observed correlation can be expressed in a convenient table format as\footnote{Dealing with more than two outcomes leads just to more rows and columns in \eqnref{eq:lossy_corr}.}
\begin{IEEEeqnarray}{rCl}
\label{eq:lossy_corr}
	\pABobs(a,b | x,y)&=&\begin{array}{|c|c|c|c|}
	\hline a \setminus b & 0 & 1 & \varnothing\\
	\hline 0 & \eta^{2}\mathsf{Q}_{00}^{xy} & \eta^{2}\mathsf{Q}_{01}^{xy} & \eta \neta \mathsf{Q}_{0}^{x}\\
	\hline 1 & \eta^{2}\mathsf{Q}_{10}^{xy} & \eta^{2}\mathsf{Q}_{11}^{xy} & \eta \neta \mathsf{Q}_{1}^{x}\\
	\hline \varnothing & \neta \eta\mathsf{Q}_{0}^{y} & \neta \eta\mathsf{Q}_{1}^{y} & \neta^2
	\\\hline \end{array}
\IEEEeqnarraynumspace
\end{IEEEeqnarray}
where by $\mathsf{Q}_{ab}^{xy} \eqdef  \qAB(a,b | x,y)$ we denote again the ideal quantum correlation observed by the users for $\eta = 1$, with its marginals of Alice and Bob reading $\mathsf{Q}_{a}^{x} = \sum_b \mathsf{Q}_{ab}^{xy}$ and $\mathsf{Q}_{b}^{y} = \sum_a \mathsf{Q}_{ab}^{xy}$, respectively. 

Note, that if one wanted to consider the effect of imperfect visibility and finite detection efficiency at the same time, it suffices to substitute $\pABobs(a,b | x,y)$ from Eq.~\eqref{eq:visibility} for $\mathsf{Q}_{ab}^{xy}$ in the correlation~\eqref{eq:lossy_corr}.

Moreover, it is worth noting that in some protocols, \emph{binning} the `no-click' outcomes is considered for security enhancement, which in the table notation corresponds to aggregating the rows and columns for the `no-click' events with the other proper outcomes. An example is provided by the CHSH protocol, which we analyze in the subsequent section.

%%%%%%%%%%%%%%%%%%%%%%%%%%%%%%%%%%%%%%%%%%%%%%%%%%%%%%%%%%%%%%%%%%%%%%%%%%%%%%%%%%%%%%%%%%%%%
%%%%%%%%%%%%%%%%%%%%%%%%%%%%%%%%%%%%%%%%%%%%%%%%%%%%%%%%%%%%%%%%%%%%%%%%%%%%%%%%%%%%%%%%%%%%%
\section{Applications to DIQKD protocols}
\label{sec:appl_to_DIQKD}
%%%%%%%%%%%%%%%%%%%%%%%%%%%%%%%%%%%%%%%%%%%%%%%%%%%%%%%%%%%%%%%%%%%%%%%%%%%%%%%%%%%%%%%%%%%%%
%%%%%%%%%%%%%%%%%%%%%%%%%%%%%%%%%%%%%%%%%%%%%%%%%%%%%%%%%%%%%%%%%%%%%%%%%%%%%%%%%%%%%%%%%%%%%
In the following, we apply the CC attack to derive upper bounds on the one-way and two-way key rates in noisy scenarios, i.e.~as functions of detection efficiency $\eta$ and visibility $V$, for a range of DIQKD protocols. Most importantly, as a result, we determine \emph{critical} visibilities $\crit{V}$ and detection efficiencies $\crit{\eta}$, below which our upper bounds on the key rates become negative and preclude a secure experimental realisation of a given protocol. As these critical values signify then 
lower bounds on minimal robustness parameters that the protocol can tolerate---below these values there exists an explicit attack, the CC attack, that invalidates the security---by comparing them with the ones obtained from the state-of-the-art security proofs, one can judge how much room there exists for potential improvement of the latter.

In order to apply the CC attack and upper-bound the key rate in a noisy one-way \eqref{eq:oneway_UB_ind_attack} or two-way \eqref{eq:r_twoway_UB_ind_attack} DIQKD protocol, one must first specify the correlation $\qAB(a, b|x,y)$ that would be shared by the parties in the absence of imperfections. The true noisy correlation being observed, $\pABobs(a,b | x,y)$, is then decomposed within the attack into the local and nonlocal parts. Although the local contribution is determined via the linear program~\eqref{eq:linear_program}, the eavesdropper must specify in advance the set of nonlocal correlations $\pvec^{\cNL}$ to be used within the convex decomposition. In this work, we choose $\pvec^{\cNL}$ to consist of only one correlation, namely, the noiseless $\qAB(a, b|x,y)$. We find this choice to be optimal for our purposes by heuristic methods, however, we leave it open whether the upper bounds on key rates derived under this choice can be further improved by performing a rigorous optimisation of $\pvec^\cNL$.

In this section, we consider the application of the CC attack to particular DIQKD protocols, which exhibit state-of-the-art robustness to noise. In particular, we summarise the experimentally relevant bounds on critical visibilities and detection efficiencies below which the protocols become vulnerable to the CC attack and thus insecure. As an example, an explicit derivation of the CC-based upper bound on the key rate and the resulting lower bounds on tolerable noise levels are presented for the CHSH-based protocol with deterministic binning of the non-detection events, while similar derivations applicable to the other protocols considered are relegated to the \nameref{sec:Appendices}.

%%%%%%%%%%%%%%%%%%%%%%%%%%%%%%%%%%%%%%%%%%%%%%%%%%%%%%%%%%%%%%%%%%%%%%%%%%%%%%%
\subsection{Protocols based on the CHSH violation}
\label{sec:chsh-based}
Within the canonical CHSH-based protocol~\cite{ABG+07,PAB+09}, the parties strive to obtain correlations maximally violating the CHSH inequality~\cite{CHSH69}. The value of the CHSH violation may then be used to construct a lower bound on the DW rate \eref{eq:DW_rate}~\cite{ABG+07,PAB+09,WAP20}, which, if the violation is high enough, may be positive and thus certify the possibility of distilling a secure cryptographic key. For this to be possible, Alice uses two binary-outcome measurements, labelled by her input $x \in \{0, 1\}$, while Bob uses three labelled by the inputs $y \in \{0, 1, 2\}$, corresponding to a scenario with $m_A = 2$ and $m_B = 3$. In each round they select their inputs randomly, and only the rounds with $x,y \in \{0, 1\}$ are used to estimate the CHSH violation, whereas only the rounds with $(x^*, y^*) \eqdef  (0, 2)$, i.e.~with the \emph{key settings} chosen, are used to distil the key. This formally constitutes the 2322-scenario%
\footnote{Recall that by $\mA\mB\nA\nB$ we denote a scenario with $\mA$ ($\mB$) measurement settings and $\nA$ ($\nB$) possible outcomes for each measurement setting on Alice's (Bob's) side.}, also when finite visibility ($V<1$ in \eqnref{eq:visibility}) is accounted for, while in case of imperfect detection ($\eta<1$ in \eqnref{eq:lossy_corr}) it becomes the 2333-scenario with the third extra outcome corresponding to the `no-click' event observed by any of the parties.

Within the canonical protocol~\cite{ABG+07,PAB+09} the security relies on the CHSH-scenario with binary outcomes. Therefore, in case of imperfect detection and the 2333-scenario, the standard technique used in security proofs is to have the parties bin the third `no-click' outputs, i.e.~assign them to one of the two `proper' measurement outcomes ($0$ or $1$), for the inputs $x,y \in \{0, 1\}$ used to estimate the CHSH-violation, with $x^*=0$ being also used by Alice to distil the key. Although the construction of the CC attack may be performed for any given correlation, it must include all the steps conducted within the protocol being considered, in particular, also the binning procedure. 

In what follows, we assume the typical choice of binning~\cite{WAP20,brown_device-independent_2021,HST+20,TLR20}, i.e.~the \emph{deterministic} assignment of all the 'no-clicks' $\varnothing$ to one of `proper' outcomes, say $0$, by each party. Nonetheless, within \nameref{sec:Appendices} we consider the option of not binning at all, as well as other binning strategies, which in combination with any preprocessing applied by Alice on her outcome for $x^*=0$, i.e.~$p_{A'|A}$ in \eqnref{eq:oneway_rate}, correspond to just instances of stochastic maps that may be further optimised over to determine a preprocessing-independent upper bound on the one-way key rate.

%

%%%%%%%%%%%%%%%%%%%%%%%%%%%%%%%%%%%%%%%%%
\subsubsection{Generating the observed non-local correlations.}
\label{subsec:gen_non-loc_corr}

Here, we study a family of protocols inspired by the original CHSH construction and adopt the convention of~\cite{WAP20}, in which the ideal correlation $\qAB(a, b|x, y)$ available to the parties should be understood as the one obtained by them when sharing a \emph{partially entangled state} of two qubits:
\begin{equation}
|\psi_{\theta}\rangle \eqdef \cos\left(\frac{\theta}{2}\right)|00\rangle+\sin\left(\frac{\theta}{2}\right)|11\rangle,
\label{eq:part_ent_state}
\end{equation}
with the measurements of Alice and Bob, $M^x_a$ and $M^y_b$ in \eqnref{eq:correlation}, corresponding to eigenstate projectors of the dichotomic observables $A_x$ and $B_y$, respectively:~$A_0 = B_2 = \sigma_z$ for the key settings $(x^*,y^*)$, while $A_1$ and $B_{0/1}$ are chosen to maximise the CHSH-type functional:
\begin{IEEEeqnarray}{rCl}
S_\t{det}(\eta,\theta) & \eqdef  &\eta^{2}\left\langle B_{0}(A_{0}+A_{1})+B_{1}(A_{0}-A_{1})\right\rangle \nonumber\\&& + \,2 \eta \neta\left\langle A_{0}+B_{0}\right\rangle +2\neta^{2},
\label{eq:S_det}
\end{IEEEeqnarray}
which accounts already for the finite detection efficiency, $\eta < 1$ in \eqnref{eq:lossy_corr}, and assumes deterministic binning of the `no-click' events. Note that, due to the linearity of the expression \eref{eq:S_det}, the above choice of measurements remains optimal when also the finite visibility, $V<1$ in \eqnref{eq:rhoV}, is considered\footnote{With $\langle X \rangle=\tr{\psi_\theta X}$ in \eqnref{eq:S_det} being then replaced by $\langle X \rangle=\tr{\left[V\psi_\theta+(1-V)\frac{\openone}{4}\right] X}$ according to \eqnref{eq:rhoV}.}.

%%%%%%%%%%%%%%%%%%%%%%%%%%%%%%%%%%%%%%%%
\subsection{One-way CHSH protocols involving maximally entangled states}
\label{sec:DIQKD_max_ent}

\subsubsection{Finite detection efficiency}

Firstly, we sketch the calculation of the upper bound on the one-way key rate~\eref{eq:oneway_UB_ind_attack} based on the CC attack (see Apps.~\ref{sec:key_rates_CC} and \ref{sec:UBs_thresholds_CC} for a more detailed derivation) for the above CHSH-based protocol with finite detection efficiency $\eta$ and deterministic binning.
%in which the parties deterministically bin the non-detection events $\varnothing$ to the output $0$ and Alice doesn't make use of the publicly announced variable $M$. 
For protocols in which Alice does not announce publicly any variable $M$ and bins her key setting outcome $A$ deterministically, the bound~\eqref{eq:oneway_UB_ind_attack} reads
\begin{IEEEeqnarray}{rCl}
\roneway{det}(A\to B|A') &\le& H(A'|E)_\t{det} - H(A'|B)_\t{det},
\IEEEeqnarraynumspace
\label{eq:r_1way_det}
\end{IEEEeqnarray}
where the binary variable $A'$ is obtained by transforming the ternary outcome $A$ of Alice's measurement with $x^*=0$ by the stochastic map
\begin{equation}
\label{eq:det_bin}
\mathcal{S_{\text{det}}} \eqdef \left(\begin{array}{ccc}
1 & 0 & 1\\
0 & 1 & 0
\end{array}\right),
\end{equation}
responsible for binning the `no-click' events (last column) deterministically onto the `0' outcome. For instance, the marginal probability of Alice then reads
\begin{equation}
\pobs_{A'}(a'|x^*) = \sum_a \mathcal{S}_{\text{det}}(a'|a) \; \pobs_{A}(a|x^*),
\end{equation}
describing now the distribution of $A'$ rather than $A$. Although it is the full correlation $\pABobs(a, b|x,y)$ that determines the value of the local weight $q^\cL$ within the CC attack (see \eqnref{eq:linear_program}), the key is distilled only from the $( x^*, y^* )$-rounds. As a consequence, both entropies in \eqnref{eq:r_1way_det} are computed for the key settings and we may drop for convenience the conditioning on $( x^*, y^* )$ in all the following expressions, so that, e.g.,~$\mathsf{Q}_{ab} \eqdef  \mathsf{Q}_{ab}^{x^* y^*}\!$, $\QA{a} \eqdef  \mathsf{Q}_a^{x^*\!}$, $\QB{b} \eqdef  \mathsf{Q}_b^{y^*}$, or $\pobs_{AB}(a,b) \eqdef \pobs_{AB}(a,b|x^*, y^*)$.

We calculate first the EC-term $H(A'|B)_\trm{det}$ in \eqnref{eq:r_1way_det}, which depends solely on the correlation being observed. After applying the stochastic map~\eqref{eq:det_bin} on the outcome of Alice in \eqnref{eq:lossy_corr}, we obtain the resulting shared correlation as
\begin{IEEEeqnarray}{rCl}
\label{eq:simplified_lossy_det}
\pobs_{A'B}(a',b) &=&
\begin{array}{|c|c|c|c|}
\hline a'\setminus b & 0 & 1 & \varnothing\\
\hline 0 & 
\begin{tabular}{@{}c@{}}
$\eta^{2}\,\mathsf{Q}_{00}$\\$+\eta\neta \QB{0}$
\end{tabular} & 
\begin{tabular}{@{}c@{}}
$\eta^{2}\,\mathsf{Q}_{01}$\\$+\eta\neta \QB{1}$
\end{tabular} & 
\begin{tabular}{@{}c@{}}
$\neta \eta\QA{0}$\\$+\neta^{2}$
\end{tabular}\\ \hline 
1 & \eta^{2}\,\mathsf{Q}_{10} & \eta^{2}\,\mathsf{Q}_{11} & \eta\neta \QA{1} \\ \hline 
\end{array}
\,,\IEEEeqnarraynumspace
\end{IEEEeqnarray}
whose first row is obtained by summing the first and third rows in \eqnref{eq:lossy_corr}. Now, the conditional probability distribution of Alice is obtained by dividing the columns in \eqnref{eq:simplified_lossy_det} by the corresponding marginal probabilities of Bob, $(\pobs_B(0), \pobs_B(1), \pobs_B(\varnothing)) = (\eta \mathsf{Q}_0^{B}, \eta \mathsf{Q}_1^{B}, \neta)$ obtained by summing the columns in \eqnref{eq:simplified_lossy_det}, i.e.:\footnote{Note that although it leads to an easy calculation of the formula for the conditional entropy $H(A'|B)$, this division, and hence Eq.~\eqref{eq:first_cond_prob}, is incorrect for the extremal values $\eta = 0$ and $\eta = 1$ due to dividing by $0$. In these edge cases, to calculate $H(A'|B)$ one should rather use the chain rule $H(A'|B) = H(A', B) - H(B)$ on the joint probability distribution~\eqref{eq:simplified_lossy_det}, which ultimately yields the same result as in Eq.~\eqref{eq:hab_det}. This applies as well to other instances of conditional entropy calculated in the paper.}
\begin{multline}
\label{eq:first_cond_prob}
\pobs_{A'|B}(a'|b)=\frac{\pobs_{A'B}(a',b)}{\pobs_{B}(b)} \\
= \begin{array}{|c|c|c|c|}
\hline a'\setminus b & 0 & 1 & \varnothing\\
\hline 0 & 
\begin{tabular}{@{}c@{}}
$\eta\,\frac{\mathsf{Q}_{00}}{\QB{0}}+ \neta$\end{tabular} & \begin{tabular}{@{}c@{}} $\eta\,\frac{\mathsf{Q}_{01}}{\QB{1}} + \neta$\end{tabular} & \begin{tabular}{@{}c@{}} $\eta\QA{0}+\neta$\end{tabular}\\
\hline 1 & \eta\,\frac{\mathsf{Q}_{10}}{\QB{0}} &\eta\,\frac{\mathsf{Q}_{11}}{\QB{1}} & \eta\QA{1}\\\hline 
\end{array}\,.
\end{multline}
As a result, we may directly compute the relevant conditional entropy as
\begin{IEEEeqnarray}{l}
H(A'|B)_{\text{det}}= \sum_b \pobs_\sub{B}(b) H(A'|B=b) \label{eq:hab_det} \\ 
\qquad = \eta\,\QB{0}\,h\!\left[\frac{\eta\,\mathsf{Q}_{10}}{\QB{0}}\right]+\eta\,\QB{1}\,h\!\left[\frac{\eta\,\mathsf{Q}_{11}}{\QB{1}}\right]+\neta\,h\!\left[\eta\QA{1}\right]\!,
\nonumber
\end{IEEEeqnarray}
where $h[x]\eqdef -x\log_{2}x-(1-x)\log_{2}(1-x)$ is the binary entropy function. 

On the contrary, in order to determine the PA-term $H(A'|E)_\trm{det}$ in~\eqref{eq:r_1way_det}, one needs to find the value of $q^\cL$ for the specified correlation.
%, signifying the percentage of rounds in which Eve distributes deterministic correlations to Alice and Bob and knows their outputs $a$ and $b$. 
This can be done by means of linear programming, but often also analytically, as discussed in~\secref{sec:CC}. Let us also recall that whenever Eve distributes a local correlation within the CC attack, she possesses full knowledge about the outcome of Alice, $A$, and hence of $A'$, since it is obtained via a deterministic transformation. Therefore, the PA term is completely determined by the non-local rounds in which the noiseless correlation $Q(a,b|x,y)$ is distributed, so that
\begin{equation}
H(A'|E)_\trm{det} = (1-q^\cL) h[\mathsf{Q}_0^{\mathrm{A}}]
\label{eq:hae_det}
\end{equation}
corresponds to the entropy of the marginal $\QA{a}$ multiplied by the probability of a round being nonlocal.

Finally, we obtain an upper bound on the key rate as a function of detection efficiency $\eta$ by subtracting the EC-term \eqref{eq:hab_det} from the PA-term \eqref{eq:hae_det}, i.e.:
\begin{IEEEeqnarray}{l}
\label{eq:rdet_prefinal}
\roneway{det}
(A\to B|A') \le (1-q^\cL) h[\mathsf{Q}_0^{\mathrm{A}}] \\ 
\qquad 
- \eta\,\QB{0}\,h\!\left[\frac{\eta\,\mathsf{Q}_{10}}{\QB{0}}\right]\! - \eta\,\QB{1}\,h\!\left[\frac{\eta\,\mathsf{Q}_{11}}{\QB{1}}\right]\!-\neta\,h\!\left[\eta\QA{1}\right]\!,
\nonumber
\end{IEEEeqnarray}
which applies for any correlation \eref{eq:lossy_corr}, given the deterministic binning of no-clicks and one-way communication in the protocol.

However, recall that the CC-based upper bound \eref{eq:rdet_prefinal} requires the local weight $q^\cL$ to be determined for a particular $\pABobs$. We present first the solution when Alice and Bob share a maximally entangled Bell state, $\ket{\Phi^+}$, i.e.~set $\theta = \pi/2$ in~\eqnref{eq:part_ent_state}, which yields $\mathsf{Q}_{ab} = \delta_{ab}/2$ and $\QA{a} = \QB{b} = 1/2$ for the key measurements $A_0 = B_2 = \sigma_z$.
The other measurements maximising the expression \eref{eq:S_det} turn out then to be the
standard CHSH-optimal observables, i.e.
\begin{IEEEeqnarray}{c}
\label{eq:standard_measurements}
%A_0 = B_2 = \sigma_z,\; 
A_1 = \sigma_x, 
\quad
B_{0/1} = \frac{1}{\sqrt{2}}\left( \sigma_z \pm  \sigma_x\right).
\IEEEeqnarraynumspace
\end{IEEEeqnarray}
We show in \appref{app:qL_maxent_detbin_eta} how to determine then the maximal local weight analytically, which reads
\begin{equation}
\label{eq:qL_maxCHSH_eta}
q^\cL = (1-\eta)\left(1 + \left(3 + 2 \sqrt2 \right) \eta\right) \quad \text{for} \quad \eta > \eta_\text{loc} 
\end{equation}
where $\eta_\trm{loc}\eqdef2(\sqrt2 - 1) \approx 82.8\%$ is the detection efficiency below which the resulting correlation \eref{eq:lossy_corr} becomes local~\cite{E93,M86} and, thus, disallows any DIQKD to be possible.
Finally, we can write \eqnref{eq:rdet_prefinal} as
\begin{align}
\label{eq:rdet_final}
& \roneway{det}(A\to B|A') \le \left(3+2 \sqrt2 \right)\eta^2 \\ 
& \qquad\qquad
- 2\left(1+\sqrt2\right)\eta  - \frac{\eta}{2} \, h\!\left[\eta \right]\!-\neta\,h\!\left[\frac{\eta}{2}\right]\!,
\nonumber
\end{align}
which becomes negative below $\crit{\eta} \approx 89.16\%$. This formally demonstrates that for detection efficiencies $\eta_\trm{loc}\le\eta\le\crit{\eta}$, no positive key is possible despite the correlation \eref{eq:lossy_corr} being non-local~\cite{Farkas2021}. We include $\crit{\eta}\approx 89.16\%$ in \tabref{tab:critical} (see the penultimate column for the 2333-scenario) presenting it against the best-known efficiency threshold, $\UB{\DW{\eta}} \approx 90.78\%$, above which the DW rate \eref{eq:DW_rate} is assured to be positive~\cite{WAP20}. Hence, it follows that the true\footnote{\label{footnote_detbin}Given deterministic binning of `no-click' events.} DW-threshold fulfils $89.16\%\le\DW{\eta}\le90.78\%$, with the CC attack leaving less than $2\%$ for the improvement of $\UB{\DW{\eta}}$ by devising stronger lower bounds on the DW rate. 

Note that the above efficiency window applies when considering the most general eavesdropping attacks. On one hand, the DW rate~\eref{eq:DW_rate} is valid for coherent attacks despite the deterministic binning, as the EAT still holds, see the discussion above~\eqnref{eq:oneway_UB_ind_attack}. On the other, as we consider a particular attack, by improving its strength $\crit{\eta}$ can only be increased.

\begin{figure}[t!]
\centering
\includegraphics[width=\columnwidth]{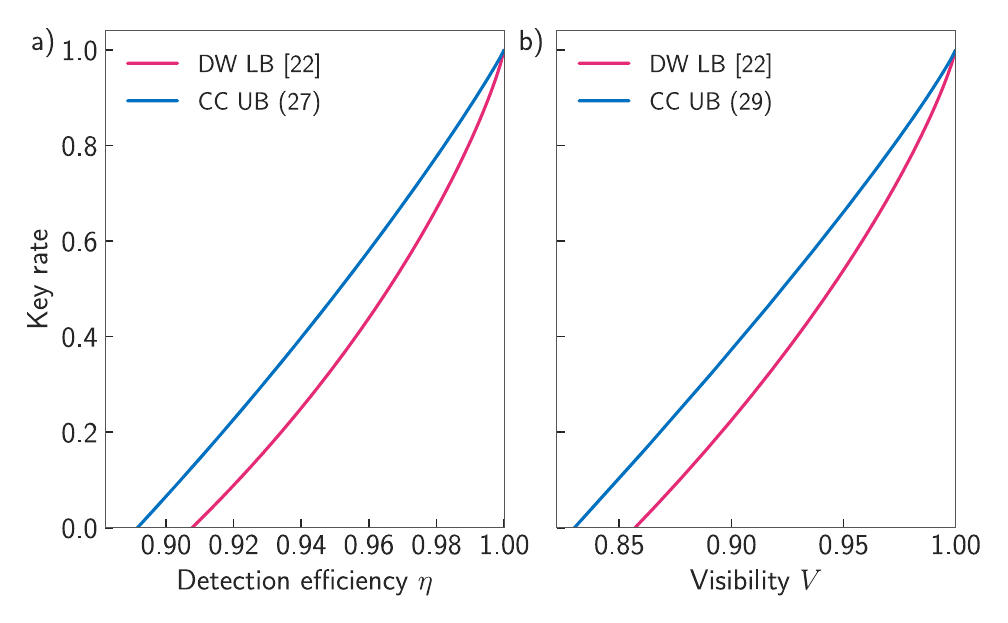}
\caption{\textbf{CC-based upper bounds vs lower bounds on the DW rate~\eqref{eq:DW_rate} for the CHSH-based protocols involving maximally entangled states}, as a function of (a):~the detection efficiency $\eta$ in the 2333 scenario; and (b):~the visibility $V$ in the 2322 scenario. \emph{Blue lines} correspond to the analytic CC-based upper bounds on key rates~\eqref{eq:rdet_final} and \eqref{eq:r_oneway_UB_V}, whereas \emph{red lines} are the corresponding lower bounds on the DW rate~\eqref{eq:DW_rate} derived in~\citeref{WAP20}. The points at which the curves cross the zero in (a) and (b) are the critical detection efficiencies and visibilities cited for the 2333- and 2322-scenarios, respectively, in the \emph{`none'} column of~\tabref{tab:critical}.
}
\label{fig:woodhead-lb-comp}
\end{figure}

In \figref{fig:woodhead-lb-comp}a) we explicitly compare the upper bound~\eqref{eq:rdet_final} with the analytic lower bound on the DW rate~\eqref{eq:DW_rate} established in~\citeref{WAP20} as a function of $\eta$. It can be seen that the CC-based upper bound remains relatively tight in the whole region of positive key rates, with the maximal difference between the two bounds never exceeding $0.15$ of a bit per round.

\subsubsection{Finite visibility}

From the above analysis, it is now straightforward to determine the CC-based upper bound on the key rate if instead the finite visibility ($V<1$) is considered. It is obtained by letting $\eta=1$ within the EC-term \eref{eq:hab_det} and replacing therein the noiseless correlation $\mathsf{Q}_{ab}$ with $\mathsf{P}_{ab} \eqdef V \mathsf{Q}_{ab} + (1-V)/4$ in accordance with \eqnref{eq:visibility}. On the other hand, the PA term \eref{eq:hae_det} is left intact, as Eve again distributes the noiseless correlation $\mathsf{Q}_{ab}$ within the non-local rounds of our CC attack. Hence, focusing again on $\mathsf{Q}_{ab}$ maximally violating the CHSH inequality, we have $\mathsf{P}_{ab} = (V/2) \delta_{ab} + (1-V)/4$ and $\PA{a} = \PB{b} = 1/2$ for $( x^*, y^* )$, while measurements \eref{eq:standard_measurements} remain optimal for the other settings. The complete correlation $\pABobs(a,b | x,y)$ of \eqnref{eq:visibility} constrains then the maximal local weight to (see \cite{Farkas2021} and \appref{app:ql_maxent_V}):
\begin{equation}
\label{eq:visibility_local_weight0}
	q^\cL = (1-V)/(1-1/\sqrt{2}) \quad\text{for}\quad V>1/\sqrt{2},
\end{equation}
with $1/\sqrt{2}$ being the well-known locality threshold for Werner states. As a result, we obtain for this 2322-scenario%
\footnote{For finite visibility ($V \leq 1$) but perfect detection efficiency ($\eta=1$), no binning strategy is required due to outcomes of measurements remaining binary.}\addtocounter{footnote}{-1}\addtocounter{Hfootnote}{-1} 
the following CC-based upper bound:
\begin{equation}
\roneway (A\to B) \le V (2 + \sqrt2) - h\!\left[ \frac{1+V}{2} \right] - \sqrt2 - 1,
\label{eq:r_oneway_UB_V}
\end{equation}
which ceases to be positive at $\crit{V} \approx 83.00\%$. Hence, given the best-known value of $V$ above which the DW rate is positive, $\UB{\DW{V}}\approx85.70\%$~\cite{WAP20}, the CC attack constrains narrowly any further improvement of this threshold to about $2.7\%$, i.e.~$83.00\%\le\DW{V}\le85.70\%$, see the first row of column `\emph{none}' in \tabref{tab:critical}. 
Again, the above window in which the true visibility threshold lies is valid for coherent attacks of the eavesdropper, by the same arguments as discussed in the finite efficiency case.

Furthermore, we explicitly compare in \figref{fig:woodhead-lb-comp}b) the upper bound~\eqref{eq:r_oneway_UB_V} with the corresponding analytic lower bound on the DW rate~\eqref{eq:DW_rate} found in~\citeref{WAP20} as a function of $V$. Similarly to the case of finite detection efficiency and \eqnref{eq:rdet_final}, we observe that the CC-based upper bound remains relatively tight in the whole region of positive key rates, and the difference between the two bounds again does not exceed $0.15$ of a bit per round.

\begin{table*}[t]\centering
\begin{tabular}{@{}lccccccc@{}}
\multicolumn{8}{c}{\textbf{DIQKD protocols involving maximally entangled states}} \\
\midrule[\heavyrulewidth]
 & \multicolumn{2}{c}{\textbf{two-way}} &  \multicolumn{5}{c}{\textbf{one-way}} \\ 
\cmidrule[\heavyrulewidth](lr){2-3} \cmidrule[\heavyrulewidth](lr){4-8} 
\textit{Preprocessing:} & &  & \textit{any} &\multicolumn{2}{c}{\textit{noisy}}  & \multicolumn{2}{c}{\textit{none}} \\ 
\cmidrule(lr){4-4} \cmidrule(lr){5-6} \cmidrule(lr){7-8}
\midrule
&\multicolumn{7}{c}{%uniform noise model (
Finite visibility $V$}
\\
\cmidrule(lr){2-8}
Scenario:& $\crit{V}$ & $\UB{\ad{V}}$ & $\critmin{V}$ & $\crit{V}$ & $\UB{\DW{V}}$ & $\crit{V}$ & $\UB{\DW{V}}$\\
2322 & $74.45$ & $88.0$ & $80.85$ & $80.85$ & $83.83$ & $83.00$ & $85.70$\\
2222 & $78.36$ & $84.6$ & $88.52$ & $88.52$ & $92.38$ & $90.61$ & $93.76$\\
\midrule
&\multicolumn{7}{c}{%losses model (
Finite detection efficiency $\eta$} \\
\cmidrule(lr){2-8}
Scenario:& $\crit{\eta}$ & $\UB{\ad{\eta}}$ & $\critmin{\eta}$ & $\crit{\eta}$ & $\UB{\DW{\eta}}$ 
& $\crit{\eta}$ & $\UB{\DW{\eta}}$\\
2333 & $85.36^{*}$ & $93.7$ & $85.36$ & $88.52^*$ & $90.30$ & $89.16^*$ & $90.78$\\
2233 & $87.87^{*}$ & $91.7$ & $92.64$ & $93.59^*$ & $95.84$ & $94.80^*$ & $96.62$\\
\midrule
\multicolumn{8}{r}{\scriptsize{* with deterministic binning of `no-click' events.}} \\
%\bottomrule
\end{tabular}
\caption{\textbf{Critical visibilities $\crit{V}$ and detection efficiencies $\crit{\eta}$} (in \%) derived with help of the CC attack, below which no DIQKD protocol can be made secure when relying on measurements of maximally entangled states within each of the $\mA\mB\nA\nB$-scenarios listed. %in the first column. 
The left-most critical values apply to all DIQKD protocols and are compared against the thresholds attained by two-way protocols involving advantage distillation (a.d.)~\cite{TLR20}. The critical noise parameters are tightened for one-way protocols, for which a stricter upper bound on key rate \eref{eq:oneway_UB_ind_attack} applies and varies between various strategies of data preprocessing:~\emph{any} (found by heuristic search), \emph{noisy}~\cite{WAP20,HST+20,SBV+20} or \emph{none}. The latter two cases are compared against the thresholds determined by lower-bounding the Devetak-Winter (DW) rate \eref{eq:DW_rate}, which can be computed for all the scenarios considered~\cite{WAP20}. All the stated values are valid in presence of coherent attacks, apart from the a.d.-based thresholds~\cite{TLR20} (second column) that are derived in presence of collective attacks only.}
\label{tab:critical}
\end{table*}

\subsubsection{22- scenarios}

We further repeat the above derivation for the less favourable situation in which Bob, similarly to Alice, uses the same measurement setting to distil the key as for the CHSH violation. This yields then 2233- and 2222-scenarios for finite detection efficiency and visibility, respectively, with $\mathsf{Q}_{ab}=(2+(-1)^{a\oplus b}\sqrt{2})/8$. Following the procedure of \citeref{WAP20}, we compute the thresholds above which the DW rate is guaranteed to be positive and include them in \tabref{tab:critical}. These are then higher
but so are the CC-based critical values (see also \appref{app:1way_maxent_UBs})%
---with the CC attack proving itself again to be very effective.

\subsubsection{Noisy preprocessing}

On the other hand, by performing noisy preprocessing (random bit-flip) of her key-setting outcome~\cite{RGK05}, Alice may improve the robustness of the DIQKD protocol~\cite{HST+20,WAP20}, with the tolerable noise-levels dropping then to $\UB{\DW{\eta}}\approx90.30\%$ and $\UB{\DW{V}}\approx83.83\%$~\cite{WAP20},
see the the column `\emph{noisy}' in \tabref{tab:critical} for the corresponding 2333- and 2322-scenarios, respectively (and similarly for the 2233- and 2222-scenarios). However, after incorporating the noisy preprocessing step into the CC attack---see \appref{app:1way_maxent_UBs} for analytic expressions, also for the setting in which Alice and Bob bin their `no-click' outcomes randomly---we obtain strict lower bounds on the tolerable noise-thresholds for the DW rate\footnoteref{footnote_detbin} as  
$\crit{\eta}\approx88.52\%$ and $\crit{V}\approx80.85\%$,
which again leave only a couple of percent for potential improvement.

\subsubsection{Arbitrary preprocessing}

Crucially, any binning of `no-clicks' or noisy preprocessing strategy constitutes just a special case of a stochastic map $A \to A'$ in \eqnref{eq:oneway_rate}, while the CC attack allows us, in fact, to determine an upper bound on the one-way key rate \eref{eq:oneway_rate} that applies for \emph{any preprocessing}. In particular, by resorting to heuristic methods (see \appref{app:1way_maxent_opt_maps} for details) we maximise the upper bound \eref{eq:oneway_UB_ind_attack} over all maps $A \to A'$ and $A \to M$, as in \eqnref{eq:oneway_rate}, so that it yields critical thresholds on detection efficiency and visibility, $\critmin{\eta}$ and $\critmin{V}$, that are universally valid for any one-way protocol given a particular correlation $\pABobs$. 

As shown in the column `\emph{any}' of \tabref{tab:critical}, we observe that for both 2322- and 2222-scenarios the universal values $\critmin{V}$ coincide with the CC-based critical visibilities obtained for noisy preprocessing, which can thus be considered optimal in terms of robustness against the CC attack.
For finite detection efficiency, we are able to find analytically the minimal $\critmin{\eta}=\frac14 \left( 2 + \sqrt{2} \right)\approx85.36\%$ for the 2333-scenario, as it is attained by a preprocessing strategy in which only the map $A \to M$ in \eqnref{eq:oneway_rate} is required, with Alice publicly announcing in each round under the binary variable $M$ whether or not she observes a `no-click'. From the perspective of our CC attack, as `no-clicks' may happen only in the rounds in which Eve distributes a local correlation and has perfect knowledge of $A$, announcing $M$ does not provide any ``extra'' information to Eve, but helps Bob perform the EC---diminishing $H(A|B)$. 
In the 2233-scenario the situation is slightly different:~although a similar strategy of signalling `no-clicks' proves better than deterministic binning followed by noisy preprocessing, the optimal preprocessing that is most robust against the CC attack, yielding $\critmin{\eta}\approx92.64\%$, actually corresponds to randomly binning the `no-click' events before performing noisy preprocessing. We elaborate on these findings in \appref{app:1way_maxent_opt_maps}.

\subsection{Two-way CHSH protocols involving maximally entangled states}
%%%%%
\label{sec:DIQKD_max_ent_2way}
\subsubsection{Finite detection efficiency}
In order to establish critical noise thresholds that hold for \emph{all DIQKD protocols}, i.e.~also ones exploiting two-way communication, we resort to the CC-based upper bound \eref{eq:r_twoway_UB_ind_attack} based on the intrinsic information \eref{eq:intrinsic_info}. However, we write it in terms of the conditional mutual information as
\begin{equation}
  \rtwoway(A\leftrightarrow B) \;\le\;  I(A \!:\!B | F),
  \label{eq:r_twoway_UB_mapEF}
\end{equation}
for any given mapping $p(F|E)$ that Eve applies on her random variable, $E\to F$, within the CC attack. 

As in the one-way case, we consider firstly the noisy correlations \eref{eq:lossy_corr} that incorporate finite detection efficiency $\eta<1$, with $\mathsf{Q}_{ab}^{xy}$ violating again maximally the CHSH inequality. However, we are then unable to find (via an extensive numerical search) a map such that $I(A \!:\!B | F)=0$ for any $\eta>\eta_\trm{loc}\approx 82.8\%$, as defined below \eqnref{eq:qL_maxCHSH_eta}. In particular, for every choice of $p(F|E)$ we make, the upper bound \eref{eq:r_twoway_UB_mapEF} can be made vanishing only trivially, i.e.~when the noisy correlation \eref{eq:lossy_corr} becomes local. 

We then consider a simpler version of the protocol in which, again, Alice and Bob bin deterministically their `no-click' outcomes before performing any two-way processing of their bits, which is the case in current two-way DIQKD protocols involving an \emph{advantage distillation} (a.d.) procedure~\cite{TLR20}. Then, by using the map $p(F|E)$ proposed by us in~\cite{Farkas2021}, see also \appref{app:2way_maxent_UBs_eta}, we arrive at an upper bound \eref{eq:r_twoway_UB_mapEF} of the form:
\begin{align}
& \rtwoway{det}(A'\leftrightarrow B') \;\le\;
  \tilde{p} \sum_{a' \ne b'} h\!\left[ \frac{q^\cNL \mathsf{Q}_{00} + \mathsf{Q}_{a'b'}^\eta}{\tilde{p}} \right] \nonumber \\
& \quad\quad
- \tilde{p} \; H\!\left\{\frac{\mathsf{Q}_{01}^\eta}{\tilde{p}},\frac{\mathsf{Q}_{10}^\eta}{\tilde{p}}, \frac{q^\cNL \mathsf{Q}_{00}}{\tilde{p}}, \frac{q^\cNL \mathsf{Q}_{11}}{\tilde{p}} \right\}
  \label{eq:r_twoway_UB_detbin_Qab} 
  \IEEEeqnarraynumspace
\end{align}
where $A'$ ($B'$) is now the key-setting outcome of Alice (Bob) after deterministic binning of `no-clicks', $\mathsf{Q}_{ab}^\eta\eqdef  \eta\mathsf{Q}_{ab} + \eta\neta \mathsf{Q}_{11}$, 
$\tilde{p} \eqdef 
\eta+2\eta\neta\mathsf{Q}_{11}+(q^\cNL-\eta) (\mathsf{Q}_{00}+\mathsf{Q}_{11})$,
and we define the entropy of any probability vector $(p_i)_i$, satisfying $\forall_i\!:\,p_i\ge0$ and $\sum_i p_i = 1$, as $H\!\left\{ (p_{i})_i\right\} \eqdef-\sum_{i}p_{i}\log_2 p_{i}$. Substituting for the correlation $\mathsf{Q}_{ab}^{xy}$ that yields:~maximal CHSH violation, $\mathsf{Q}_{ab} = \delta_{ab}/2$, and the maximal local weight \eref{eq:qL_maxCHSH_eta}; we obtain the two-way equivalent of \eqnref{eq:rdet_final} as 
\begin{IEEEeqnarray}{C}
  \rtwoway{det}(A'\leftrightarrow B') \;\le\; \eta \left( 2 \left (1+\sqrt2 \right)\eta -2 \sqrt2 - 1\right) \nonumber \\
   \qquad\times \left( 1 - h\left[\frac{\neta }{1-2 \left(1+\sqrt2\right)\neta}\right]\right)\!,
   \IEEEeqnarraynumspace
  \label{eq:r_twoway_UB_detbin_eta}
\end{IEEEeqnarray}
which exhibits a zero at $\crit{\eta}=\frac14\left(2 + \sqrt{2}\right)\approx85.36\%$. 

Hence, by convexity of the two-way upper bound \eref{eq:r_twoway_UB_ind_attack} (see \cite{Farkas2021} and \appref{sec:ind_attack_explicit}), this implies that $\rtwoway{det}=0$ for any $\eta_\trm{loc}\approx 82.8\%\le\eta\le\crit{\eta}$, and no DIQKD protocol is possible\footnoteref{footnote_detbin} within this range of $\eta$ despite the shared correlation being non-local. We include the above CC-based $\crit{\eta}$ in the first column of \tabref{tab:critical}, where it consistently lower-bounds all the best-known tolerable detection efficiencies derived for one-way protocols\footnoteref{footnote_detbin}, $\UB{\DW{\eta}}$~\cite{WAP20}, as well as for two-way protocols involving the a.d.~procedure, $\UB{\ad{\eta}}$~\cite{TLR20}%
\footnote{Note that the two-way advantage distillation (a.d.) procedure~\cite{TLR20} is superior over the one-way protocol with noisy preprocessing~\cite{WKB+20} only for the $2222$ and $2233$ scenarios considered in \tabref{tab:critical}, for which indeed $\UB{\ad{V}}\!<\UB{\DW{V}}$ and $\UB{\ad{\eta}}\!<\UB{\DW{\eta}}$.}. 
Moreover, it coincides exactly with $\critmin{\eta}$, so that for the above $2333$-scenario no two-way protocol\footnoteref{footnote_detbin} may be more robust against our CC attack than the one-way protocol with optimal preprocessing. This is not the case for the $2233$-scenario with $x^*, y^*\in\{0,1\}$ and $\mathsf{Q}_{ab}=(2+(-1)^{a\oplus b}\sqrt{2})/8$, which upon being substituted into \eqnref{eq:r_twoway_UB_detbin_Qab} leads to $\crit{\eta}=3(1-1/\sqrt{2})\approx87.87\%$ that, however, is only $4\%$ away from best-known threshold $\UB{\ad{\eta}}\approx91.7\%$~\cite{TLR20}. The derivation can be found in \appref{app:2way_maxent_UBs_eta}, where we also deal with the special case of $x^*=y^*=1$, for which a different $p(F|E)$ must be chosen for the upper bound \eref{eq:r_twoway_UB_mapEF} to provide a non-trivial $\crit{\eta}\approx87.47\%\,(>\!\eta_\trm{loc})$.

\subsubsection{Finite visibility}

As in the case of one-way protocols, we repeat the above construction when finite visibility ($V<1$) is considered instead, and no binning procedure is necessary. However, this corresponds to the special case of correlations considered by us already in \cite{Farkas2021} (see Eq.~(14) therein with $\theta=\pi/4$), which leads to
\begin{IEEEeqnarray}{C}
  \label{eq:r_twoway_UB_V}
  \rtwoway(A\leftrightarrow B) \;\le\;  \frac{\left(2+\sqrt{2}\right) V-3 \sqrt{2}+2}{2 \left(2-\sqrt{2}\right)} \nonumber \\
  \quad\qquad \times \left(1-h\left[\frac{4 V-2 \sqrt{2}}{\left(2+\sqrt{2}\right) V-3 \sqrt{2}+2}\right]\right)
\IEEEeqnarraynumspace
\end{IEEEeqnarray}
that ceases to be positive below $\crit{V} \approx 74.45\%$ (see the top-left entry of \tabref{tab:critical}). As a result, within the range $1/\sqrt{2}\le V\le\crit{V}$ there is strictly \emph{no} possibility for any standard DIQKD protocol to yield positive keys, while the correlations remain non-local~\cite{Farkas2021}. We also construct the equivalent of the upper bound \eref{eq:r_twoway_UB_V} for the 2222-scenario, in which case $\crit{V} \approx 78.36\%$---see \appref{app:2way_maxent_UBs_V} but also \tabref{tab:critical} where the threshold value for the a.d.-protocol is also listed, $\UB{\ad{V}}\approx84.6\%$~\cite{TLR20}, which according to the CC attack may thus be improved only by at most $\approx\!6\%$.

%%%%%%%%%%%%%%%%%%%%%%%%%%%%%%%%%%%%%%%%%%%%%%%%%%%%%%%%%%%%%%%%%%%%%%%%%%%%%%%
\subsection{One-way CHSH protocols involving partially entangled states}
\label{sec:DIQKD_part_ent}
Since the seminal work of Eberhard \cite{E93} it is well known that in order to get the highest robustness to finite detection efficiency in observing Bell-violation, one should consider correlations obtained by measuring partially entangled states, i.e.~as in \eqnref{eq:part_ent_state} with $\theta\neq\pi/2$ and, in particular, in the limit $\theta\to0$. On the contrary, this is not the case when finite visibility $V<1$ is considered instead, as then setting $\theta = \pi/2$ always yields the highest CHSH violation. That is why, we repeat the above one-way key analysis for $\eta<1$ where (as already stated in \secref{subsec:gen_non-loc_corr}) we choose the measurements of Alice and Bob such that the CHSH functional \eref{eq:S_det} is maximised for a given value of $\eta$ and $\theta$. Each maximal value of the functional, on the other hand, allows us then to directly compute valid\footnoteref{footnote_detbin} lower-bounds on the attainable DW rate \eref{eq:DW_rate}, also when accounting for noisy preprocessing~\cite{WAP20}.

\begin{figure}[t!]
\centering
\includegraphics[width=\columnwidth]{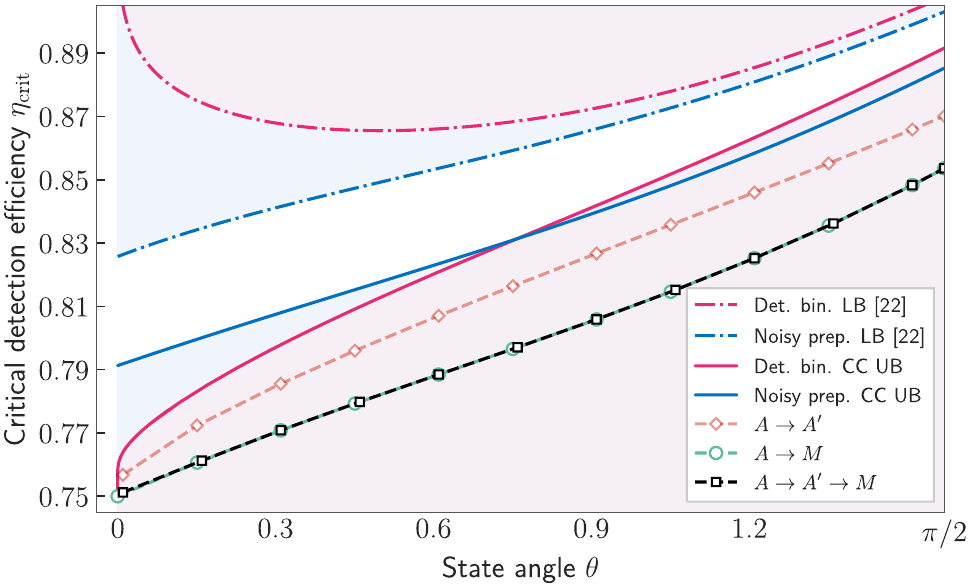}
\caption{\textbf{Critical detection efficiencies as a function of the $\theta$-angle parametrising the partially entangled state} involved in a one-way DIQKD protocol. \emph{Dot-dashed curves} describe thresholds, $\UB{\DW{\eta}}$, above which the DW rate \eref{eq:DW_rate} is guaranteed to be positive~\cite{WAP20}, while \emph{solid curves} denote critical values, $\crit{\eta}$, below which the CC attack excludes the possibility of key distillation; in both cases the inconclusive outcomes are binned deterministically without (\emph{red}) or with (\emph{blue}) inclusion of noisy (with $\pp\to1/2$) preprocessing. The critical efficiencies may be further diminished by optimising heuristically over the preprocessing strategies, i.e.~stochastic maps in \eqnref{eq:oneway_rate} that generally encompass the operations of Alice:~manipulating somehow her ternary output ($A\to A'$), publicly announcing some form of her preprocessed variable ($A\to M$), or both ($A\to A'\to M$).}
\label{fig:partial_thresholds}
\end{figure}

In \figref{fig:partial_thresholds}, we present the corresponding thresholds, $\UB{\DW{\eta}}$~\cite{WAP20}, above which the DW rate is guaranteed to be positive (\emph{dot-dashed lines}) as a function of the angle $\theta$ defining the the partially entangled state \eref{eq:part_ent_state}. Crucially, we compare these with the critical detection efficiencies, $\crit{\eta}$, obtained with help of the CC attack (\emph{solid lines}). We compute the latter by resorting to the upper bound \eref{eq:rdet_prefinal} and substituting for the correlation $\mathsf{Q}_{ab}^{xy}(\theta,\eta)$ maximising \eqnref{eq:S_det}, which in turn specifies the maximal local weight $q^\cL(\theta,\eta)$ obtained via a linear program. However, in the limit of the partially entangled state approaching its separable form (see \appref{app:ql_part_ent}), we evaluate analytically $q^\cL(\theta\to0,\eta)=1-\eta(3\eta-2)$. As this is the limit in which the highest robustness to imperfect detection is exhibited, this allows us to determine analytically (see \appref{app:1way_partent_UBs}) the minimal $\crit{\eta}$ allowed by the CC attack as $3/4=75\%$ and $(\sqrt{21}-3)/2\approx79.13\%$ for deterministic binning without and with inclusion of noisy preprocessing, respectively---see \emph{red} and \emph{blue solid curves} in \figref{fig:partial_thresholds} and their values at $\theta\to0$, while the values at $\theta=\pi/2$ consistently coincide with the ones stated in \tabref{tab:critical}.

We observe that the noisy preprocessing that introduces bit-flip errors onto the bit-string of Alice in an almost uniform manner ($\pp\to1/2$), which is known to significantly lower the threshold $\UB{\DW{\eta}}$~\cite{WAP20,HST+20,SBV+20}, actually improves the effectiveness of the CC attack for small $\theta$-angles---note the \emph{blue line} crossing the \emph{red line} in \figref{fig:partial_thresholds}. As a result, for deterministic binning and noisy preprocessing the CC attack provides a very stringent restriction, $79.13\%\le\DW{\eta}\le82.57\%$, on any potential improvement of the minimal tolerable detection efficiency---see the narrow $\approx\!3\%$ gap at $\theta\to0$ between \emph{blue solid and dot-dashed curves} in \figref{fig:partial_thresholds}. 

\begin{figure}[b!]
\centering
\includegraphics[width=\columnwidth]{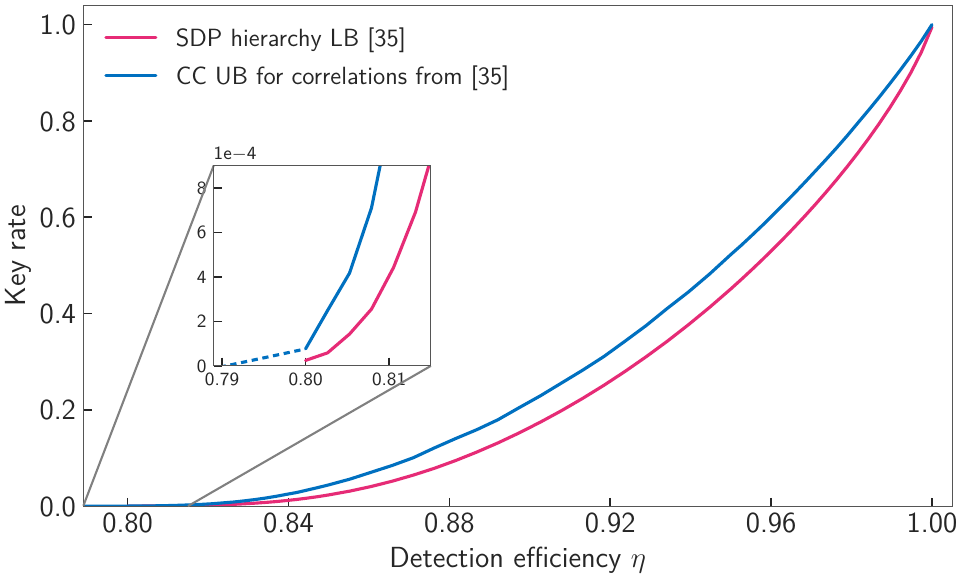}
\caption{\textbf{CC-based upper bound derived for the protocol of~\citeref{brown_device-independent_2021} with noisy preprocessing, as compared with the lower bound on the DW rate~\eqref{eq:DW_rate} established therein.} For each optimal correlation and bit-flip probability $\mathsf{p}$ that maximises the lower bound (LB) at a given $\eta$ (\emph{red line})~\cite{brown_device-independent_2021}, we compute the upper bound (UB) on the rate above which the CC attack invalidates the security (\emph{blue line}). Within the inset we magnify the region of critical detection efficiencies that appear in~\tabref{tab:critical_partial}. Note that due to the correlations of~\citeref{brown_device-independent_2021} being provided only for the region of positive LBs, $\eta \geq \UB{\DW{\eta}} = 80.00\%$, the CC-based UB for the region $\eta < \UB{\DW{\eta}}$ (and, hence, the critical value $\crit{\eta} = 79.04\%$ in~\tabref{tab:critical_partial}) is computed (\emph{dashed line} within the inset) using the same correlation and bit-flip probability $\mathsf{p}$ as for $\UB{\DW{\eta}}$.}
\label{fig:brown-lb-comp}
\end{figure}

As the attainable threshold $\UB{\DW{\eta}}=82.57\%$ of \cite{WAP20} has recently been improved by Brown \emph{et al.} \cite{brown_device-independent_2021} and Masini \emph{et al.} \cite{masini_simple_2021}, we present the corresponding two best-known thresholds in \tabref{tab:critical_partial}. We compare them explicitly against the critical efficiencies allowed by the CC attack, which we are able to evaluate having access to the exact correlations used, and the particular bit-flip strength $\pp$ employed at the noisy-preprocessing stage in \cite{brown_device-independent_2021,masini_simple_2021}, thanks to the courtesy of the authors. Strikingly, the CC attack leaves only a $\approx\!1\%$ gap for potential improvement, while the CC-based upper bound remains tight for the whole region of detection efficiencies with positive key rates---see \figref{fig:brown-lb-comp} in which we compare it explicitly with the lower bound on the DW rate~\eqref{eq:DW_rate} established in \citeref{brown_device-independent_2021}.

\begin{table}[t]\centering
\begin{tabular}{@{}ccl@{}}
\multicolumn{3}{c}{\textbf{DIQKD protocols}} \\
\multicolumn{3}{c}{\textbf{involving partially entangled states}} \\
\midrule[\heavyrulewidth]
$\crit{\eta}$ & $\UB{\DW{\eta}}$ & \textit{Reference}\\ 
\cmidrule(lr){1-1} \cmidrule(lr){2-3}
$79.04\%$ & $80.00\%$ & Brown \emph{et al.} \cite{brown_device-independent_2021}\\
$79.15\%$ & $80.26\%$ & Masini \emph{et al.} \cite{masini_simple_2021}\\
\midrule
%\bottomrule
\end{tabular}
\caption{\textbf{Critical detection efficiencies $\crit{\eta}$} (in \%) determined by the CC attack for the shared correlations (and rates of bit-flip errors applied within noisy preprocessing) that lead to the best-known thresholds, $\UB{\DW{\eta}}$, above which the DW rate \eref{eq:DW_rate} is guaranteed to be positive~\cite{masini_simple_2021,brown_device-independent_2021}. The CC attack proves that there is hardly any room for improvement of these state-of-the-art threshold values, given the data preprocessing (binning `no-clicks' + bit-flip errors) employed.}
\label{tab:critical_partial}
\end{table}

However, the state-of-the-art proofs of the DW rate \eref{eq:DW_rate}, e.g.~\cite{brown_device-independent_2021,masini_simple_2021}, require one to somehow bin the `no-click' events and perform noisy preprocessing on the binary raw-data. Hence, one may still ask the question by how much could the thresholds in \tabref{tab:critical_partial} be still improved, if novel derivations of one-way key rates were possible that allow Alice to perform any preprocessing map on her ternary variable $A$. That is why, for the correlations considered in \figref{fig:partial_thresholds}, we also compute the critical thresholds determined by the CC attack that are, however, minimised over all meaningful preprocessing strategies, i.e.~$A\to A'\to M$ appearing in \eqnref{eq:oneway_rate}. We observe that, as in the case of maximally entangled states, from the point of view of the CC attack it is always optimal for Alice to announce the inconclusive rounds (via the map $A\to M$)---see \emph{black-squared} and \emph{green-circled curves} coinciding in \figref{fig:partial_thresholds}---while in the limit $\theta\to0$ it is sufficient to solely bin the `no-clicks'. As a result, bearing in mind that \figref{fig:partial_thresholds} considers particular $\theta$-parametrised family of correlations, we conclude that the analytic value $\crit{\eta}=3/4$ constitutes then a fundamental bound on the detection efficiency, below which no one-way DIQKD protocol may be possible. Note that it is strictly larger than $\eta_\trm{loc}=2/3$ below which the correlations cease to be non-local~\cite{E93}.

%%%%%%%%%%%%%%%%%%%%%%%%%%%%%%%%%%%%%
\subsubsection{Robustness improvement by postselection}
\label{sec:postselection}
\begin{figure}[t!]%[h]
\centering
\includegraphics[width=\columnwidth]{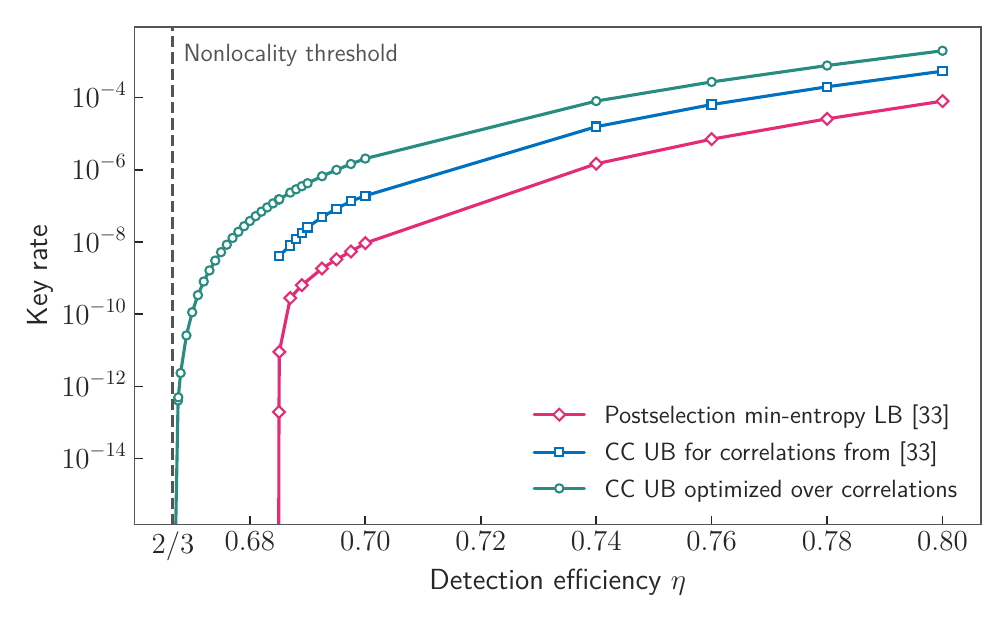}
\caption{
\textbf{Lower bound on the DW rate \eref{eq:DW_rate} compared with the CC-based upper bound for the protocol of \citeref{xu_device-independent_2021} involving postselection.} For each optimal correlation and the acceptance probability of postselection determined by Xu \emph{et al.} \cite{xu_device-independent_2021} to 
maximise the lower bound (LB) at a given $\eta$ (\emph{red diamonds}), we compute the upper bound (UB) on the rate above which the CC attack invalidates the security (\emph{blue squares}). Moreover, by performing brute-force optimisation of the shared correlation and the acceptance probability to maximise the CC-based UB instead (\emph{green circles}), we observe that the CC attack in principle allows for positive rates within the whole non-local range $\eta\ge\eta_\trm{loc}=2/3$.}
\label{fig:postselect}
\end{figure}
Nonetheless, it has been recently demonstrated that the thresholds stated in \tabref{tab:critical_partial} can be decreased to $\approx\!68.5\%$, if Alice and Bob perform postselection of their raw data for the key settings~\cite{xu_device-independent_2021}. However, it is unclear if the derived bounds using this postselection are valid for general attacks where Eve exploits correlations among realisations of the protocol. In fact, there are situations in which some entropy is left in the postselected data when Eve applies any i.i.d.~attack, but there is an attack using correlations between two realisations of the experiment for which Eve can perfectly predict the postselected outputs~\cite{thinh2015}.

In \figref{fig:postselect}, we present for the protocol of Xu \emph{et al.} \cite{xu_device-independent_2021} lower bounds on the rates accompanied by upper bounds based on the CC attack, as a function of the detection efficiency. The protocol corresponds to the 2333-scenario in which the `no-click' events are binned again onto a predetermined outcome, say `1', while Alice and Bob then separately decide whether to accept or discard each of `1's contained within their bit-string of key-generation rounds with probability $\qq$ or $1-\qq$, respectively. This does not open the detection loophole, as no postselection is performed within the rounds used to assure the nonlocality of the correlation. Moreover, Alice and Bob reveal publicly only the information whether each bit is accepted, irrespectively of its actual value (accepting simply all `0's). 

The \emph{red curve} in \figref{fig:postselect} corresponds to the lower bound on the DW rate obtained by the authors of \citeref{xu_device-independent_2021} by approximating the von Neumann entropy with min-entropy in \eqnref{eq:DW_rate} and optimising the acceptance probability $\qq$. For each point (\emph{dots/squares} in \figref{fig:postselect}), see \appref{app:1way_postselect}, we evaluate the upper bound that follows from the CC attack for the corresponding correlation and the optimal value of $\qq$ utilised by Xu \emph{et al.} \cite{xu_device-independent_2021}---see the \emph{blue curve} in \figref{fig:postselect}. Moreover, we also maximise by brute-force heuristic methods the so-determined CC-based upper bound over all correlations (and the acceptance probability $\qq$ for each correlation), in order to determine the \emph{green curve} in \figref{fig:postselect}. We observe that the CC attack disallows any significant improvement of the already very low rate, $\lesssim10^{-4}$ for any $\eta\le80\%$, however, it suggests that positive key rates could be potentially attained with postselection for the whole non-local range $\eta\ge2/3$ (disallowing any non-i.i.d.~attacks though~\cite{thinh2015}).

%%%%%%%%%%%%%%%%%%%%%%%%%%%%%%%%%%%%%%%%%%%%%%%%%%%%%%%%%%%%%%%%%%%%%%%%%%%%%%%%%%
\subsection{One-way protocols with more than two settings and outcomes}
\label{sec:beyond_two_set}
\begin{table}[b!]\centering
\begin{tabular}{@{}lccccc@{}}
%\multicolumn{5}{l}{\textbf{DIQKD protocols with more settings and outcomes}} \\
\multicolumn{6}{c}{\begin{tabular}{@{}c@{}}\textbf{DIQKD protocols} \\ \textbf{with more settings and outcomes}\end{tabular}} \\
\midrule[\heavyrulewidth]
Scenario & Correlation & $\crit{V}$ & $\UB{\DW{V}}$ & $\crit{\eta}$ & $\UB{\DW{\eta}}$ \\ 
\cmidrule(r){1-1} \cmidrule(lr){2-2} \cmidrule(lr){3-4} \cmidrule(lr){5-6}
4522 & $Q_{4422}$~\cite{Vertesi2010} & $85.35$ & $93.96$ & $86.78$ & $94.74$\\
3444 & $Q_{234}$~\cite{Gonzales2021} & $85.29$ & $91.04$ & $88.28$ & $92.18$\\
\midrule
%\bottomrule
\end{tabular}
\caption{\textbf{Critical visibilities $\crit{V}$ and detection efficiencies $\crit{\eta}$} (in \%) below which no DIQKD is possible due to the CC attack, for protocols considered by Gonzales-Ureta \emph{et al.}~\cite{Gonzales2021} that employ correlations being obtained by measuring two maximally entangled ququarts and violating $I_{4422}^4$ \cite{BRUNNER2008} and $I_{234}$ \cite{Cabello2001} Bell inequalities within 4522- and 3444-scenarios, respectively. The CC-based critical values are compared against the thresholds, $\UB{\DW{V}}$ and $\UB{\DW{\eta}}$, above which the DW rate \eref{eq:DW_rate} has been proven to be positive~\cite{Gonzales2021}.
}
\label{tab:critical_more}
\end{table}
In this last section, we would like to emphasise that the CC attack can be applied efficiently to any protocol by just following its consecutive stages, also when it involves correlations with larger number of settings and outcomes on both sides. In order to do so, we consider the two DIQKD schemes recently analysed by Gonzales-Ureta \emph{et al.}~\cite{Gonzales2021} with correlations obtained by measuring a
maximally entangled two-ququart state ($\tfrac{1}{2}\sum_{i=1}^4\ket{ii}$) within the 4522- and 3444-scenarios. In particular, the correlations employed in \citeref{Gonzales2021} correspond then to ones that exhibit robustness to noise (again, finite detection efficiency and visibility) when violating $I_{4422}^4$ \cite{BRUNNER2008} and $I_{234}$ \cite{Cabello2001} Bell inequalities, respectively. In the former case the correlation $Q_{4422}$ introduced in \citeref{Vertesi2010} is used, while in the latter the correlation $Q_{234}$ that leads to the maximal quantum violation of $I_{234}=9$~\cite{Gonzales2021}. In \tabref{tab:critical_more}, we compare the resulting thresholds, $\UB{\DW{V}}$ and $\UB{\DW{\eta}}$, above which the DW rate \eref{eq:DW_rate} has been proven to be positive by Gonzales-Ureta \emph{et al.}~\cite{Gonzales2021}, against the corresponding critical values on visibility and detection efficiency that the CC attack allows for, $\crit{V}$ and $\crit{\eta}$. The CC-based values suggest that the thresholds obtained by Gonzales-Ureta \emph{et al.}~\cite{Gonzales2021} may be potentially improved below $90\%$, but not beyond $85\%$.

%%%%%%%%%%%%%%%%%%%%%%%%%%%%%%%%%%%%%%%%%%
\section{Conclusions}
\label{sec:conclusions}
We have introduced the convex-combination (CC) attack as an easy-to-use tool to compute upper bounds on asymptotic rates in one-way and two-way DIQKD protocols. This in turn allows one to quickly establish critical noise parameters (here, detection efficiency and visibility) below which these upper bounds vanish and, hence, the CC attack disallows any DIQKD to be possible.

By applying the CC attack to one-way and two-way protocols involving either maximally or partially entangled states, as well as ones including a postselection stage or relying on correlations with more than two measurement settings and outcomes, we have demonstrated that despite its simple construction---decomposition of a given quantum correlation into a `local' and a `more non-local' part---the CC attack turns out to be very efficient in proving that the current thresholds on noise tolerance established with help of state-of-the-art security proofs are already very close to critical noise values, below which the CC attack invalidates the security. It is worth stressing that computing the upper bounds on the key rates, or equivalently, on Eve's entropy, in the CC attack is very simple, in particular, much simpler than computing lower bounds. In light of the heuristic results derived in this work, the CC attack appears to be also a versatile tool to benchmark lower bounds to entropies obtained with existing techniques, such as the hierarchies of~\cite{brown_device-independent_2021,brown_computing_2021}.

We have successfully applied the CC attack in its simplest form, in which the `non-local' (but quantum) correlation within the convex decomposition is fixed, while the `local' contribution can then be chosen to maximise its weight within the decomposition by means of a linear program. On one hand, it may not be generally true that the strategy of maximising the probability of distributing the local correlation, for which the eavesdropper perfectly knows the outcomes, is actually optimal from the point of view of providing the tightest upper bound on the key rate. On the other, we have pessimistically assumed the eavesdropper not to possess any information about the outcomes in case the `more non-local' correlation is distributed. Although the linear program can be straightforwardly adapted to utilise multiple non-local point within the decomposition, it would be desirable to generalise the construction, so that it actually includes optimisation over the non-local points e.g.~by approximating the quantum set of correlations with sufficient accuracy from outside by means of a convergent hierarchy of relaxations~\cite{navascues2007,navascues2008}. We leave the above interesting developments of our CC attack for future work.

\paragraph*{Note Added.}
Upon completion of this manuscript, we have learned that the CC attack has been applied to a two-way protocol by Yu-Zhe Zhang \emph{et al.}~\cite{Zhang2022}, while incorporating the optimisation of the non-local point(s) in the CC decomposition with help of the NPA-hierarchy~\cite{navascues2007}, as suggested above.

%%%%%%%%%%%%%%%%%%%%%%%%%%%%%%%%%%%%%%%%%%%%%%%%%%%%%%%%%%%%%%%%%%%%%
\section*{Acknowledgements}
We thank Erik Woodhead, Stefano Pironio, Feihu Xu, Peter Brown, J\k{e}drzej Kaniewski, Karol Horodecki and Filip Rozp\k{e}dek for fruitful comments. We are also very grateful to the authors of Refs.~\cite{brown_device-independent_2021,masini_simple_2021} and \cite{xu_device-independent_2021} for providing us with an explicit numerical form of the correlations employed in the respective protocols, especially to Yu-Zhe Zhang and Yi-Zheng Zhen for an extensive discussion about the application of the CC attack to the protocol involving post-selection~\cite{xu_device-independent_2021}. 

We acknowledge support from the Government of Spain (TRANQI, FUNQIP, NextGeneration EU/PRTR and Severo Ochoa CEX2019-000910-S), Fundaci\'{o} Cellex, Fundaci\'{o} Mir-Puig, Generalitat de Catalunya (CERCA Program), the ERC AdG CERQUTE, the AXA Chair in Quantum Information Science, the EU Quantum Flagship project QRANGE and QSNP, and the Foundation for Polish Science within the “Quantum Optical Technologies” project carried out within the International Research Agendas programme cofinanced by the European Union under the European Regional Development Fund. MBJ acknowledges funding from the European Union’s Horizon 2020 research and innovation programme under the Marie Sk\l{}odowska-Curie grant agreement No.~847517. This  project  has  received  funding  from  the  European  Union’s  Horizon~2020  research and innovation programme under the Marie Sk\l{}odowska-Curie grant agreement No.~754510.

\bibliographystyle{myquantum}
\bibliography{CCattack}

%%%%%%%%%%%%%%%%%%%%%%%%%%%%%%%%%%%%%%%%%%%%%%%%%%%%%%%%%%%%%%%%%%%%%
\newpage
\onecolumngrid
\appendix
\part*{Appendices}
\label{sec:Appendices}
%%%%%%%%%%%%%%%%%%%%%%%%%%%%%%%%%%%%%%%%%%%%%%%%%%%%%%%%%%%%%%%%%%%%%%%%%%%%%%%%%%%
Within the appendices, we firstly provide in \appref{sec:ind_attack_explicit} an explicit construction of the tripartite state and the measurements allowing the eavesdropper Eve to implement any individual attack. 
In \appref{sec:max_loc_weight}, we then show how to obtain analytic expressions for the maximal local weight $q^\cL$ utilised within the CC attack for the CHSH-based protocols subject to finite visibility ($V<1$) and detection efficiency ($\eta<1$) that involve maximally entangled states, but also partially entangled states (with $\theta\to0$ in \eqnref{eq:part_ent_state}) when $\eta<1$. Throughout our work---see the beginning of~\secref{sec:appl_to_DIQKD} for a discussion of this choice---we consider the version of the CC attack in which Eve uses only one nonlocal correlation, $\pAB^\cNL(a,b|x,y) \equiv \qAB(a,b|x,y)$, which corresponds to the probability distribution of Alice and Bob registering measurement outcomes $a$ and $b$ in the noiseless scenario of $\eta = V = 1$.
In \appref{sec:key_rates_CC}, we demonstrate how to construct the upper bounds on one-way key rates based on this choice of the CC attack for the two noise models considered and various preprocessing strategies:~\textit{random} and \textit{deterministic} binning of non-detection events, with and without \textit{noisy preprocessing}. This allows us to explicitly derive in \appref{sec:UBs_thresholds_CC} the thresholds on the tolerable noise parameters, in particular, in \appref{app:1way_maxent_UBs} for the protocols involving maximally entangled states with the resulting (numerical) values presented in Tab.~\ref{tab:critical} of the main text. We achieve this analytically for all particular preprocessing strategies considered, and semi-analytically when including the optimisation over all potential preprocessing maps.
We then generalise the above analysis to two-way protocols in \appref{app:2way_maxent_UBs} (also ones that involve multiple key settings~\cite{SGP+20} for $V<1$), as well as one-way protocols involving partially entangled states in \appref{app:1way_partent_UBs}. The latter case we study in more detail in \appref{app:1way_postselect}, where we allow further for postselection of some events, as proposed in~\citeref{xu_device-independent_2021}.

%%%%%%%%%%%%%%%%%%%%%%%%%%%%%%%%%%%%%%%%%%%%%%%%%%%%%%%%%%%%%%%%%%%%%%%%%%%%%%%%%%%%%%%%%%%%%%%%%%%%%%%%%%%%%%%%%
\section{Explicit form of the state and measurements in individual attacks}
\label{sec:ind_attack_explicit}
%%%%%%%%%%%%%%%%%%%%%%%%%%%%%%%%%%%%%%%%%%%%%%%%%%%%%%%%%%%%%%%%%%%%%%%%%%%%%%%%%%%%%%%%%%%%%%%%%%%%%%%%%%%%%%%%%%

In this section, we provide an explicit construction for a shared tripartite state, and measurements for Alice, Bob and Eve for achieving the individual attack in Eq.~\eqref{eq:class_ind_attacks}. We start from the observed correlation, and write it as a convex combination of quantum correlations:
\begin{equation}\label{eq:cvx_decomp_app}
\pABobs(a,b|x,y) = \sum_\lambda q(\lambda) \pAB(a,b|x,y,\lambda).
\end{equation}
That is, there exist a state $\rho_{AB}$ on a Hilbert space $\cH_A \otimes \cH_B$ and measurements $\{M^x_a\}$ and $\{N^y_b\}$ on $\cH_A$ and $\cH_B$, respectively, such that
\begin{equation}
\pABobs(a,b|x,y) = \tr{ \rho_{AB} ( M^x_a \otimes N^y_b ) }.
\end{equation}
Moreover, $q(\lambda)$ is a probability distribution and there exist states $\rho^\lambda_{AB}$ on a Hilbert space $\cH^\lambda_A \otimes \cH^\lambda_B$ and measurements $\{(M^\lambda)^x_a\}$ and $\{(N^\lambda)^y_b\}$ on $\cH^\lambda_A$ and $\cH^\lambda_B$, respectively, such that
\begin{equation}
\pAB(a,b|x,y,\lambda) = \tr{ \rho^\lambda_{AB} [ (M^\lambda)^x_a \otimes (N^\lambda)^y_b ] }.
\end{equation}
It is clear from the convex structure of the quantum set of correlations that for every valid convex decomposition of the form Eq.~\eqref{eq:cvx_decomp_app}, if $q(\lambda)$, the states $\rho^\lambda_{AB}$, and the measurements $\{(M^\lambda)^x_a\}$ and $\{(N^\lambda)^y_b\}$ are known, then one can build the state $\rho_{AB}$ and the measurements $\{M^x_a\}$ and $\{N^y_b\}$. In particular, one can pick the Hilbert spaces $\cH_A = \bigoplus_\lambda \cH^\lambda_A$ and $\cH_B = \bigoplus_\lambda \cH^\lambda_B$, and define the state $\rho_{AB} = \bigoplus_\lambda q(\lambda) \rho^\lambda_{AB}$ on $\cH_A \otimes \cH_B$. This state together with the measurements $M^x_a = \bigoplus_\lambda (M^\lambda)^x_a$ on $\cH_A$ and $N^y_b = \bigoplus_\lambda (N^\lambda)^y_b$ on $\cH_B$ gives rise to the correlation $\pABobs(a,b|x,y)$.

We now construct a tripartite state $\rho_{ABE}$ on $\cH_A \otimes \cH_B \otimes \cH_E$ and measurement operators $E_e$ on $\cH_E$ such that the resulting tripartite correlation in the individual attack reads
\begin{equation}
\pABE(a,b,e | x,y) = \tr{ \rho_{ABE} (M^x_a \otimes N^y_b \otimes E_e) } = \sum_\lambda q(\lambda) p(e|\lambda) \pAB(a,b|x,y,\lambda),
\end{equation}
for some arbitrary distributions $p(e|\lambda)$. We simply choose the state
\begin{equation}
\rho_{ABE} = \bigoplus_\lambda q(\lambda) \rho^\lambda_{AB} \otimes \ketbraq{\lambda}_E,
\end{equation}
where $\{\ket{\lambda}\}$ is an orthonormal basis on $\cH_E$, and the measurement operators
\begin{equation}
E_e = \sum_\lambda p(e|\lambda) \ketbraq{\lambda}
\end{equation}
on $\cH_E$.  It follows that
\begin{equation}
\tr{ \rho_{ABE} (M^x_a \otimes N^y_b \otimes E_e) } = \pABE(a,b,e|x,y) = \sum_\lambda q(\lambda) p(e|\lambda) \pAB(a,b|x,y,\lambda)
\end{equation}
as required, and clearly, $\sum_e \pABE(a,b,e|x,y) = \pABobs(a,b|x,y)$ for all $a,b,x,y$. Last, we note that $p(e|\lambda)$ can be chosen in a way that $e$ preserves the information about $\lambda$ (given that the alphabet size of $e$ is large enough).

%%%%%%%%%%%%%%%%%%%%%%%%%%%%%%%%%%%%%%%%%%%%%%%%%%%%%%%%%%%%%%%%%%%%%%%%%%%%%%%%%%%%%%%%%%%%%%%%%%%%%%%%%%%%%%%%%
\section{Analytical evaluation of the maximal local weight $q^\cL$ in the CC attack}
\label{sec:max_loc_weight}
%%%%%%%%%%%%%%%%%%%%%%%%%%%%%%%%%%%%%%%%%%%%%%%%%%%%%%%%%%%%%%%%%%%%%%%%%%%%%%%%%%%%%%%%%%%%%%%%%%%%%%%%%%%%%%%%%%
%
%%%%%%%%%%%%%%%%%%%%%%%%%%%%%%%%%%%%%%%%%%%%%%%%%%%%%%%
\subsection{Correlations yielding maximal CHSH-violation subject to finite visibility}
\label{app:ql_maxent_V}
In this section, we provide the analytical form of the local weight~\eqref{eq:visibility_local_weight0} in the CC attack for the 2322 CHSH-based protocol subject to finite visibility. The correlation in the noise-free case is the one obtained by Alice and Bob sharing the state $\ket{\Phi^{+}}$, i.e.~setting $\theta = \pi/2$ in~\eqnref{eq:part_ent_state}, on which they perform CHSH-optimal dichotomic measurements~\eqref{eq:standard_measurements} with outcomes $a,b \in \{0,1\}$, and reads
\begin{equation}\label{eqs:biased_CHSH}
\qAB(a,b | x,y ) = 
\begin{cases}
\frac14 \left[ 1 + \frac{(-1)^{a+b+xy}}{\sqrt2} \right] & \text{if } x,y \in \{0,1\} \\
\frac12 \, \delta_{a,b} & \text{if } (x,y) = (0,2) \\
\frac14 & \text{if } (x,y) = (1,2).
\end{cases}
\end{equation}
We consider noisy versions of this correlation with finite visibility $V \in [0,1]$, i.e.~the uniform noise as specified in \eqnref{eq:visibility} with $\nA = \nB = 2$, i.e.,
\begin{equation}
\pABobs(a,b | x,y) = V \, \qAB(a,b | x,y) + \frac{1 - V}{4}.
\end{equation}
The CC attack for this protocol consists of the convex combination of a local correlation $p_{AB}^\cL(a,b|x,y)$ and the noise-free correlation $\qAB(a,b|x,y)$, such that the observed correlation is of the form:
\begin{equation}\label{eqs:simple_attack}
\pABobs(a,b|x,y) = q^\cL \, \pAB^\cL(a,b|x,y) + (1 - q^\cL) \, \qAB(a,b|x,y).
\end{equation}
From equating the above expression for $\pABobs(a,b|x,y)$ it follows that
\begin{equation}\label{eqs:local_correlation_visibility}
\pAB^\cL(a,b|x,y) = \tilde{V} \, \qAB(a,b | x,y ) + \frac{1-\tilde{V}}{4},
\end{equation}
where $\tilde{V} \in [0,1]$ and $q^\cL = (1-V)/(1-\tilde{V})$. Therefore, maximising $q^\cL$ simply corresponds to maximising $\tilde{V}$ in Eq.~\eqref{eqs:local_correlation_visibility}, such that $\pAB^\cL(a,b|x,y)$ is local. The result of this maximisation is the \textit{local visibility} $V^\cL$, which in turn fully characterises the CC attack for this protocol with $q^\cL = (1-V)/(1-V^\cL)$ when $V \ge V^\cL$ and otherwise $q^\cL = 1$. Even though the maximisation is a linear program, it is possible to solve it explicitly as all the facets of the polytope in the 2322-scenario are known analytically~\cite{CG04}. In particular, all the facets that correspond to non-trivial constraints (i.e.~do not correspond to the positivity and normalisation of the probabilities) are of the CHSH-type:
\begin{equation}\label{eqs:CHSH}
-2 \le \langle A_{x_0} B_{y_0} \rangle + \langle A_{x_0} B_{y_1} \rangle + \langle A_{x_1} B_{y_0} \rangle - \langle A_{x_1} B_{y_1} \rangle \le 2 \quad x_0,x_1 \in \{0,1\}, \, y_0, y_1 \in \{0,1,2\},
\end{equation}
where $\langle A_x B_y \rangle = p(0,0|x,y) + p(1,1|x,y) - p(0,1|x,y) - p(1,0|x,y)$ are the correlators of a given $p(a,b|x,y)$. That is, a correlation in the 2322-scenario is local if and only if it satisfies all the inequalities in Eq.~\eqref{eqs:CHSH}.

Let us denote the correlators of $\qAB(a,b|x,y)$ by $\langle A_x B_y \rangle^\cNL$, and those of $\pAB^\cL(a,b|x,y)$ by $\langle A_x B_y \rangle^\cL$. It is easy to see that 
\begin{equation}
\langle A_x B_y \rangle^\cL = \tilde{V} \, \langle A_x B_y \rangle^\cNL.
\end{equation}
Therefore, finding the maximal $\tilde{V}$ such that $\pAB^\cL(a,b|x,y)$ is local corresponds to finding the maximal $\tilde{V}$ such that
\begin{equation}\label{eqs:CHSH_v}
-\frac{2}{\tilde{V}} \le \cS^\cNL_{x_0, x_1, y_0, y_1} \le \frac{2}{\tilde{V}} \quad \forall x_0,x_1 \in \{0,1\}, \, \forall y_0, y_1 \in \{0,1,2\},
\end{equation}
where we have defined
\begin{equation}
\cS^\cNL_{x_0, x_1, y_0, y_1} = \langle A_{x_0} B_{y_0} \rangle^\cNL + \langle A_{x_0} B_{y_1} \rangle^\cNL + \langle A_{x_1} B_{y_0} \rangle^\cNL - \langle A_{x_1} B_{y_1} \rangle^\cNL.
\end{equation}
A straightforward computation yields
\begin{align}
\cS^\cNL_{0,1,0,1} &=  2 \sqrt2   
&\cS^\cNL_{0,1,0,2} &=  1+\sqrt2   
&\cS^\cNL_{0,1,1,0} &=  -\sqrt2  
&\cS^\cNL_{0,1,1,2} &=  1 \nonumber \\
\cS^\cNL_{0,1,2,0} &=  1  
&\cS^\cNL_{0,1,2,1} &=  1+\sqrt2  
&\cS^\cNL_{1,0,0,1} &=   0  
&\cS^\cNL_{1,0,0,2} &=   \sqrt2 - 1 \nonumber \\
\cS^\cNL_{1,0,1,0} &=   0    
&\cS^\cNL_{1,0,1,2} &=   -1  
&\cS^\cNL_{1,0,2,0} &=   1  
&\cS^\cNL_{1,0,2,1} &=   1-\sqrt2. 
\end{align}
It is clear from the first equation that the maximal $\tilde{V}$ (denoted by $V^\cL$) is bounded by $V^\cL \le 1/\sqrt2$. Furthermore, substituting $\tilde{V} = 1/\sqrt2$ into Eq.~\eqref{eqs:CHSH_v} also implies that $\pAB^\cL(a,b|x,y)$ with visibility $\tilde{V} = 1/\sqrt2$ is local, and therefore $V^\cL \ge 1/\sqrt2$. Hence, we get that $ V^\cL = 1/\sqrt2$, fully characterising the CC attack for this protocol and determinig the local weight as
\begin{equation}
\label{eq:visibility_local_weight}
q^\cL = \min \left \{ 1, \frac{1-V}{1-\frac{1}{\sqrt2}} \right \}.
\end{equation}

%%%%%%%%%%%%%%%%%%%%%%%%%%%%%%%%%%%%%%%%%%%%%%%%%%%%%%%%%%%%%%%%%%%%%%%%%%%%%%%%%%
\subsection{Correlations yielding maximal CHSH-violation subject to finite detection efficiency}
\label{app:qL_maxent_detbin_eta}
In this section, we determine explicitly the maximal local weights in the CC attack for the CHSH-based 2333-protocol subject to finite detection efficiency. As in \eqnref{eq:lossy_corr} of the main text, the lossy observed correlation is given by
\begin{IEEEeqnarray}{rCl}
\label{eq:lossy_corr2}
  \pABobs(a,b | x,y)&=&\begin{array}{|c|c|c|c|}
  \hline a \setminus b & 0 & 1 & \varnothing\\
  \hline 0 & \eta^{2}\mathsf{Q}_{00}^{xy} & \eta^{2}\mathsf{Q}_{01}^{xy} & \eta \neta \mathsf{Q}_{0}^{x}\\
  \hline 1 & \eta^{2}\mathsf{Q}_{10}^{xy} & \eta^{2}\mathsf{Q}_{11}^{xy} & \eta \neta \mathsf{Q}_{1}^{x}\\
  \hline \varnothing & \neta \eta\mathsf{Q}_{0}^{y} & \neta \eta\mathsf{Q}_{1}^{y} & \neta^2
  \\\hline \end{array},
\IEEEeqnarraynumspace
\end{IEEEeqnarray}
where $\mathsf{Q}_{ab}^{xy} \eqdef Q(a,b|x,y)$ denotes the ideal, noiseless correlation with marginals $\mathsf{Q}_{a}^{x} = \sum_b \mathsf{Q}_{ab}^{xy}$ and $\mathsf{Q}_{b}^{y} = \sum_a \mathsf{Q}_{ab}^{xy}$ for Alice and Bob, respectively. Specifically, we calculate the maximal local weight $q^{\cL}(\theta = \pi/2, \eta)$ for protocols involving maximally entangled states, discussed in~Secs.~\ref{sec:DIQKD_max_ent}\&\ref{sec:DIQKD_max_ent_2way}, for which $Q(a,b|x,y)$ is given by~\eqnref{eqs:biased_CHSH} above, as well as $q^{\cL}(\theta \to 0, \eta)$ for protocols involving partially entangled states, discussed in~\secref{sec:DIQKD_part_ent}, where the $\mathsf{Q}$-probabilities take a more complicated form discussed in \appref{app:ql_part_ent} below.

Consider first the 2233-scenario. The complete characterisation of the local polytope in terms of facet (Bell) inequalities becomes more complicated than in the 2222-scenario as the number of such inequalities is 1116~\cite{CG04}. However, they may still be checked for violation with the help of some symbolic computation software. In general, the corresponding facet inequalities can be cast into three categories~\cite{CG04}: 36 ``trivial'' inequalities ensuring non-negativity of probabilities, 648 CHSH-like inequalities (resulting from the original CHSH inequality by some relabelling of measurements, outcomes and parties), and 432 CGLMP-like inequalities~\cite{CGLMP02} (also all equivalent under some choice of relabelling). All of them impose constraints on conditional probabilities, assuring the resulting correlation to admit a local hidden-variable model.

On the other hand, we note that for a given correlation $\pABobs(a,b|x,y)$ observed by Alice and Bob, and a particular lossless correlation $\pAB^{\cNL}(a,b|x,y)$ distributed by Eve in the ``nonlocal'' rounds of the CC attack, the local correlation in Eq.~\eqref{eq:CC_constr} must satisfy the following equality:
\begin{equation}\label{eqs:local_lossy_corr}
\pAB^\cL(a,b|x,y) = \frac{\pABobs(a,b|x,y) - (1 - q^\cL) \, \qAB(a,b|x,y)}{q^\cL},
\end{equation}
where $q^\cL$ is to be maximised and we set $\pAB^\cNL = \qAB$. Although the maximal value of $q^\cL$ can always be determined numerically by the linear program, one may equivalently treat $q^\cL$ as a free parameter and verify what is its maximal value such that none of the aforementioned inequalities is violated by the correlation in Eq.~\eqref{eqs:local_lossy_corr}. Importantly, in this way not only the maximal value of $q^\cL$ is determined, but also the particular (facet) inequality may be identified, i.e.~the facet of the local polytope on which the correlation \eqref{eqs:local_lossy_corr} then resides in the correlation space when $q^\cL$ is maximal. As all the Bell inequalities are linear in probabilities~\cite{BCP+14}, i.e. 
\begin{equation}\label{eq:bell_inequality}
\sum_{a, b, x, y} b_{a,b,x,y} \; \pAB^\cL(a,b|x,y) \leq 0,
\end{equation}
with $b_{a,b,x,y}\in\mathbb{R}$, it can be rearranged into an inequality for $q^\cL$ using Eq.~\eqref{eqs:local_lossy_corr}, and expressing $\pABobs(a,b|x,y)$ and $\qAB(a,b|x,y)$ as functions of $\eta$ and $\theta$.

\subsubsection{Maximally entangled states}
Focusing first on the 2233-scenario with $\theta = \pi/2$ in \eqref{eq:part_ent_state} and standard CHSH-optimal measurements $x, y \in \{0, 1\}$ \eqref{eq:standard_measurements}, we observe that for all values of $\eta$, the relevant inequality imposing locality of the correlation $\pAB^\cL$ in Eq.~\eqref{eqs:local_lossy_corr} is of the CHSH type, does not involve non-detection events, and after simplifying reads:
\begin{equation}
  -\pA^\cL(1|0)-\pB^\cL(1|0)+\pAB^{\cL}(1,1|0,0)+\pAB^{\cL}(1,1|0,1)\\+\pAB^{\cL}(1,1|1,0)-\pAB^{\cL}(1,1|1,1)\leq0
\label{eq:max_ent_ineq}
\end{equation}
with $\pA^\cL$ and $\pB^\cL$ being the marginal distributions. We can evaluate the relevant conditional probabilities by calculating the corresponding terms for the correlation $\pABobs(a,b|x,y)$ observed by Alice and Bob from~\eqref{eq:lossy_corr2} as
\begin{equation}
\begin{aligned}\label{eq:probs_maxent}
\pAobs(1|0) = \pBobs(1|0) = & \; \frac{\eta}{2},\\
\pABobs(1,1|0,0) = \pABobs(1,1|0,1) = \pABobs(1,1|1,0) = & \; \eta^{2} \frac{2+\sqrt2}{8},\\
\pABobs(1,1|1,1) = & \; \eta^{2} \frac{2-\sqrt2}{8}.
\end{aligned}
\end{equation}
Hence, substituting for the local distribution $\pAB^\cL$ and its marginals into \eqnref{eq:max_ent_ineq} according to \eqnref{eqs:local_lossy_corr}, with observed and ideal ($\eta=1$) correlations specified as above, we obtain the desired upper bound on the local weight within the CC attack, i.e.~$q^\cL \leq (1-\eta) \left(1+\left(3+2\sqrt{2}\right)\eta\right)$, so that we can write explicitly the maximal local weight in the lossy 2233-scenario utilising maximally entangled states as
\begin{equation}\label{eqs:local_weight_lossy}
q^\cL = \min \left\{1, (1-\eta) \left(1+\left(3+2\sqrt{2}\right)\eta\right)\right\}.
\end{equation}

In the 2333-scenario Bob uses an additional measurement $B_2 = \sigma_z$ identical to $A_0$, so that these are correlated and, hence, most efficient in generating the key. Although the dimensionality of the correlation space is then formally increased, such an added setting does not impose any further locality constraints on the resulting shared correlation. In particular, as the inequality \eref{eq:max_ent_ineq} remains then the only relevant, the above analysis similarly applies. For completeness, however, we verify this numerically by running explicitly the linear program that consistently outputs maximal values of $q^\cL$ according to \eqnref{eqs:local_weight_lossy} also in the 2333-scenario considered.

%%%%%%%%%%%%%%%%%%%%%%%%%%%%%%%%%%%%%%%%%%%%%%%%%%%%%%%%%%%%%%%%%%%%%%%%%%%%%%%%%%
\subsubsection{Partially entangled states}
\label{app:ql_part_ent}
Moreover, focusing further on the 2233-scenario but examining correlations determined by the partially entangled states~\eqref{eq:part_ent_state} and measurements chosen to maximise the CHSH functional \eqref{eq:S_det}, in the limiting case $\theta \to 0$, we observe the relevant inequality imposing locality to be the same one as for the maximally entangled states~\eqref{eq:max_ent_ineq}. In this case, we have
\begin{equation}
\begin{aligned}\label{eq:probs_partial}
\pAobs(1|0) = & \;\frac{\eta}{2}(1-\cos\theta),\\
\pBobs(1|0)= & \; \frac{\eta}{2}(1-z_{+}\cos\theta),\\
\pABobs(1,1|0,0)= & \; \frac{\eta^{2}}{4}(1+z_{+})(1-\cos\theta),\\
\pABobs(1,1|0,1)= & \; \frac{\eta^{2}}{4}(1+z_{-})(1-\cos\theta),\\
\pABobs(1,1|1,0)= & \; \frac{\eta^{2}}{4}\left[1+z_{+}\cos\phi_{A}-(z_{+}+\cos\phi_{A})\cos\theta+\sqrt{1-z_{+}^{2}}\sin\phi_{A} \sin\theta\right],\\
\pABobs(1,1|1,1)= & \; \frac{\eta^{2}}{4}\left[1+z_{-}\cos\phi_{A}-(z_{-}+\cos\phi_{A})\cos\theta-\sqrt{1-z_{-}^{2}}\sin\phi_{A}\sin\theta\right],
\end{aligned}
\end{equation}
with 
\begin{align}\label{eq:meas_params}
P  :&\!=\alpha\eta+\alpha\neta\cos\theta, & Q  :&\!=\eta\cos\phi_{A}+\neta\cos\theta ,\nonumber\\ R  :&\!=\eta\sin\phi_{A}\sin\theta, &
 z_{\pm} & \!=\frac{P\pm Q}{\sqrt{(P\pm Q)^{2}+R^{2}}},
\end{align}
and $\phi_{A}$ characterising the optimal measurement $A_1$ (cf.~\cite{WAP20} where this notation is introduced)
\begin{equation}
A_{1}=\cos(\phi_{A})\;\sigma_{z}+\sin(\phi_{A})\;\sigma_{x}.\label{eq:A_1_opt}
\end{equation}
The corresponding conditional probabilities $\qAB$ can be obtained by setting $\eta = 1$ in~\eqref{eq:probs_partial}, and depend on $\theta$ and on $\phi_{A}$ due to the optimisation of measurements. Using these expressions we arrive at:
\begin{multline}
-\eta_{\cL}\left(1-\frac{1+z_{+}}{2}\cos\theta\right)+\frac{\chi_{\cL}^{2}}{4}\left[(2+z_{+}+z_{-})(1-\cos\theta)\right.\\\left.+(z_{+}-z_{-})(\cos\phi_{A}-\cos\theta)+\left(\sqrt{1-z_{+}^{2}}+\sqrt{1-z_{-}^{2}} \sin\theta\sin\phi_{A}\right)\right]\leq0,
\label{eq:facet_ineq_qL}
\end{multline}
where
\begin{equation}
\label{eq:eta_cL_xi_cL}
\eta_\cL\eqdef \frac{\eta+q^\cL-1}{q^\cL}\;\text{ and }\;\chi_{\cL}^{2}\eqdef \frac{\eta^{2}+q^\cL-1}{q^\cL}.
\end{equation}
Expanding the inequality (\ref{eq:facet_ineq_qL}) in the lowest
order of $\theta$ and $\phi_{A}$ we have
\begin{equation}
\left[\frac{q^\cL-1-\eta(2-3\eta)}{4 q^\cL} + \mathcal{O}\!\left(\phi_A^2\right) \right ] \theta^2 + \mathcal{O}\!\left(\theta^3\right) \leq 0.
\end{equation}
Hence, in the limit $\theta\to0$, in which also $\phi_{A}\to0$~\cite{WAP20} and the lowest-order term $\sim\!\theta^2$ dominates, we obtain a general upper bound on local weight:~$q^\cL\leq 1 - \eta (3\eta - 2)$. Thus, we conclude that the maximal local weight for the CC attack in the lossy 2233 protocol utilising partially entangled states with $\theta \to 0$ and measurements chosen to maximise the CHSH-violation is given by
\begin{equation}
\label{eq:local_weight_partial}
q^\cL = \min \left \{1, 1 - \eta (3\eta - 2)\right\},
\end{equation}
with $q^{\cL} = 1$ certifying the observed correlation to be local for $\eta \leq \eta_\trm{loc} = 2/3$---in consistency with~\citeref{E93}.

Similarly to the $\theta = \pi/2$-case above, we confirm for completeness that adding an extra key setting of Bob, $B_2 = \sigma_z$, in the 2333-scenario does not affect the above analysis. In particular, we compute numerically the maximal local weights with the linear program~\eqref{eq:linear_program}, which match then exactly the expression \eqref{eq:local_weight_partial}, as expected.

%%%%%%%%%%%%%%%%%%%%%%%%%%%%%%%%%%%%%%%%%%%%%%%%%%%%%%%%%%%%%%%%%%%%%%%%%%%%%%%%%%%%%%%%%%%%%%%%%%%%%%%%%%%%%%%%%%
%%%%%%%%%%%%%%%%%%%%%%%%%%%%%%%%%%%%%%%%%%%%%%%%%%%%%%%%%%%%%%%%%%%%%%%%%%%%%%%%%%%%%%%%%%%%%%%%%%%%%%%%%%%%%%%%%%
\section{Constructing the upper bounds on one-way key rates with help of the CC attack}
\label{sec:key_rates_CC}
%%%%%%%%%%%%%%%%%%%%%%%%%%%%%%%%%%%%%%%%%%%%%%%%%%%%%%%%%%%%%%%%%%%%%%%%%%%%%%%%%%%%%%%%%%%%%%%%%%%%%%%%%%%%%%%%%%
%%%%%%%%%%%%%%%%%%%%%%%%%%%%%%%%%%%%%%%%%%%%%%%%%%%%%%%%%%%%%%%%%%%%%%%%%%%%%%%%%%%%%%%%%%%%%%%%%%%%%%%%%%%%%%%%%%

In this section, we calculate the \textit{error correction} (EC) and \textit{privacy amplification} (PA) terms appearing in the upper bound on the one-way key rate~\eqref{eq:oneway_UB_ind_attack} for the finite visibility ($V<1$) and detection efficiency ($\eta<1$) noise models, while considering particular preprocessing strategies that Alice may apply to her raw data. Specifically, we consider here three types of them referenced in \tabref{tab:critical}:~one trivial case, i.e.~in which Alice does not transform her outcome at all; and two cases in which she converts the ternary variable $A$ into a binary variable $A'$, so that the preprocessing map $p_{A'|A}$ corresponds then to a $2\times3$ stochastic matrix $\cS$, i.e.~\emph{deterministic binning} of the non-detection event $\varnothing$ with and without \emph{noisy preprocessing} (performing also a bit-flip with some probability on the resulting binary variable). Moreover, we discuss two additional cases not shown in \tabref{tab:critical} that involve \emph{random binning}, i.e.~the non-detection event $\varnothing$ is randomly binned to one of the two measurement outcomes, with and without noisy preprocessing. While these preprocessing strategies appear most commonly in literature, the methodology described here may naturally be adapted to other protocols and preprocessing schemes. 

The preprocessing strategies considered here make no use of the publicly announced random variable $M$ appearing in the upper bound \eref{eq:oneway_UB_ind_attack} and, hence, we drop for our purposes the $M$-conditioning and rewrite the r.h.s.~of \eqnref{eq:oneway_UB_ind_attack} as
\begin{equation}
\UB{r_{\text{1-way},\bullet}} \eqdef H(A'|E)_{\bullet} - H(A'|B)_{\bullet},
\label{eq:appendix-oneway-UB}
\end{equation}
where for simplicity we also omit the notation $(A \to B|A')$ and instead introduce another subscript $\bullet$, within which we will denote the particular preprocessing map $A \to A'$ being employed---e.g.~``det''/``rand'' or ``n.p.'' for deterministic/random binning or noisy preprocessing, respectively. In the following, we refer to $H(A'|B)$ as the EC-term and to $H(A'|E)$ as the PA-term.

Having determined the local weight of the correlation $\pABobs(a,b|x,y)$---which includes all the possible inputs $x$ and $y$---the calculation of the EC- and PA-terms depends only on the tripartite distribution conditioned the key settings $\xk$ and $\yk$, i.e.~$\pABE(a,b,e|\xk, \yk)$. Therefore, for simplicity, we adopt the following notation, dropping the  \{$\xk$, $\yk$\} labels:
\begin{equation}
\begin{gathered}
\label{eq:Pab_Qab}
\mathsf{P}_{ab} = V Q(a, b|\xk, \yk) + \frac{1-V}{4}, \\ \mathsf{P}_{a}^{\mathrm{A}} = V Q(a|\xk) + \frac{1-V}{2}, \quad \mathsf{P}_{b}^{\mathrm{B}} = V Q(b|\yk) + \frac{1-V}{2},
\end{gathered}
\end{equation}
where the corresponding marginals satisfy $\mathsf{P}_{a}^{\mathrm{A}}=\sum_{b}\mathsf{P}_{ab}$ and $\mathsf{P}_{b}^{\mathrm{B}}=\sum_{a}\mathsf{P}_{ab}$, and $Q(a, b|\xk, \yk)$, $Q(a|\xk)$, $Q(b|\yk)$ denote the ideal probabilities of obtaining measurement outcomes $a$ and $b$ in case of perfect detection efficiency and visibility, $\eta = V = 1$, after Alice and Bob have chosen the key settings $\xk$ and $\yk$. 

The introduction of the $\mathsf{P}$-probabilities allows us to consider finite detection efficiency and finite visibility at the same time, as we can write the shared correlation \eqref{eq:lossy_corr2} for the key settings, $\pABobs(a,b|\xk,\yk)$ as
\begin{equation}
\label{eqs:simplified_lossy_corr}
\pAB(a,b)\quad=\quad{}\begin{array}{|c|c|c|c|}
\hline a\setminus b & 0 & 1 & \varnothing\\
\hline 0 & \eta^{2}\,\mathsf{P}_{00} & \eta^{2}\,\mathsf{P}_{01} & \eta \neta \PA{0}\\
\hline 1 & \eta^{2}\,\mathsf{P}_{10} & \eta^{2}\,\mathsf{P}_{11} & \eta \neta \PA{1}\\
\hline \varnothing & \neta \eta\PB{0} & \neta \eta\PB{1} & \neta^2
\\\hline \end{array}\,,
\end{equation}
where $\neta \eqdef  1 - \eta$ and we have dropped the `obs' superscript for simplicity. To recover the purely noisy correlation it suffices to set $\eta = 1$ in~\eqref{eqs:simplified_lossy_corr}, in which case the outcomes $\varnothing$ don't occur, whereas to obtain the purely lossy correlation~\eqref{eq:lossy_corr2} it suffices to replace the $\mathsf{P}$-probabilities with $\mathsf{Q}$-probabilities as they become equal if one sets $V=1$ in~\eqref{eq:Pab_Qab}. Lastly, note that this always yields the marginal distribution of Bob as 
\begin{equation}
\label{eq:bob_marginals}
\pB(b)=\sum_{a}\pAB(a,b)=
\begin{cases}
  \eta\PB{0}, &\text{ if }b=0\\ 
  \eta\PB{1}, &\text{ if }b=1\\
  \neta, &\text{ if }b=\varnothing
\end{cases}\,,
\end{equation}
being trivially independent of the preprocessing map $p_{A'|A}$ applied by Alice.

%%%%%%%%%%%%%%%%%%%%%%%%%%%%%%%%%%%%%%%%%%%%%%%%%%%%%%%%%%%%%%%%%%%%%%%%%%%%%%%%%%%%%%%%%%%%%%%%%%%%%%%%%%%%%%%%%%
\subsection{Calculation of the EC-term \texorpdfstring{$H(A'|B)$}{Lg}}
%%%%%%%%%%%%%%%%%%%%%%%%%%%%%%%%%%%%%%%%%%%%%%%%%%%%%%%%%%%%%%%%%%%%%%%%%%%%%%%%%%%%%%%%%%%%%%%%%%%%%%%%%%%%%%%%%%
%

%%%%%%%%%%%%%%%%%%%%%%%%%%%%%%%%%%%%%%%%%%%%%%%%%%%%%%%%
\subsubsection{No preprocessing}
If Alice performs no preprocessing, then simply $A\equiv A'$ and
\begin{align}
H(A'|B)_{\text{no-prep}}&=H(A|B) = \sum_b \pB(b) H(A|B=b) \nonumber\\ &= \eta\,\PB{0}\,H(A|B=0)+\eta\,\PB{1}\,H(A|B=1)+ \neta \,H(A|B=\varnothing),
\end{align}
where $H(A|B=b)$ is the entropy of Alice's outcome conditioned on Bob measuring $b$. Each $H(A|B=b)$ can be evaluated with the help of the conditional probability 
\begin{equation}
p_{A|B}(a|b)=\frac{\pAB(a,b)}{\pB(b)}
\quad=\quad
\begin{array}{|c|c|c|c|}
\hline a\setminus b & 0 & 1 & \varnothing\\
\hline 0 & \eta\,\frac{\mathsf{P}_{00}}{\PB{0}} & \eta\,\frac{\mathsf{P}_{01}}{\PB{1}} & \eta\,\PA{0}\\
\hline 1 & \eta\,\frac{\mathsf{P}_{10}}{\PB{0}} & \eta\,\frac{\mathsf{P}_{11}}{\PB{1}} & \eta\,\PA{1}\\
\hline \varnothing & \neta & \neta & \neta
\\\hline \end{array}\,,
\label{eqs:simplified_lossy_corr_conditional}
\end{equation}
which is obtained by dividing each column of Tab.~\eqref{eqs:simplified_lossy_corr} by the corresponding marginal probability of Bob in Eq.~\eqref{eq:bob_marginals}.
The columns of Tab.~\eqref{eqs:simplified_lossy_corr_conditional} determine then the conditional entropy $H(A|B)$, i.e.~the EC-term, as after defining the entropy of a probability vector as $H\!\left\{ (p_{i})_i\right\}\eqdef -\sum_{i}p_{i}\log_2 p_{i}$ for any $\sum_i p_i = 1$, it can be just written as a sum of entropies for each of the columns, i.e.:
\begin{equation}
\label{eq:ec_noprep}
H(A'|B)_{\text{no-prep}}=\eta\,\PB{0}\;H\!\left\{\eta\,\frac{\mathsf{P}_{00}}{\PB{0}},\eta\,\frac{\mathsf{P}_{10}}{\PB{0}},\neta\right\}+\eta\,\PB{1}\;H\!\left\{\eta\,\frac{\mathsf{P}_{01}}{\PB{1}},\eta\,\frac{\mathsf{P}_{11}}{\PB{1}},\neta\right\}+\neta\;H\!\left\{\eta\PA{0},\eta\PA{1},\neta\right\}.
\end{equation}

%%%%%%%%%%%%%%%%%%%%%%%%%%%%%%%%%%%%%%%%%%%%%%%%%%%%%%%%
\subsubsection{Deterministic binning}
We now consider the case when Alice deterministically bins every no-click event $\varnothing$. Without loss of generality, we may assume she always interprets~it as the $0$-outcome. This formally corresponds to her applying a stochastic map, see also \eqnref{eq:det_bin} of the main text, of the form
\begin{equation}
\mathcal{S_{\text{det}}}=
\left(\begin{array}{ccc}
1 & 0 & 1\\
0 & 1 & 0
\end{array}\right)
\label{eq:det_bin_appendix}
\end{equation}
to Tab.~\eqref{eqs:simplified_lossy_corr}, so that the resulting shared correlation then reads
\begin{equation}
\p{A' B}(a', b)_{\text{det}}
\quad=\quad
\begin{array}{|c|c|c|c|}
\hline a'\setminus b & 0 & 1 & \varnothing\\
\hline 0 & 
\begin{tabular}{@{}c@{}}
$\eta^{2}\,\mathsf{P}_{00} +\eta\neta\PB{0}$
\end{tabular} & 
\begin{tabular}{@{}c@{}}
$\eta^{2}\,\mathsf{P}_{01}+\eta\neta\PB{1}$
\end{tabular} & 
\begin{tabular}{@{}c@{}}
$\neta\eta\PA{0}+\neta^{2}$
\end{tabular}\\
\hline 1 & \eta^{2}\,\mathsf{P}_{10} & \eta^{2}\,\mathsf{P}_{11} & \eta\neta\PA{1}\\\hline 
\end{array}
\,,
\end{equation}
whose first row is obtained by summing the first and the third row of Tab.~\eqref{eqs:simplified_lossy_corr}.

Again, in order to determine the EC-term we compute Alice's conditional probability distribution by dividing the columns of Tab.~\eqref{eq:simplified_lossy_det} by the corresponding marginal probabilities~\eqref{eq:bob_marginals}, i.e.:
\begin{equation}
\label{eq:simplified_lossy_det_cond}
\p{A'|B}(a'|b)_{\mathrm{det}}\quad=\quad\frac{\p{A'B}(a',b)}{\pB(b)}
\quad=\quad\begin{array}{|c|c|c|c|}
\hline a'\setminus b & 0 & 1 & \varnothing\\
\hline 0 & 
\begin{tabular}{@{}c@{}}
$\eta\,\frac{\mathsf{P}_{00}}{\PB{0}}+ \neta$\end{tabular} & \begin{tabular}{@{}c@{}} $\eta\,\frac{\mathsf{P}_{01}}{\PB{1}} + \neta$\end{tabular} & \begin{tabular}{@{}c@{}} $\eta\PA{0}+\neta$\end{tabular}\\
\hline 1 & \eta\,\frac{\mathsf{P}_{10}}{\PB{0}} &\eta\,\frac{\mathsf{P}_{11}}{\PB{1}} & \eta\PA{1}\\\hline 
\end{array}\,,
\end{equation}
which allows to directly compute the relevant conditional entropy in case Alice bins deterministically:
\begin{equation}
%\begin{gathered}
\label{eq:ec_det}
H(A'|B)_{\text{det}}=\eta\,\PB{0}\,h\!\left[\frac{\eta\,\mathsf{P}_{10}}{\PB{0}}\right]+\eta\,\PB{1}\,h\!\left[\frac{\eta\,\mathsf{P}_{11}}{\PB{1}}\right]+\neta\,h\!\left[\eta\PA{1}\right],
%\end{gathered}
\end{equation}
where $h[x]\eqdef -x\log_{2}x-(1-x)\log_{2}(1-x)$ is the binary entropy function. 

%%%%%%%%%%%%%%%%%%%%%%%%%%%%%%%%%%%%%%%%%%%%%%%%%%%%%%%
\subsubsection{Deterministic binning with noisy preprocessing (bit-flip)}
In case Alice applies further noisy preprocessing~\cite{HST+20,WAP20,SBV+20} to her bit-string, she simply flips the value of each bit with probability $\pp$ after having binned them deterministically. This corresponds to her applying instead a stochastic matrix:
\begin{equation}
\label{eq:noisy_prep_map}
\mathcal{S_{\text{det+n.p.}}}=\left(
\begin{array}{ccc}
1-\pp & \pp & 1-\pp\\
\pp & 1-\pp & \pp
\end{array}\right),
\end{equation}
which consistently reproduces the one of deterministic binning in \eqnref{eq:det_bin_appendix} (and \eqnref{eq:det_bin}) when letting $\pp\to0$. On the other hand, it follows that the conditional distribution of Alice can thus be obtained by ``mixing'' the two rows of Tab.~\eqref{eq:simplified_lossy_det_cond} with weights $1-\pp$ and $\pp$, respectively, accounting for the bit-flip errors, i.e.:
\begin{equation}
\label{eq:simplified_lossy_det_pp_cond}
\p{A'|B}(a'|b)_{\text{det+n.p.}}\quad=\quad
\begin{array}{|c|c|c|c|}
\hline a'\setminus b & 0 & 1 & \varnothing\\
\hline 0 & \begin{tabular}{@{}c@{}}$(1-\pp)\left(\eta\,\frac{\mathsf{P}_{00}}{\PB{0}}+ \neta\right)$\\$ + \pp \left(\eta\,\frac{\mathsf{P}_{10}}{\PB{0}}\right)$\end{tabular} & \begin{tabular}{@{}c@{}}$(1-\pp)\left(\eta\,\frac{\mathsf{P}_{01}}{\PB{1}}+ \neta\right)$\\$ + \pp \left(\eta\,\frac{\mathsf{P}_{11}}{\PB{1}}\right)$\end{tabular} & \begin{tabular}{@{}c@{}}$ (1-\pp)\left(\eta\PA{0}+\neta\right)$\\$+\pp \left( \eta\PA{1} \right)$\end{tabular}\\
\hline 1 & \begin{tabular}{@{}c@{}}$\pp\left(\eta\,\frac{\mathsf{P}_{00}}{\PB{0}}+ \neta\right)$\\$ + (1-\pp) \left(\eta\,\frac{\mathsf{P}_{10}}{\PB{0}}\right)$\end{tabular} &\begin{tabular}{@{}c@{}}$ \pp \left(\eta\,\frac{\mathsf{P}_{01}}{\PB{1}}+ \neta\right)$\\$ + (1-\pp) \left(\eta\,\frac{\mathsf{P}_{11}}{\PB{1}}\right)$\end{tabular} & \begin{tabular}{@{}c@{}}$ \pp\left(\eta\PA{0}+\neta\right)$\\$+ (1-\pp) \left( \eta\PA{1} \right)$\end{tabular}\\\hline \end{array}\,.
\end{equation}
Evaluating now the conditional entropy based on the above conditional distribution, we obtain the EC-term as 
\begin{align}
\label{eq:ec_np}
H(A'|B)_{\text{det+n.p.}} &= \eta\,\PB{0}\,h\!\left[(1-\pp)\left(\eta\,\frac{\mathsf{P}_{00}}{\PB{0}}+\neta\right)+\pp\left(\frac{\eta\,\mathsf{P}_{10}}{\PB{0}}\right)\right]+\eta\,\PB{1}\,h\!\left[(1-\pp)\left(\eta\,\frac{\mathsf{P}_{01}}{\PB{1}}+\neta\right)+\pp\left(\frac{\eta\,\mathsf{P}_{11}}{\PB{1}}\right)\right]\nonumber\\&\quad+\neta\,h\!\left[(1-\pp)\left(\eta\PA{0}+\neta\right)+\pp\left(\eta\PA{1}\right)\right],
\end{align}
which, as expected, reproduces $H(A'|B)_\text{det}$ in Eq.~\eqref{eq:ec_det} after letting $\pp\to0$.

%%%%%%%%%%%%%%%%%%%%%%%%%%%%%%%%%%%%%%%%%%%%%%%%%%%%%%%
\subsubsection{Random binning}
We also consider the case when Alice rather randomly bins her ternary variable $A$, in particular, she assigns each outcome $\varnothing$ with equal probability to either outcome ‘0’ or ‘1’. The corresponding stochastic matrix applied by Alice to Tab.~\eqref{eqs:simplified_lossy_corr} then reads 
\begin{equation}
\label{eq:rand_bin_map}
\mathcal{S_{\text{rand}}}=\left(\begin{array}{ccc}
1 & 0 & \frac{1}{2}\\
0 & 1 & \frac{1}{2}
\end{array}\right),
\end{equation}
so that the third row of Tab.~\eqref{eqs:simplified_lossy_corr} gets redistributed equally (with a factor of $1/2$) over the first two rows, i.e.:
\begin{equation}
\label{eq:simplified_lossy_rand}
p_{A' B}(a', b)_{\text{rand}}\quad=\quad\begin{array}{|c|c|c|c|}
\hline a'\setminus b & 0 & 1 & \varnothing\\
\hline 0 & \eta^{2}\,\mathsf{P}_{00}+\frac{1}{2}\eta\neta\PB{0} & \eta^{2}\,\mathsf{P}_{01}+\frac{1}{2}\eta\neta\PB{1} & \neta\eta\PA{0}+\frac{1}{2}\neta^{2}\\
\hline 1 & \eta^{2}\,\mathsf{P}_{10}+\frac{1}{2}\eta\neta\PB{0} & \eta^{2}\,\mathsf{P}_{11}+\frac{1}{2}\eta\neta\PB{1} & \eta\neta\PA{1}+\frac{1}{2}\neta^{2}\\\hline \end{array}\,.
\end{equation}
As before, we determine then the probability distribution of Alice's outcomes conditioned on Bob's as
\begin{equation}
\p{A'|B}(a'|b)_{\text{rand}}\quad=\quad\begin{array}{|c|c|c|c|}
\hline a'\setminus b & 0 & 1 & \varnothing\\
\hline 0 & \eta\,\frac{\mathsf{P}_{00}}{\PB{0}}+\frac{1}{2}\neta & \eta\,\frac{\mathsf{P}_{01}}{\PB{1}}+\frac{1}{2}\neta & \eta\PA{0}+\frac{1}{2}\neta\\
\hline 1 & \eta\,\frac{\mathsf{P}_{10}}{\PB{0}}+\frac{1}{2}\neta & \eta\,\frac{\mathsf{P}_{11}}{\PB{1}}+\frac{1}{2}\neta & \eta\PA{1}+\frac{1}{2}\neta\\\hline 
\end{array}
\,,
\end{equation}
with the help of which we calculate the EC-term applicable to the case of random binning:
\begin{equation}
\label{eq:ec_rand}
H(A'|B)_{\text{rand}}=\eta\PB{0}\,h\!\left[\frac{\eta\,\mathsf{P}_{00}}{\PB{0}}+\frac{\neta}{2}\right]+\eta\PB{1}\,h\!\left[\frac{\eta\,\mathsf{P}_{01}}{\PB{1}}+\frac{\neta}{2}\right]+\neta\,h\!\left[\eta\PA{0}+\frac{\neta}{2}\right].
\end{equation}

%%%%%%%%%%%%%%%%%%%%%%%%%%%%%%%%%%%%%%%%%%%%%%%%%%%%%%%
\subsubsection{Random binning with noisy preprocessing (bit-flip).}
Finally, as before for deterministic binning, we consider the case in which Alice, apart from randomly binning the $\varnothing$-outcome, applies also noisy preprocessing~\cite{HST+20,WAP20} to the resulting bit, i.e.~flips its value with probability $\pp$. This then corresponds to her applying to Tab.~\eqref{eqs:simplified_lossy_corr} the stochastic matrix 
\begin{equation}
\label{eq:noisy_prep_map_rand}
\mathcal{S_{\text{rand+n.p.}}}=\left(\begin{array}{ccc}
1-\pp & \pp & \frac12\\
\pp & 1-\pp & \frac12
\end{array}\right),
\end{equation}
which consistently reproduces the one of random binning \eqref{eq:rand_bin_map} when letting $\pp\to0$. As before, the conditional probability distribution of Alice can then be obtained by just ``mixing'' the two rows of Tab.~\eqref{eq:simplified_lossy_rand} (describing the case of random binning) with probabilities $\pp$ and $1-\pp$, i.e.:
\begin{equation}
\label{eq:simplified_lossy_rand_np}
\p{A'|B}(a'|b)_{\text{rand+n.p.}}\quad=\quad
\begin{array}{|c|c|c|c|}
\hline a'\setminus b & 0 & 1 & \varnothing\\
\hline 0 & \begin{tabular}{@{}c@{}}$(1-\pp) \eta\,\frac{\mathsf{P}_{00}}{\PB{0}} $\\$+ \pp \eta\,\frac{\mathsf{P}_{10}}{\PB{0}} + \frac{1}{2}\neta$ \end{tabular} & \begin{tabular}{@{}c@{}}$(1-\pp) \eta\,\frac{\mathsf{P}_{01}}{\PB{1}} $\\$+ \pp \eta\,\frac{\mathsf{P}_{11}}{\PB{1}} + \frac{1}{2}\neta$ \end{tabular} & \begin{tabular}{@{}c@{}}$(1-\pp)\eta\PA{0}$\\$+\pp\eta\PA{1}+\frac{1}{2}\neta$\end{tabular}\\
\hline 1 & \begin{tabular}{@{}c@{}}$(1-\pp) \eta\,\frac{\mathsf{P}_{10}}{\PB{0}} $\\$+ \pp \eta\,\frac{\mathsf{P}_{00}}{\PB{0}} + \frac{1}{2}\neta$ \end{tabular} & \begin{tabular}{@{}c@{}}$(1-\pp) \eta\,\frac{\mathsf{P}_{11}}{\PB{1}} $\\$+ \pp \eta\,\frac{\mathsf{P}_{01}}{\PB{1}} + \frac{1}{2}\neta$ \end{tabular} & \begin{tabular}{@{}c@{}}$(1-\pp)\eta\PA{1}$\\$+\pp\eta\PA{0}+\frac{1}{2}\neta$\end{tabular}\\\hline 
\end{array}\,,
\end{equation}
so that the relevant conditional entropy constituting the EC-term reads:
\begin{eqnarray}
\label{eq:ec_np_rand}
H(A'|B)_{\text{rand+n.p.}}
&=& 
\eta\PB{0}\,h\!\left[(1-\pp) \eta\,\frac{\mathsf{P}_{00}}{\PB{0}} + \pp \eta\,\frac{\mathsf{P}_{10}}{\PB{0}} + \frac{1}{2}\neta\right]\\
&&+\;\eta\PB{1}\,h\!\left[(1-\pp) \eta\,\frac{\mathsf{P}_{01}}{\PB{1}} + \pp \eta\,\frac{\mathsf{P}_{11}}{\PB{1}} + \frac{1}{2}\neta\right]+\neta\,h\!\left[(1-\pp)\eta\PA{0}+\pp\eta\PA{1}+\frac{1}{2}\neta\right].
\nonumber
\end{eqnarray}

%%%%%%%%%%%%%%%%%%%%%%%%%%%%%%%%%%%%%%%%%%%%%%%%%%%%%%%%%%%%%%%%%%%%%%%%%%%%%%%%%%%%%%%%%%%%%%%%%%%%%%%%%%%%%%
\subsection{Calculation of the PA-term \texorpdfstring{$H(A'|E)$}{Lg}}
%%%%%%%%%%%%%%%%%%%%%%%%%%%%%%%%%%%%%%%%%%%%%%%%%%%%%%%%%%%%%%%%%%%%%%%%%%%%%%%%%%%%%%%%%%%%%%%%%%%%%%%%%%%%%%
%
As within the CC attack Eve knows whether a local or a nonlocal correlation is being distributed to Alice and Bob, the entropy of the variable $A$ (which describes the output of the measurement used by Alice for key distribution) conditioned on Eve's knowledge is given by the convex mixture of local and non-local contributions. Moreover, Eve not only knows when a local distribution is shared by the parties, but also knows then perfectly the outcomes $A$ and $B$ of Alice and Bob, respectively. Hence, the contribution of the local distribution to the conditional entropy of $A$ is zero, unless Alice performs a non-deterministic preprocessing $p_{A'|A}\equiv\cS$ of the outcome $A$ that introduces some randomness, so that the knowledge of Eve about the resulting variable $A'$ is no longer perfect. 

In order to determine the conditional entropy $H(A'|E)$, we must only track Eve's knowledge of Alice's outputs. Hence, without loss of generality, we can assume Eve to hold a random variable $E$ taking four values $e\in \{ \{\tilde{e}\},?\}$, where $?$ means that she distributed a nonlocal correlation and has no knowledge of Alice's output, while values $\tilde{e} \in \{0, 1, \varnothing \}$ correspond to the perfect knowledge of Alice's output $A$, which Eve possesses after distributing a local correlation (so that always $\tilde{e}=a$). 

As a consequence, we can generally write the PA-term for the CC attack as
\begin{align}
H(A'|E) &=\sum_{e} p(E=e) H\left(A'|E=e\right) = (1-q^\cL) H(A'|E=\,?) + q^\cL \sum_{\tilde{e}} p(E=\tilde{e}|\cL) H\left(A'|E=\tilde{e}\right),
\label{eq:H(A':E)}
\end{align}
where $q^\cL$ is the local weight, so that $p(E=\,?) = 1-q^\cL$ and $p(E=\tilde{e}) = q^\cL \, p(E=\tilde{e}|\cL)$, with $p(E=\tilde{e}|\cL)$ denoting the probability of Eve recording $\tilde{e}$ given she has distributed a local correlation.

As Eve has perfect knowledge of Alice's outcome, the conditional entropy within the ``local rounds'' is
\begin{align}
H\left(A'|E=\tilde{e}\right) &= -\sum_{a'} \p{A'}(a'|E=\tilde{e}) \log_2\!\left[ \p{A'}(a'|E=\tilde{e}) \right] = -\sum_{a',a} \cS_{a'a} \, \pA(a|E=\tilde{e}) \,\log_2\!\left [ \sum_{a} \cS_{a'a} \, \pA(a|E=\tilde{e}) \right ] \nonumber\\
&= - \sum_{a'} \cS_{a' \tilde{e}} \log_2 \cS_{a' \tilde{e}} 
=: H\!\left(\cS_{\cdot \tilde{e}}\right),
\end{align}
where we have used the fact that $\pA(a|E=\tilde{e}) = \delta_{a, \tilde{e}}$, and defined above $H\!\left(\cS_{\cdot \tilde{e}}\right):= H[A']_{p(A'|A=\tilde{e})}$ as the entropy of the distribution described by the $\tilde{e}$-column of the stochastic matrix $\cS$, which equivalently represents the randomness (entropy) of the preprocessed variable $A'$ when $A=\tilde{e}$.

We can further simplify the expression \eqref{eq:H(A':E)} by expanding the probability $p(E=\tilde{e}|\cL)$, after realising that $p(E=\tilde{e}|\cL)=\pA^\cL(a=\tilde{e})$, where $\pA^\cL(a)=\sum_b \pAB^\cL(a,b|\xk,\yk)$ is the Alice's marginal of the local correlation. As the convex decomposition of the observed correlation \eref{eq:CC_constr} naturally carries over onto the marginal, i.e.:
\begin{equation}
\label{eq:conv_decomp_marginA}
\pA(a) = q^\cL \, \pA^\cL(a) + (1-q^\cL) \, \pA^\cNL(a),
\end{equation}
we can then explicitly compute
\begin{equation}
\label{eq:Alice_local_prob}
\pA^\cL(a) = 
\begin{cases}
\eta_\cL\,\PA{a} - \frac{1-q^\cL}{q^\cL} (1-V)\left(\QA{a}-\frac12\right) & \text{if } a \in \{0,1\} \\
1-\eta_\cL =: \neta_\cL & \text{if } a = \varnothing
\end{cases},
\end{equation}
after substituting for the observed Alice's marginal, $\pA(a)=\sum_b\pAB(a,b)$, according to the lossy correlation~\eqref{eqs:simplified_lossy_corr}, while the nonlocal contribution in \eqnref{eq:conv_decomp_marginA} corresponds to the noiseless $\pA^\cNL(a) = \QA{a}$. We also define $\eta_\cL$ as above in \eqnref{eq:eta_cL_xi_cL}, which should be understood as the effective ``local'' detection efficiency.

Finally, we arrive at the expression for the PA-term as
\begin{equation}
\label{eq:PA-term}
H(A'|E) = (1-q^\cL)\, H(A'|E=\,?) + q^\cL \left[ \pA^\cL(0) \,H\!\left(\cS_{\cdot 0}\right)  + \pA^\cL(1) \,H\!\left(\cS_{\cdot 1}\right) + \pA^\cL(\varnothing) \,H\!\left(\cS_{\cdot \varnothing}\right) \right],
\end{equation}
where we should recall that $H\left(A'|E=?\right)$ is the conditional entropy of Alice's outputs applicable whenever Eve distributes the nonlocal correlation within the CC attack, i.e.~$\pAB^{\cNL}(a) = \QA{a}$, and Alice preprocesses the outcome $A$ onto $A'$ according to the map $\cS$.

In what follows, we calculate in detail the PA-term \eqref{eq:PA-term} for the preprocessing strategies of Alice listed in Tab.~\ref{tab:critical}, as considered above in the evaluation of the EC-term, as well as the other two cases of random binning with and without noisy preprocessing.

%%%%%%%%%%%%%%%%%%%%%%%%%%%%%%%%%%%%%%%%%%%%%%%%%%%%%%%%
\subsubsection{No preprocessing or any deterministic binning}
Whenever the matrix $\mathcal{S}$ describes a stochastic map that is deterministic, i.e.~contains only 0s or 1s as its entries, the whole second term in \eqnref{eq:PA-term} identically vanishes. On the other hand, as within nonlocal rounds inconclusive outcomes never occur, and so any operations on $\varnothing$-outcomes are never performed, any binning strategy does not affect the first term in \eqnref{eq:PA-term}. Thus, we can write the PA-term in absence of preprocessing or for any deterministic binning as
\begin{equation}
\label{eq:pa_noprep}
H(A'|E)_\text{no-prep} = H(A'|E)_\text{det} = (1-q^\cL) \, H(A'|E=\,?) = (1-q^\cL)\, h\!\left[\mathsf{Q}_0^{\mathrm{A}}\right].
\end{equation} 

%%%%%%%%%%%%%%%%%%%%%%%%%%%%%%%%%%%%%%%%%%%%%%%%%%%%%%%%
\subsubsection{Deterministic binning with noisy preprocessing (bit-flip)}
In case Alice decides to further ``noisy preprocess'' her outcomes after having binned $\varnothing$ to $0$, then the overall preprocessing map she applies corresponds to the $\cS$-matrix introduced in Eq.~\eqref{eq:noisy_prep_map}. As a result, the first term in Eq.~\eqref{eq:PA-term} can be obtained from Eq.~\eqref{eq:pa_noprep} after including a bit-flip occurring with probability $\pp$, while the second term in Eq.~\eqref{eq:PA-term} is then no longer zero, as the entropy for each column of $\cS$ equals now $h[\pp]$. Thus, the full PA-term \eqref{eq:PA-term} then reads
\begin{equation}
\label{eq:pa_np}
H(A'|E)_\text{det+n.p.} = (1-q^\cL)\, h\!\left[(1-\pp)\,\mathsf{Q}_0^{\mathrm{A}} + \pp\,\mathsf{Q}_1^{\mathrm{A}}\right] + q^\cL \, h[\pp].
\end{equation}

%%%%%%%%%%%%%%%%%%%%%%%%%%%%%%%%%%%%%%%%%%%%%%%%%%%%%%%%
\subsubsection{Random binning}
As before, the first term in \eqnref{eq:PA-term} is unaffected by any binning of the $\varnothing$-outcomes and, hence, also when Alice bins these randomly. However, the second term in \eqnref{eq:PA-term} must now be evaluated based on the stochastic matrix $\cS$ given in \eqnref{eq:rand_bin_map}, which is no longer deterministic---its last column yields a non-trivial contribution. Hence, for random binning of inconclusive outcomes we obtain
\begin{equation}
\label{eq:pa_rand}
H(A'|E)_\text{rand} = (1-q^\cL)\, h\!\left[\mathsf{Q}_0^{\mathrm{A}}\right] + q^\cL \neta_\cL \, h\!\left[\frac{1}{2}\right] = (1-q^\cL)\, h\!\left[\mathsf{Q}_0^{\mathrm{A}}\right] + 1 - \eta.
\end{equation}

\subsubsection{Random binning with noisy preprocessing (bit-flip)}
In case Alice decides to further ``noisy preprocess'' her outcomes after having binned $\varnothing$ randomly to 0 and 1, she, in fact, implements the stochastic matrix $\cS$ given in Eq.~\eqref{eq:noisy_prep_map_rand}. Within the first term of Eq.~\eqref{eq:PA-term} one has to account for the bit-flip occurring with probability $\pp$ and arrives at the same expression as in Eq.~\eqref{eq:pa_np}. Whereas for the second term, we note that the entropy of the first two columns of $\cS$ in Eq.~\eqref{eq:noisy_prep_map_rand} is then $h[\pp]$, while the entropy of the last column is 1. Therefore, we have
\begin{equation}
\label{eq:pa_rand_np}
H(A'|E)_\text{rand+n.p.} = (1-q^\cL) h[(1-\pp)\,\mathsf{Q}_0^{\mathrm{A}} + \pp\,\mathsf{Q}_1^{\mathrm{A}}] + (\eta - 1 + q^\cL) \, h[\pp] + 1 - \eta.
\end{equation}

%%%%%%%%%%%%%%%%%%%%%%%%%%%%%%%%%%%%%%%%%%%%%%%%%%%%%%%%%%%%%%%%%%%%%%%%%%%%%%%%%%%%%%%%%%%%%%%%%%%%%%%%%%%%%%%%%%
%%%%%%%%%%%%%%%%%%%%%%%%%%%%%%%%%%%%%%%%%%%%%%%%%%%%%%%%%%%%%%%%%%%%%%%%%%%%%%%%%%%%%%%%%%%%%%%%%%%%%%%%%%%%%%%%%%
\section{CC-based upper bounds on key rates and the resulting noise thresholds}
\label{sec:UBs_thresholds_CC}
%%%%%%%%%%%%%%%%%%%%%%%%%%%%%%%%%%%%%%%%%%%%%%%%%%%%%%%%%%%%%%%%%%%%%%%%%%%%%%%%%%%%%%%%%%%%%%%%%%%%%%%%%%%%%%%%%%
%%%%%%%%%%%%%%%%%%%%%%%%%%%%%%%%%%%%%%%%%%%%%%%%%%%%%%%%%%%%%%%%%%%%%%%%%%%%%%%%%%%%%%%%%%%%%%%%%%%%%%%%%%%%%%%%%%

%%%%%%%%%%%%%%%%%%%%%%%%%%%%%%%%%%%%%%%%%%%%%%%%%%%%%%%%%%%%%%%%%%%%%%%%%%%%%%%%%%%%%%%%%%%%%%%%%%%%%%%%%%%%%%%%
\subsection{One-way protocols involving maximally entangled states}
\label{app:1way_maxent_UBs}
%%%%%%%%%%%%%%%%%%%%%%%%%%%%%%%%%%%%%%%%%%%%%%%%%%%%%%%%%%%%%%%%%%%%%%%%%%%%%%%%%%%%%%%%%%%%%%%%%%%%%%%%%%%%%%%%
%
In this section we utilise the formulae derived in \appref{sec:key_rates_CC} for the EC- and PA-terms under particular preprocessing strategies of Alice, in order to determine the corresponding upper bounds~\eqref{eq:appendix-oneway-UB} on the one-way key rates for the standard CHSH-based 2333- and 2233-protocols (in the finite detection efficiency model), as well as 2322- and 2222-protocols (in the finite visibility model). The goal is to determine analytically the tolerable noise thresholds below which no key rate can be distilled, some of which are listed in Table~\ref{tab:critical} for specific preprocessing strategies. 

In this section we assume that the parties ideally measure the pure, maximally entangled state with $\theta = \pi/2$ in~\eqref{eq:part_ent_state} via projective measurements~\eqref{eq:standard_measurements} with the measurement settings $\xk,\yk=\{0,2\}$ being used for key generation in the 2333- and 2322-protocols, and any settings $\xk,\yk \in \{0,1\}$ in the 2233- and 2222-protocols. While the obtained results for the EC- and PA-terms hold generally for any $\eta$ and $V$, we consider here specifically two cases of purely lossy correlations (with $V=1$) and of purely noisy correlations (with $\eta=1$).

%%%%%%%%%%%%%%%%%%%%%%%%%%%%%%%%%%%%%%%%%%%%%%%%%%%%%%%
\subsubsection{Finite detection efficiency}
Given perfect visibility ($V=1$) but imperfect detection efficiency ($\eta<1$), the correlation \eref{eqs:simplified_lossy_corr} used for the key generation simplifies to the purely lossy one~\eqref{eq:lossy_corr2} with all $\mathsf{P}_{ab} = \mathsf{Q}_{ab}$. Moreover, in case of the 2333-protocol we have from~\eqnref{eqs:biased_CHSH} that $\mathsf{Q}_{ab}=\frac{1}{2} \delta_{a,b}$ with marginals $\QA{a} = \QB{b} = \frac{1}{2}$, whereas for the 2233-protocol $\mathsf{Q}_{00} = \mathsf{Q}_{11} = (2-\sqrt{2})/8$ and $\mathsf{Q}_{01} = \mathsf{Q}_{10} = (2+\sqrt{2})/8$ if $\xk = \yk = 1$, and $\mathsf{Q}_{00} = \mathsf{Q}_{11} = (2+\sqrt{2})/8$, $\mathsf{Q}_{01} = \mathsf{Q}_{10} = (2-\sqrt{2})/8$ otherwise (with marginal probabilities also always equal to $1/2$).

%%%%%%%%%%%%%%%%%%%%%%%%%%%
\paragraph{No preprocessing.}
In absence of any preprocessing map, we use Eqs.~\eref{eq:ec_noprep} and \eref{eq:pa_noprep} to calculate $\UB{\roneway{no-prep}} = H(A'|E)_\text{no-prep} - H(A'|B)_{\text{no-prep}}$, which after substituting also for the $\mathsf{Q}$-probabilities of the 2333-protocol and the optimal local weight \eref{eqs:local_weight_lossy} reads
\begin{equation}
\UB{\roneway{no-prep}}(\eta) = \eta  \left[4 \eta + \log_2 \eta  - 2 \sqrt{2} (1-\eta)\right] + 2 \log_2 (1 - \eta ) -3 \eta,
\end{equation}
and leads to the critical detection efficiency
\begin{equation}
\crit{\eta}^\text{no-prep} \approx 91.85\%.
\label{eq:eta_crit_noprep}
\end{equation}
Following the same steps for the 2233-protocol one arrives at the formula for $\UB{\roneway{no-prep}}$ with a zero at $\crit{\eta}^\text{no-prep} \approx 96.90\%$ 
irrespectively of the particular choice of key settings $\xk,\yk\in\{0,1\}$.

%%%%%%%%%%%%%%%%%%%%%%%%%%%
\paragraph{Deterministic binning.}
In case Alice applies deterministic binning as her preprocessing strategy, we use Eqs.~\eref{eq:ec_det} and \eref{eq:pa_noprep} to calculate $\UB{\roneway{det}} = H(A'|E)_\text{det} - H(A'|B)_{\text{det}}$ instead, which after substituting for the $\mathsf{Q}$-probabilities of the 2333-protocol and the optimal local weight \eref{eqs:local_weight_lossy} reads
\begin{equation}
\UB{\roneway{det}}(\eta) = \eta  \left(3\eta  + 2\sqrt{2} \eta + \frac{\log_2 \eta }{2}- 2\sqrt{2}\right) - \frac{1-\eta }{2} \left[-(2-\eta) \log_2 (2-\eta ) - \eta  \log_2 (1-\eta) \right] - 1 -\eta,
\label{eq:UB_roneway_det}
\end{equation}
and leads to the critical detection efficiency
\begin{equation}
\crit{\eta}^\text{det} \approx 89.16\%.
\end{equation}
The fact that $\crit{\eta}^\text{det}<\crit{\eta}^\text{no-prep}$ suggests that the binning procedure of the inconclusive outcomes is indeed beneficial for the parties to be able to tolerate lower detection efficiencies. Following the same steps for the 2233-protocol one arrives at the formula for $\UB{\roneway{det}}$ with a zero at $\crit{\eta}^\text{det} \approx 94.80\%$ irrespectively of the choice of key settings.

%%%%%%%%%%%%%%%%%%%%%%%%%%%
\paragraph{Deterministic binning with noisy preprocessing.}
\label{app:det_bin_noisy_prep}
If Alice decides to apply noisy preprocessing apart from binning deterministically her inconclusive outcomes $\varnothing$, we have $\UB{\roneway{det+n.p.}} = H(A'|E)_\text{det+n.p.} - H(A'|B)_{\text{det+n.p.}}$ that can be calculated using Eqs.~\eref{eq:ec_np} and \eref{eq:pa_np}, so that after substituting for the $\mathsf{Q}$-probabilities of the 2333-protocol it reads
\begin{equation}
\UB{\roneway{det+n.p.}}(\eta) 
= 1-q^\cL + q^\cL \, h[\pp] - \frac{\eta}{2}\,h\!\left[\pp\right] 
-\frac{\eta}{2}\,h\!\left[(1-\pp)\left(1-\eta\right)+\pp \eta \right] - (1-\eta)\,h\!\left[(1-\pp)\left(1-\frac{\eta}{2}\right)+\pp\left(\frac{\eta}{2}\right)\right].
\label{eq:UB_roneway_det_np}
\end{equation}
Substituting then for the optimal local weight, $q^\cL$ in Eq.~\eqref{eqs:local_weight_lossy}, one can verify that the critical detection efficiency gets smaller with the bit-flip probability approaching $\pp \to \frac{1}{2}{}_\pm$. Although in such a regime the upper bound and, hence, any attainable rate is severely suppressed, in order to determine its lowest possible positive value we expand $\UB{\roneway{det+n.p.}}$ in $\delta$ after substituting for $\pp = \frac{1}{2} \pm \delta$, i.e.:
\begin{equation}
\UB{\roneway{det+n.p.}}(\eta) 
= \frac{2 \eta \left(\eta ^2+2 \sqrt{2} \eta + 4 \eta -2 \sqrt{2} -4 \right)}{\ln 2} \delta ^2 +O\left(\delta ^3\right),
\end{equation}
which allows us to locate the zero at 
\begin{equation}
\crit{\eta}^\text{det+n.p.} = \sqrt{10+6 \sqrt{2}}-2-\sqrt{2} \approx 88.52\%.
\end{equation}
Following the same steps for the 2233-protocol one arrives at the formula for $\UB{\roneway{det+n.p.}}$ with a zero at 
$\crit{\eta}^\text{det+n.p.} = 2 \left(\sqrt{8 + 5 \sqrt{2}} - 2 - \sqrt{2}\right) \approx 93.59\%$
irrespectively of the choice of key settings.

%%%%%%%%%%%%%%%%%%%%%%%%%%%
\paragraph{Random binning.}
In case Alice applies random binning as her preprocessing strategy, we use Eqs.~\eref{eq:ec_rand} and \eref{eq:pa_rand} to calculate $\UB{\roneway{rand}} = H(A'|E)_\text{rand} - H(A'|B)_{\text{rand}}$, which after substituting for the $\mathsf{Q}$-probabilities of the 2333-protocol and the optimal local weight \eref{eqs:local_weight_lossy} reads
\begin{equation}
\UB{\roneway{rand}}(\eta) 
= \frac{\eta}{2}   \left(\eta  \left( \log_2 (1+\eta)-\log_2 (1-\eta )\right) + \log_2 (1-\eta )+\log_2 (1+\eta) -2 (1-\eta ) \left(3+2 \sqrt{2}\right)  \right),
\end{equation}
and leads to the critical detection efficiency
\begin{equation}
\crit{\eta}^\text{rand} \approx 88.34\%.
\end{equation}
Following the same steps for the 2233-protocol one arrives at the formula for $\UB{\roneway{rand}}$ with a zero at $\crit{\eta}^\text{rand} \approx 94.03\%$ irrespectively of the choice of key settings.

%%%%%%%%%%%%%%%%%%%%%%%%%%%
\paragraph{Random binning with noisy preprocessing.}
\label{par:random_binn_np}
If Alice decides to apply noisy preprocessing apart from randomly binning her inconclusive outcomes $\varnothing$, we have $\UB{\roneway{rand+n.p.}} = H(A'|E)_\text{rand+n.p.} - H(A'|B)_{\text{rand+n.p.}}$ that can be calculated using Eqs.~\eref{eq:ec_np_rand} and \eref{eq:pa_rand_np}, so that after substituting for the $\mathsf{Q}$-probabilities of the 2333-protocol it reads
\begin{equation}
\UB{\roneway{rand+n.p.}}(\eta)
=1-q^\cL + (\eta - 1 + q^\cL) \, h[\pp]-\eta \,h\!\left[\pp \eta  + \frac{1}{2}(1-\eta)\right].
\end{equation}
Substituting then for the optimal local weight, $q^\cL$ in Eq.~\eqref{eqs:local_weight_lossy}, one can verify that the critical detection efficiency gets smaller with the bit-flip probability approaching $\pp \to \frac{1}{2}_\pm$. Hence, similarly to the ``det+n.p.'' case, we can determine its lowest possible positive value by expanding $r^\uparrow_\text{rand+n.p.}$ in $\delta$ after substituting also for $\pp = \frac{1}{2} \pm \delta$, i.e.:
\begin{equation}
\UB{\roneway{rand+n.p.}}(\eta)
= \frac{2  \eta  \left(\eta  \left(\eta +2 \sqrt{2}+3\right)-2 \sqrt{2}-3\right) \delta ^2}{\log (2)}+O\left(\delta^3\right),
\end{equation}
which implies
\begin{equation}
\label{eq:crit_efficiency_rand_np}
\crit{\eta}^\text{rand+n.p.}= \frac{1}{2} \left(\sqrt{29+20 \sqrt{2}} -3-2 \sqrt{2}\right) \approx 87.01\%.
\end{equation}
Following the same steps for the 2233-protocol one arrives at the formula for $r^\uparrow_\text{rand+n.p.}$ with a zero at 
$\crit{\eta}^\text{rand} = \sqrt{23 + 16 \sqrt{2}} -3 - 2 \sqrt{2}  \approx 92.64\%$
irrespectively of the choice of key settings.

%%%%%%%%%%%%%%%%%%%%%%%%%%%%%%%%%%%%%%%%%%%%%%%%%%%%%%
\subsubsection{Finite visibility}
Given perfect detection efficiency ($\eta=1$) but imperfect visibility ($V<1$), one should consider for key generation the correlation~\eqref{eqs:simplified_lossy_corr} after setting $\eta = 1$ instead, where now in the case of the 2333-protocol:~$\mathsf{P}_{ab}=V \frac{1}{2} \delta_{a,b} + \frac{1-V}{4}$ with marginals $\PA{a} = \PB{b} = \frac12$; whereas for the 2233-protocol:~$\mathsf{P}_{00} = \mathsf{P}_{11} = V (2-\sqrt{2})/8 + \frac{1-V}{4}$ and $\mathsf{P}_{01} = \mathsf{P}_{10} = V (2+\sqrt{2})/8 + \frac{1-V}{4}$ if $\xk = \yk = 1$, and $\mathsf{P}_{00} = \mathsf{P}_{11} = V (2+\sqrt{2})/8 + \frac{1-V}{4}$, $\mathsf{P}_{01} = \mathsf{P}_{10} = V (2-\sqrt{2})/8 + \frac{1-V}{4}$ otherwise (with marginal probabilities also always equal to $\frac12$). Consistently, all the noiseless $\mathsf{Q}$-probabilities specified previously when dealing with purely lossy correlations can be recovered by setting $V=1$ in all the $\mathsf{P}$-probabilities listed above.

%%%%%%%%%%%%%%%%%%%%%%%%%%%
\paragraph{No preprocessing.}
In absence of any preprocessing map the upper bound on the one-way rate can be calculated again using Eqs.~\eref{eq:ec_noprep} and \eref{eq:pa_noprep} as $\UB{\roneway{no-prep}} = H(A'|E)_\text{no-prep} - H(A'|B)_{\text{no-prep}}$, which after substituting for the $\mathsf{P}$- and $\mathsf{Q}$-probabilities of the 2333-protocol, $\eta = 1$, and the optimal local weight \eref{eq:visibility_local_weight} reads
\begin{equation}
\begin{gathered}
\UB{\roneway{no-prep}}(V) 
= V (2+\sqrt2) - h\!\left[\frac{1+V}{2}\right] - 1 - \sqrt2,
\end{gathered}
\end{equation}
and vanishes at the critical visibility:
\begin{equation}
\crit{V}^\text{no-prep} \approx 83.00\%.
\label{eq:v_crit_noprep}
\end{equation}
Following the same steps for the 2222-protocol one arrives at the formula for $r^\uparrow_\text{no-prep}$ with a zero at 
$\crit{V}^\text{no-prep} \approx 90.61\%$,
irrespectively of the particular choice of key settings $\xk,\yk\in\{0,1\}$.

%%%%%%%%%%%%%%%%%%%%%%%%%%%
\paragraph{Noisy preprocessing.}
If Alice decides to apply noisy preprocessing to her binary variable, the upper bound ~\eqref{eq:appendix-oneway-UB} on the one-way rate can be evaluated by setting $\eta=1$ in either Eqs.~\eref{eq:ec_np}\&\eref{eq:pa_np} or Eqs.~\eref{eq:ec_np_rand}\&\eref{eq:pa_rand_np} as $\UB{\roneway{n.p.}} = H(A'|E)_{\bullet\text{+n.p.}} - H(A'|B)_{\bullet\text{+n.p.}}$. Because for the perfect detection efficiency the inconclusive outcomes never occur, formulae derived assuming any binning of the $\varnothing$-outcomes are valid upon setting $\eta = 1$. Hence, substituting also for the $\mathsf{P}$- and $\mathsf{Q}$--probabilities of the 2322-protocol, we arrive at
\begin{equation}
\UB{\roneway{n.p.}}(V) = 1-q^\cL + q^\cL h[\pp] - h\!\left[\left(\pp-\frac{1}{2}\right) V+\frac{1}{2}\right].
\end{equation}
Substituting then for the optimal local weight, $q^\cL$ in Eq.~\eqref{eq:visibility_local_weight}, one can again verify that the critical visibility gets smaller with the bit-flip probability approaching $\pp \to \frac{1}{2}_\pm$. Although in such a regime the upper bound, and hence any attainable rate, is severely suppressed, in order to determine its lowest possible positive value we expand $r^\uparrow_\text{n.p.}$ in $\delta$ after substituting also for $\pp = \frac{1}{2} \pm \delta$, i.e.:
\begin{equation}
\UB{\roneway{n.p.}}(V) = \frac{2  \left( V^2 - \left(1-V \right ) \left(2+\sqrt2 \right) \right)}{\ln 2} \delta ^2 +O\left(\delta ^3\right),
\end{equation}
which allows us to locate the zero at 
\begin{equation}
\label{eq:visibility_np}
\crit{V}^\text{n.p.} = \sqrt{\frac{7}{2}+2 \sqrt{2}} -1-\frac{1}{\sqrt{2}} \approx 80.85\%.
\end{equation}
Following the same steps for the 2222-protocol, one arrives at the formula for $r^\uparrow_\text{n.p.}$ with a zero at 
$\crit{V}^\text{n.p.} = \sqrt{10+6 \sqrt{2}}-2-\sqrt{2} \approx 88.52\%$,
irrespectively of the choice of key settings.

%%%%%%%%%%%%%%%%%%%%%%%%%%%%%%%%%%%%%%%%%%%%%%%%%%%%%%
\subsubsection{Optimisation over all preprocessing maps}
\label{app:1way_maxent_opt_maps}
In this section, we discuss the results obtained for the computation of the \emph{general} upper bounds applicable to one-way key rates~\eqref{eq:oneway_rate}---the asymptotic one-way key rates optimised over all preprocessing strategies including not just the stochastic mapping applied by Alice, $p_{A'|A}$, on her variable $A$, but also the extra message $M$ she prepares by applying $p_{M'|A'}$ on $A'$ and sends publicly to Bob, i.e.:
\begin{equation}
\roneway(A\to B)
\qquad\leq\qquad 
\UB{\roneway} \eqdef \max\limits_{p_{A'|A},\;p_{M|A'}} I(A' \! :\! B|M) - I(A' \! :\! E|M),
\label{eq:gen_up_bound}
\end{equation}
where the so-defined $r_\text{1-way}^\uparrow$ corresponds to the upper bound \eqref{eq:oneway_UB_ind_attack} being now crucially maximised over all the $p_{A'|A}$ and $p_{M|A'}$ preprocessing maps, including ones with $|A'| > 2$ or $|M| > 2$. In particular, we perform the maximisation in Eq.~\eqref{eq:gen_up_bound} numerically by means of heuristic methods, despite dealing with a non-convex optimisation problem. This allows us to, at least numerically, determine $r_\text{1-way}^\uparrow$ while firstly accounting for finite detection efficiency ($\eta<1$ with purely lossy correlation \eqref{eq:lossy_corr2} being shared) within the CHSH-based 2333- and 2233-protocols.
In a similar manner, we then consider the visibility to be finite instead ($V<1$), and determine $r_\text{1-way}^\uparrow$ for the corresponding CHSH-based 2322- and 2222-protocols.
These upper bounds can then be used to determine \emph{universal} noise thresholds $\critmin{\eta}$ and $\critmin{V}$, below which \emph{no} key can be distilled with one-way communication---these appear in Table~\ref{tab:critical} (column `\emph{any}') and in~\figref{fig:partial_thresholds} of the main text. However, in order to perform the numerical optimisation, we must decide on the number of outcomes for the discrete random variables $A'$ and $M$, which, in principle, can be as large as possible. We proceed phenomenologically, i.e.~in each case we raise the outcome-number by one, until the moment we can conclude that no further increase is necessary. 

Finally, let us emphasise that we perform the optimisation here over preprocessing strategies for the same CHSH-optimal correlations, for which the thresholds in~\appref{app:1way_maxent_UBs} were derived, utilising maximally entangled states $\ket{\Phi^+}$ and standard CHSH measurements~\eqref{eq:standard_measurements}.

%%%%%%%%%%%%%%%%%%%%%%%%%%%
\paragraph{Finite detection efficiency.}
In our optimisation we directly seek the minimal detection efficiency, $\eta$, such that $r_\text{1-way}^\uparrow=0$, which then constitutes the desired universal threshold, $\critmin{\eta}$, below which no key can be extracted with one-way communication. We first consider the 2333-protocol with measurement settings $\xk,\yk=\{0,2\}$, and then briefly discuss the 2233-protocol case with $\xk,\yk \in \{0,1\}$.

We start by optimising the bound \eqref{eq:gen_up_bound} only over the $p_{A'|A}$ maps, while disregarding the $p_{M|A'}$ maps. We perform the maximisation over all corresponding stochastic matrices $\cS_{A\to A'}$ of size $|A'| \times |A|$, where we vary the outcome number  $2 \leq |A'| \leq 7$ ($|A|=3$ is fixed with $A\in\{0,1,\varnothing\}$ by the lossy correlation $\pAB(a,b)$ in \eqnref{eq:lossy_corr2} considered). Independently of the outcome number $|A'|$, we always arrive at the critical detection efficiency:
\begin{equation}
\label{eq:general_bound_AAp}
\critmin{\eta}^{A \to A'} \approx 87.0105\%,
\end{equation}
which coincides up to our best-achieved numerical precision with the critical efficiency attainable with random binning followed by noisy preprocessing, i.e.~$\crit{\eta}^\text{rand+n.p.}$ in Eq.~\eqref{eq:crit_efficiency_rand_np}. Hence, as the critical efficiency \eqref{eq:general_bound_AAp} applies to all preprocessing strategies of Alice $p_{A'|A}$, we conjecture that the strategy of random binning combined with noisy preprocessing is the optimal form of defense against the CC attack, when Alice is not utilising the message $M$ sent to Bob. However, we observe that the numerical optimisation converges to different stochastic matrices leading to the critical efficiency \eqref{eq:general_bound_AAp}, suggesting that the optimal preprocessing strategy is then not unique.

Secondly, we incorporate into the maximisation of \eqref{eq:gen_up_bound} the optimisation over both  $p_{A'|A}$ and $p_{M|A'}$, which correspond to some choice of $\cS_{A\to A'}$ and $\cS_{A'\to M}$ stochastic matrices of dimensions $|A|\times |A'|$ and $|A'|\times |M|$, respectively. Allowing the outcome numbers to range in $3 \leq |A'| \leq 5$ and $2 \leq |M| \leq 5$, we surprisingly observe that the upper bound $r_\text{1-way}^\uparrow$ can be increased thanks to inclusion of the $p_{M|A'}$ mapping. This, in turn, allows to lower the required critical efficiency to
\begin{equation}
\label{eq:general_bound_AAp_ApM}
\critmin{\eta}^{A \to A' \to M}=\critmin{\eta}^{A \to M} \approx 85.3553 \%.
\end{equation}
Furthermore, we observe this to be possible when considering already $|A'|=3$, $|M|=2$ (but not for $|A'|=2$, $|M|=2$ for which the value \eqref{eq:general_bound_AAp} is recovered). 

However, as noted in Eq.~\eqref{eq:general_bound_AAp_ApM}, we find that the inclusion of the map $p_{A'|A}$ is then unnecessary---it is sufficient for Alice to use the ``raw'' $A$-variable for the key and send some $A$-dependent message $M$ to Bob as the best preprocessing strategy. In particular, we establish the same critical efficiency \eqref{eq:general_bound_AAp_ApM} by considering now only $\cS_{A\to M}$ with $2 \leq |M| \leq 5$, implying that the outcome number $|M|=2$ is already sufficient, as the optimal maps always possess the crucial feature of effectively ``singling out'' the $\varnothing$-outcome of Alice within the message send to Bob. In particular, it is always optimal for Alice to simply announce to Bob whether she has a conclusive outcome or not, which is the bit of information to be encoded in $M$. Indeed, from the perspective of the CC attack this does not provide any extra information to Eve, as she knows whether a local or non-local correlation was distributed to the parties, while $A=\varnothing$ never occurs in the latter case. Hence, as Eve perfectly knows the outcomes of Alice whenever a local correlation is shared, she also always knows whenever Alice records any inconclusive outcome $\varnothing$. Note also that Alice announcing the $\varnothing$ outcomes does not constitute postselection---these rounds are not discarded by the parties and therefore the announcements do not lead to a violation of the detection loophole.

As we now show, the critical value \eqref{eq:general_bound_AAp_ApM} can be, in fact, analytically proven for the preprocessing strategy described above---corresponding to Alice applying on her key variable $A\in\{0,1,\varnothing\}$ the stochastic matrix
\begin{equation}
\cS_{A \to M} = \left(\begin{array}{ccc}
1 & 1 & 0\\
0 & 0 & 1
\end{array}\right),
\label{eq:S_A->M}
\end{equation}
such that the binary $M\in\{\checkmark,\varnothing\}$ takes the value $M=\varnothing$ if $A=\varnothing$, and $M=\checkmark$ if Alice records a conclusive outcome. 

To calculate the resulting bound $r_{p_{M|A}}^\uparrow =  {I(A\!:\!B|M)} - {I(A\!:\!E|M)}$ for $p_{M|A}\equiv\cS_{A \to M}$ in Eq.~\eqref{eq:S_A->M}, let us first consider the mutual information between Alice and Bob conditioned on $M$. We can construct the tripartite probability distribution $p_{MAB}(m, a, b)$ by first augmenting the lossy correlation \eqref{eq:lossy_corr2} shared by Alice and Bob in key generation rounds with an extra ``dummy'' random variable $\tilde{A}$ perfectly correlated to $A$, 
\begin{equation}
p_{\tilde{A}AB}(\tilde{a},a,b)=\begin{array}{|c|c|c|c|c|c|c|c|c|c|}
\hline \tilde{a}\setminus a, b & 0, 0 & 0, 1 & 0, \varnothing & 1, 0 & 1, 1 & 1, \varnothing & \varnothing, 0 & \varnothing, 1 & \varnothing, \varnothing\\
\hline 0 & \eta^{2}\,\mathsf{Q}_{00} & \eta^{2}\,\mathsf{Q}_{01} & \eta \neta \QA{0} & 0 & 0 & 0 & 0 & 0 & 0\\
\hline 1 & 0 & 0 & 0 & \eta^{2}\,\mathsf{Q}_{10} & \eta^{2}\,\mathsf{Q}_{11} & \eta \neta \QA{1} & 0 & 0 & 0 \\
\hline \varnothing & 0 & 0 & 0 & 0 & 0 & 0 & \neta \eta\QB{0} & \neta \eta\QB{1} & \neta^2 \\
\hline \end{array}\,,
\end{equation}
and transform it by applying $\cS_{\tilde{A} \to M}$ of Eq.~\eqref{eq:S_A->M} (adding together the first two rows) to obtain the desired
\begin{equation}
\label{eq:am_example_aab}
p_{MAB}(m,a,b)=\begin{array}{|c|c|c|c|c|c|c|c|c|c|}
\hline m\setminus a, b & 0, 0 & 0, 1 & 0, \varnothing & 1, 0 & 1, 1 & 1, \varnothing & \varnothing, 0 & \varnothing, 1 & \varnothing, \varnothing\\
\hline \checkmark & \eta^{2}\,\mathsf{Q}_{00} & \eta^{2}\,\mathsf{Q}_{01} & \eta \neta \QA{0} & \eta^{2}\,\mathsf{Q}_{10} & \eta^{2}\,\mathsf{Q}_{11} & \eta \neta \QA{1} & 0 & 0 & 0\\
\hline \varnothing & 0 & 0 & 0 & 0 & 0 & 0 & \neta \eta\QB{0} & \neta \eta\QB{1} & \neta^2 \\
\hline \end{array}\,.
\end{equation}
Consistently, the marginal distribution of the message variable $M$ reads (summing the rows in Tab.~\eqref{eq:am_example_aab}),
\begin{equation}
\label{eq:p_M}
p_M(m) =
\begin{cases}
  \eta, &\text{ if }m=\checkmark\\
  \neta, &\text{ if }m=\varnothing
\end{cases} ,
\end{equation}
as it effectively denotes whether a detection event occured or not. With help of Tab.~\eqref{eq:am_example_aab} we then calculate
\begin{align}
I(A\!:\!B|M) 
& = \eta \, I(A \! : \! B|M=\checkmark) + \neta \, I(A \!:\!B|M=\varnothing)\nonumber \\
& = \eta\left(h[\eta \QA{0}] + H\!\left\{\eta \QB{0},\eta \QB{1},\neta\right\} - H\!\left\{\eta\,\mathsf{Q}_{00} , \eta\,\mathsf{Q}_{01} , \neta\QA{0} , \eta\,\mathsf{Q}_{10} , \eta\,\mathsf{Q}_{11} , \neta\QA{1}\right\}
\right),
\end{align}
where $I(A\!:\!B|M=\varnothing)=0$, as $H(A|M=\varnothing)=0$ and $H(B|M=\varnothing)=H(A,B|M=\varnothing)$. Substituting further the $\mathsf{Q}$-distribution for the 2333-protocol, we finally obtain
\begin{equation}
I(A\!:\!B|M) = \eta \left ( h\!\left[\frac{\eta}{2} \right] + H \!\left\{\frac{\eta}{2},\frac{\eta}{2},\neta \right\}  - H \!\left\{\frac{\eta}{2} , \frac{\neta}{2} , \frac{\eta}{2} , \frac{\neta}{2} \right\} \right )  = \eta h\!\left[\frac{\eta}{2} \right] - \eta\neta.
\end{equation}

We now turn to the mutual information between Alice and Eve conditioned on $M$, which we calculate in a similar manner after identifying the tripartite distribution $p_{MAE}(m, a, e)$. We first, however, note that Eq.~\eqref{eq:Alice_local_prob} implies that
\begin{equation}
\p{AE}(a,e=\tilde{e})= q^\cL \pA^\cL (a),\quad \p{AE}(a,e=?) = (1-q^\cL) \QA{a}
\end{equation}
where $\tilde{e} \in \{ 0, 1, \varnothing \}$ is the outcome of Eve when she distributes a local correlation. Hence, by introducing again an auxiliary variable $\tilde{A}$ that is perfectly correlated to $A$, we can write the overall distribution as 
\begin{equation}
\p{\tilde{A}AE}(\tilde{a},a,e)=\begin{array}{|c|c|c|c|c|c|}
\hline \tilde{a}\setminus a, e & 0, 0  & 0, ? &  1, 1 & 1, ? &  \varnothing, \varnothing \\
\hline 0 & q^\cL \eta_\cL \mathsf{Q}_0^{\mathrm{A}}  & (1-q^\cL) \mathsf{Q}_0^{\mathrm{A}}  &  0 & 0 & 0\\
\hline 1 & 0 & 0 & q^\cL \eta_\cL \mathsf{Q}_1^{\mathrm{A}}  & (1-q^\cL) \mathsf{Q}_1^{\mathrm{A}} & 0  \\
\hline \varnothing & 0 & 0 & 0 & 0 & q^\cL (\neta_\cL)  \\
\hline \end{array}\,,
\end{equation}
and apply the stochastic map $\cS_{\tilde{A} \to M}$ of Eq.~\eqref{eq:S_A->M} onto the auxiliary variable to obtain the desired distribution:
\begin{equation}
\label{eq:am_example_aae}
\p{MAE}(m,a,e)=\begin{array}{|c|c|c|c|c|c|}
\hline m\setminus a, e & 0, 0  & 0, ? &  1, 1 & 1, ? &  \varnothing, \varnothing \\
\hline \checkmark & q^\cL \eta_\cL \mathsf{Q}_0^{\mathrm{A}}  & (1-q^\cL) \mathsf{Q}_0^{\mathrm{A}}  &  q^\cL \eta_\cL \mathsf{Q}_1^{\mathrm{A}}  & (1-q^\cL) \mathsf{Q}_1^{\mathrm{A}} & 0 \\
\hline \varnothing & 0 & 0 & 0 & 0 & q^\cL (\neta_\cL)  \\
\hline \end{array}\,.
\end{equation}
As the distribution \eqref{eq:am_example_aae} consistently yields the same marginal distribution \eqref{eq:p_M} for the message variable $M$, the conditional mutual information can be similarly split into
\begin{align}
I(A\!:\!E|M) 
& = \eta \, I(A\!:\!E|M=\checkmark)+\neta \, I(A\!:\!E|M=\varnothing)\nonumber\\
& = \eta\left(\! h \left [\eta \QA{0} \right] + H \!\left\{\frac{q^\cL \eta_\cL}{\eta} \mathsf{Q}_0^{\mathrm{A}},\frac{q^\cL \eta_\cL}{\eta} \mathsf{Q}_1^{\mathrm{A}},\frac{1-q^\cL}{\eta} \right\} \right. \nonumber\\  & \qquad \quad \left. - H \!\left\{\frac{q^\cL \eta_\cL}{\eta} \mathsf{Q}_0^{\mathrm{A}} , \frac{(1-q^\cL)}{\eta} \mathsf{Q}_0^{\mathrm{A}} , \frac{q^\cL \eta_\cL}{\eta} \mathsf{Q}_1^{\mathrm{A}} , \frac{(1-q^\cL)}{\eta} \mathsf{Q}_0^{\mathrm{A}} \right\}\right),
\end{align}
where $I(A\!:\!E|M=\varnothing) = 0$, as it is always the single outcome $\varnothing$ being transmitted between Alice and Eve when $M=\varnothing$, carrying zero information on its own. Substituting further the trivial marginals $\mathsf{Q}_0^{\mathrm{A}}=\mathsf{Q}_1^{\mathrm{A}}=\frac12$ that apply to the 2333-protocol, we have
\begin{equation}
\label{eq:mut_inf_AE_sym_marg}
I(A\!:\!E|M) =  \eta\, h\!\left[\frac{\eta}{2} \right] + q^\cL - 1 
= \eta\, h\!\left[\frac{\eta}{2} \right] + \neta(1+(3+2\sqrt{2})\eta) - 1,
\end{equation}
where we substituted already for the optimal local weight $q^\cL$ according to Eq.~\eqref{eqs:local_weight_lossy}.

Finally, we arrive at the desired upper bound for the one-way key rate based on the CC attack applicable when Alice does not preprocess her outcome $A$, but rather reveals the rounds in which she obtained $\varnothing$ by transmitting a message $M$ prepared by applying $p_{M|A}\equiv\cS_{A \to M}$ of Eq.~\eqref{eq:S_A->M} to $A$, i.e.:
\begin{equation}
r_{p_{M|A}}^\uparrow = I(A\!:\!B|M) - I(A\!:\!E|M)  = 1 - \neta(1+(3+2\sqrt{2})\eta) - \eta\neta,
\end{equation}
which vanishes at the value:
\begin{equation}
\crit{\eta}^{A\to M} = \frac14 \left( 2 + \sqrt{2} \right) \approx 85.3553 \%,
\end{equation}
which coincides, indeed, with the numerically obtained critical efficiency in Eq.~\eqref{eq:general_bound_AAp_ApM}, optimised over all preprocessing strategies of Alice. Interestingly, it further coincides with the critical efficiency~\eqref{eq:eta_crit_3323}, which we derive below by applying the CC attack to the (intrinsic--information-based) upper bound that accounts for \emph{two-way} communication, but assumes symmetric deterministic binning of inconclusive outcomes by both Alice and Bob.

For completeness, let us just summarize the results obtained for the 2233-protocol. Considering preprocessing strategies in which Alice applies only the $p_{A'|A}$ map in Eq.~\eqref{eq:gen_up_bound} (with $2 \leq |A'| \leq 5$) we obtain with help of numerical heuristic methods the following critical efficiency
\begin{equation}
\label{eq:general_bound_AAp_3322}
\critmin{\eta}^{A \to A'\to M}=\critmin{\eta}^{A \to A'} \approx 92.6380\%
\end{equation}
irrespectively of the key settings used by Alice and Bob ($\xk,\yk \in \{0,1\}$), which coincides with the expression obtained previously in Sec.~\ref{par:random_binn_np} when Alice resorts to random binning of her outcome, followed by noisy preprocessing. Nevertheless, we find numerically multiple stochastic matrices allowing to attain the value \eqref{eq:general_bound_AAp_3322}, which suggests that the optimal preprocessing strategy is not unique from the perspective of the CC attack.

As already noted in Eq.~\eqref{eq:general_bound_AAp_3322}, in contrast to the case of the 2333-protocol, we do not observe any improvement of the critical efficiency \eqref{eq:general_bound_AAp_3322} (lowering its value) by allowing Alice also to perform arbitrary maps $p_{M|A'}$ in Eq.~\eqref{eq:gen_up_bound} and letting $2 \leq |A'|, |M| \leq 5$. On the other hand, if one disregards the mapping $A \to A'$ and allows only for the map $A \to M$ with $M$ being the public message, then we observe that, similarly to the 2333-protocol, it is (numerically) optimal for Alice to just signal the occurrences of the inconclusive outcomes $\varnothing$. Such a strategy when considering the CC attack leads to the following critical efficiency:
\begin{equation}
\crit{\eta}^{A\to M} = \frac{4 \left(3+2 \sqrt{2}\right)}{2 \left(5+4 \sqrt{2}\right)+\sqrt{2} \,\log_2\!\!\left[3+2 \sqrt{2}\right]} \approx 93.5910 \%,
\end{equation}
which we obtain analytically following the same procedure as for the 2333-protocol above, irrespectively of the key settings used by Alice and Bob. Note that $\critmin{\eta}^{A \to A'\to M}<\crit{\eta}^{A\to M}$, so the CC attack suggests for the 2233-protocol that the best preprocessing strategy for Alice is to apply $p_{A'|A}$ that implements random binning of inconclusive outcomes, followed by noisy preprocessing of all the resulting outcomes (i.e.~the stochastic matrix $\cS_\text{rand+n.p.}$ in Eq.~\eqref{eq:noisy_prep_map_rand} with $\pp\to\frac{1}{2}_\pm$).

%%%%%%%%%%%%%%%%%%%%%%%%%%%
\paragraph{Finite visibility.}
In analogy to the previous section, we perform an optimisation in which we directly seek the minimal visibility, $V$, such that $r_\text{1-way}^\uparrow=0$, which then constitutes the desired universal threshold, $\critmin{V}$, below which again no key can be extracted with one-way communication. We first consider the 2322-protocol with measurement settings $\xk,\yk=\{0,2\}$, and then briefly discuss the 2222-protocol case with $\xk,\yk \in \{0,1\}$. 

We first optimise the bound \eqref{eq:gen_up_bound} only over the $p_{A'|A}$ maps, while disregarding the $p_{M|A'}$ maps. We perform the maximisation over all corresponding stochastic matrices $\cS_{A\to A'}$ of size $|A'| \times |A|$, where we vary the outcome number  $2 \leq |A'| \leq 6$ ($|A|=2$ is fixed with $A\in\{0,1\}$ by the noisy correlation $\pAB(a,b)$ considered). Independently of the outcome number $|A'|$, we always arrive at the critical visibility:
\begin{equation}
\label{eq:general_bound_AAp_noise}
\critmin{V}^{A \to A'\to M} = \critmin{V}^{A \to A'} \approx 80.8530\%,
\end{equation}
which coincides up to our best-achieved numerical precision with the critical visibility attainable with noisy preprocessing, see $\crit{V}^\text{n.p.}$ in \eqnref{eq:visibility_np}. Hence, as the critical efficiency \eqref{eq:general_bound_AAp_noise} applies to all preprocessing strategies of Alice $p_{A'|A}$, we conjecture that the strategy of noisy preprocessing is the optimal form of defense against the CC attack, when the parties observe the purely noisy correlation and Alice is not utilising the message $M$ sent to Bob. Still, note that the numerical optimisation converges to different stochastic matrices leading to the critical visibility \eqref{eq:general_bound_AAp_noise}, suggesting that the optimal preprocessing strategy is then not unique.

As already noted in Eq.~\eqref{eq:general_bound_AAp_noise}, in contrast to the case of the purely lossy 2333-protocol discussed in the previous section, we do not observe any improvement of the critical visibility \eqref{eq:general_bound_AAp_noise} (lowering its value) by allowing Alice also to perform arbitrary maps $p_{M|A'}$ in Eq.~\eqref{eq:gen_up_bound} and letting $2 \leq |A'|, |M| \leq 4$. Furthermore, allowing only for the map $p_{A|M}$ to be performed by Alice on her `raw' variable $A$, we arrive at the critical visibility
\begin{equation}
\critmin{V}^{A \to M} \approx 82.9995\%,
\end{equation}
which coincides up to our best-achieved numerical precision with the critical visibility obtained by performing no preprocessing by Alice, see $\crit{V}^\text{no-prep}$ in Eq.~\eqref{eq:v_crit_noprep}. We are thus led to the conclusion that for the lossless case of finite visibility, the inclusion of the publicly announced variable $M$ serves no purpose against the CC attack, as it cannot lower the attainable critical visibility.

For completeness, let us also cite the results obtained for the 2222-protocol. Considering preprocessing strategies in which Alice applies either only $p_{A'|A}$ map in Eq.~\eqref{eq:gen_up_bound} (with $2 \leq |A'| \leq 6$) or in which Alice applies also the $p_{M|A'}$ map in Eq.~\eqref{eq:gen_up_bound} (with $2 \leq |A'|, |M| \leq 4$), we find that the critical visibility coincides with the one obtained with noisy preprocessing---$\crit{V}^\text{n.p.}$ specified below \eqnref{eq:visibility_np}---i.e.:
\begin{equation}
\label{eq:general_bound_AAp_2222}
\critmin{V}^{A \to A'\to M}=\critmin{V}^{A \to A'} \approx 88.5238\%,
\end{equation}
again with the optimisation arriving at different optimal stochastic matrices, suggesting that they are not unique. Considering preprocessing strategies with only the $p_{A|M}$ map (letting $2 \leq |M| \leq 5$), we find
\begin{equation}
\label{eq:general_bound_AM_2222}
\critmin{V}^{A\to M} \approx 90.6075 \%,
\end{equation}
which, similarly as for the 2322-protocol, coincides to our best numerical precision with the critical visibility obtained by performing no preprocessing by Alice, see $\crit{V}^\text{no-prep}$ below Eq.~\eqref{eq:v_crit_noprep}. The above thresholds (\ref{eq:general_bound_AAp_2222}-\ref{eq:general_bound_AM_2222}) apply irrespectively of the key settings used by Alice and Bob within the 2222-protocol ($\xk,\yk \in \{0,1\}$).

%%%%%%%%%%%%%%%%%%%%%%%%%%%%%%%%%%%%%%%%%%%%%%%%%%%%%%%%%%%%%%%%%%%%%%%%%%%%%%%%%%%%%%%%%%%%%%%%%%%%%%%%%%%%%%%%
\subsection{Two-way protocols involving maximally entangled states}
\label{app:2way_maxent_UBs}
%%%%%%%%%%%%%%%%%%%%%%%%%%%%%%%%%%%%%%%%%%%%%%%%%%%%%%%%%%%%%%%%%%%%%%%%%%%%%%%%%%%%%%%%%%%%%%%%%%%%%%%%%%%%%%%%
%

%%%%%%%%%%%%%%%%%%%%%%%%%%%%%%%%%%%%%%%%%%%%%%%%%%%%%%
\subsubsection{Finite detection efficiency and deterministic binning of `no-clicks'}
\label{app:2way_maxent_UBs_eta}
As noted in the main text, for the purely lossy correlation \eqref{eq:lossy_corr2} we are unable to find a non-trivial upper bound on the two-way key rate, $\rtwoway(A\leftrightarrow B)$ in \eqnref{eq:r_twoway_UB_mapEF} of the main text, by resorting to the conditional mutual information $I(A\!:\!B|F)$ and heuristically searching over all possible stochastic maps $E \to F$ applied on Eve's variable. However, we provide a non-trivial upper bound under an additional assumption that both Alice and Bob \textit{bin} their non-detection events $\varnothing$ in a deterministic fashion, that is, whenever a detection failure $\varnothing$ occurs they simply interpret it always as the 0-outcome (or equivalently always as 1).

Such an assumption is motivated by the fact that the binning procedure of the inconclusive outcomes naturally arises in DIQKD protocols whose security is based on two-outcome Bell inequalities, both in one-way protocols~\cite{WAP20} and two-way protocols based on advantage distillation~\cite{TLR20}. As a consequence, the upper bound determined by us applies to all such protocols or, generally speaking, to any protocol in which the parties decide to deterministically bin their data before performing any other operations, also ones requiring two-way communication. Importantly, the binning is performed not only to test the Bell-violation, but also in the key-generation rounds. 

Still, note that the binning procedure does not change the nature of the protocol, which remains $2333$ or $2233$ depending whether or not, respectively, Bob uses an extra setting for the key distillation. This is because any \emph{preprocessing} of the data, an example of which is binning, is performed by the parties \emph{after} they record their strings of outcomes, which include also the $\varnothing$-events and are all correlated with the string in hands of Eve. Therefore, the decomposition \eqref{eq:CC_constr} of the CC attack must be valid before binning (or any preprocessing) and, hence, the maximal local weight allowed within the attack, $q^\cL$, remains to be given by Eq.~\eqref{eqs:local_weight_lossy}.

After determining the local weight, subsequent calculations of the upper bound on the key rate depend only on the tripartite correlation $\p{ABE}(a,b,e|\xk,\yk)$ conditioned on Alice and Bob choosing the key settings $\xk$ and $\yk$ that includes also the eavesdropper Eve performing the CC attack. In order to write down this correlation, we first notice that whenever Alice and/or Bob record $\varnothing$, this may only happen within the protocol rounds in which Eve distributes a local correlation (and knows perfectly every outcome), as whenever she distributes a nonlocal correlation Alice and Bob observe $\qAB$, which has perfect detection efficiency. Consequently, the entries in the rows 
(2-5) of Tab.~\eqref{eqs:table_abe_loss} stated below contain only non-zero diagonal elements.

In particular, the probability that both Alice and Bob register a \emph{conclusive} outcome, which we label by the variables $\tilde{a},\tilde{b} \in \{0, 1\}$, and Eve has no knowledge about the result (having distributed a nonlocal correlation) reads:
\begin{equation}
\pABE(\tilde{a},\tilde{b},?|\xk,\yk)=\p{E}(?)\,\p{AB|E}(\tilde{a},\tilde{b}|\xk,\yk,?)=(1-q^\cL)\,\mathsf{Q}_{\tilde{a}\tilde{b}},
\end{equation}
where $\p{E}(?)=q^\cNL=1-q^\cL$ is just the probability of Eve distributing a non-local correlation, while 
\begin{equation}
\p{AB|E}(\tilde{a},\tilde{b}|\xk,\yk,?)=\pAB^{\cNL}(\tilde{a},\tilde{b}|\xk,\yk)=\mathsf{Q}_{\tilde{a}\tilde{b}}
\end{equation}
is the lossless correlation producing conclusive outcomes, and we used the simplified notation $\mathsf{Q}_{ab} = \mathsf{Q}_{ab}^{\xk \yk}$.

Moreover, as for conclusive outcomes $\pAB(\tilde{a},\tilde{b}|\xk,\yk)=\p{ABE}(\tilde{a},\tilde{b},?|\xk,\yk)+\p{ABE}(\tilde{a},\tilde{b},(\tilde{a},\tilde{b})|\xk,\yk)$, we obtain the missing expression for the correlations applicable when Eve perfectly knows $\tilde{a}$ and $\tilde{b}$ as
\begin{equation}
\p{ABE}(\tilde{a},\tilde{b},(\tilde{a},\tilde{b})|\xk,\yk) = \pAB(\tilde{a},\tilde{b}|\xk,\yk) - (1-q^\cL)\,\mathsf{Q}_{\tilde{a}\tilde{b}} = (\eta^2 -1+q^\cL)\,\mathsf{Q}_{\tilde{a}\tilde{b}},
\end{equation}
which allows us then to fully write out the desired tripartite correlation:
\begin{equation} \label{eqs:table_abe_loss} 
\p{ABE}(a,b,e|\xk,\yk)\quad=\quad\begin{array}{|c|c|c|c|c|}
\hline e\setminus a, b & \tilde{a}, \tilde{b} & \tilde{a}, \varnothing & \varnothing, \tilde{b} & \varnothing, \varnothing\\
\hline (\tilde{a}, \tilde{b}) & (\eta^2 - 1 + q^\cL)\,\mathsf{Q}_{\tilde{a}\tilde{b}} & 0 & 0 & 0\\
\hline (\tilde{a}, \varnothing) & 0 & \eta \neta\,\mathsf{Q}_{\tilde{a}}^{A}  & 0 & 0
\\
\hline (\varnothing, \tilde{b}) & 0 & 0 & \eta \neta\,\mathsf{Q}_{\tilde{b}}^{B} & 0 
\\
\hline (\varnothing, \varnothing) & 0 & 0 & 0 & \neta^2 
\\
\hline ? & (1 - q^\cL)\,\mathsf{Q}_{\tilde{a}\tilde{b}} & 0 & 0 & 0 \\
\hline
\end{array}
\;.
\end{equation}
The correlations shared by Alice and Bob after they perform the binning ($\varnothing\to0$) procedure can be simply obtained from Tab.~\eqref{eqs:table_abe_loss} by adding every \emph{column} involving one (or more) $\varnothing$-outcomes to the corresponding one in which $\varnothing$ is (are) replaced by 0. 

From now on we turn our attention to Eve and propose a preprocessing strategy $E \to F$ that leads to a non-trivial upper bound on the key rate. Although what follows is the best strategy that we have found, note that any $E \to F$ map gives a valid upper bound $\rtwoway(A\leftrightarrow B) \leq {I(A\!:\!B|F)}$ and we do not exclude the possibility that there exists a map leading to a tighter bound. First, we have Eve bin her variable deterministically, just like the honest parties. On her part, this corresponds to defining a new variable $\tilde{e} \in \{ (0,0), (0,1), (1,0), (1,1), ? \}$, whose probabilities are determined by Tab.~\eqref{eqs:table_abe_loss} in the similar manner, i.e.~by adding \emph{rows} involving any $\varnothing$ to ones in which the $\varnothing$-outcome is replaced by 0. The resulting correlation obtained after all the three parties perform binning reads:
\begin{equation} \label{eqs:table_abe_loss_f1}
\p{AB\tilde{E}}(a,b,\tilde{e}|\xk,\yk)=\begin{array}{|c|c|c|c|c|}
\hline \tilde{e}\setminus a, b & 0, 0 & 0, 1 & 1, 0 & 1, 1\\
\hline (0, 0) & \begin{tabular}{@{}c@{}}($\eta^2 - 1 + q^\cL)\,\mathsf{Q}_{00} $ \\ $+\eta\neta\left(\mathsf{Q}_{0}^{A} + \mathsf{Q}_{0}^{B}\right) $ \\ $+\neta^2$\end{tabular}   & 0 & 0 & 0\\
\hline (0, 1) & 0 & \begin{tabular}{@{}c@{}}$(\eta^2 - 1 + q^\cL)\,\mathsf{Q}_{01} $\\$+\eta \neta\,\mathsf{Q}_{1}^{B}$\end{tabular}  & 0 & 0
\\
\hline (1, 0) & 0 & 0 & \begin{tabular}{@{}c@{}}$(\eta^2 - 1 + q^\cL)\,\mathsf{Q}_{10} $\\$+\eta \neta\,\mathsf{Q}_{1}^{A}$\end{tabular}  & 0 
\\
\hline (1, 1) & 0 & 0 & 0 & (\eta^2 - 1 + q^\cL)\,\mathsf{Q}_{11} 
\\
\hline ? & (1 - q^\cL)\,\mathsf{Q}_{00} & (1 - q^\cL)\,\mathsf{Q}_{01} & (1 - q^\cL)\,\mathsf{Q}_{10} & (1 - q^\cL)\,\mathsf{Q}_{11} \\
\hline
\end{array}
\;.
\end{equation}

We now transform the variable of Eve, $\tilde{E}\to F$, by applying the post-processing map proposed by us in~\citeref{Farkas2021}, which takes the form of a stochastic matrix $P_{F|\tilde{E}}$ given by
\begin{equation}
\label{eq:pfe_map}
P_{F|\tilde{E}} =
\left(\begin{array}{ccccc}
1 & 0 & 0 & 0 & 0\\
0 & 0 & 0 & 1 & 0\\
0 & 1 & 1 & 0 & 1
\end{array}\right),
\end{equation}
in order to determine the resulting tripartite correlation 
\begin{equation}
\p{ABF}(a,b,f|\xk,\yk)=\sum_{\tilde{e}}\;\p{F|\tilde{E}}(f|\tilde{e})\;\p{AB\tilde{E}}(a,b,\tilde{e}|\xk,\yk),
\end{equation}
where $\p{F|\tilde{E}}(f|\tilde{e}) = [P_{F|\tilde{E}}]_{f\tilde{e}}$ are the the entries in \eqnref{eq:pfe_map}.
By applying the map $P_{F|\tilde{E}}$, Eve keeps her outcomes $\tilde{e} \in \{ (0,0), (1,1) \}$ intact while uniformly mixing the ``other'' outcomes $\{(0,1), (1,0), ?\}$. The distribution $\p{ABF}$ is thus constructed by adding together the three relevant rows of Tab.~\eqref{eqs:table_abe_loss_f1} corresponding to the ``other'' outcomes, which gives us
\begin{equation} \label{eqs:table_abe_loss_f2}
\p{ABF}(a,b,f|\xk,\yk)=\begin{array}{|c|c|c|c|c|}
\hline f\setminus a, b & 0, 0 & 0, 1 & 1, 0 & 1, 1\\
\hline (0, 0) & \begin{tabular}{@{}c@{}}($\eta^2 - 1 + q^\cL)\,\mathsf{Q}_{00} +$ \\ $\eta\neta\left(\mathsf{Q}_{0}^{A} + \mathsf{Q}_{0}^{B}\right) + $ \\ $\neta^2$\end{tabular}   & 0 & 0 & 0\\
\hline (1, 1) & 0 & 0 & 0 & (\eta^2 - 1 + q^\cL)\,\mathsf{Q}_{11} 
\\
\hline \text{other} & (1 - q^\cL)\,\mathsf{Q}_{00} & \begin{tabular}{@{}c@{}}$\eta^2\,\mathsf{Q}_{01} + $\\$\eta \neta\,\mathsf{Q}_{1}^{B}$\end{tabular}  & \begin{tabular}{@{}c@{}}$\eta^2\,\mathsf{Q}_{10} + $\\$\eta \neta\,\mathsf{Q}_{1}^{A}$\end{tabular}  & (1 - q^\cL)\,\mathsf{Q}_{11} \\
\hline
\end{array}
\;.
\end{equation}

The above choice allows us to calculate a non-trivial upper bound on the two-way key rate introduced in \eqnref{eq:r_twoway_UB_mapEF} of the main text, i.e.~$\rtwoway(A\leftrightarrow B) \leq {I(A\!:\!B|F)}$, by evaluating the conditional mutual information for the distribution $\p{ABF}$ in Eq.~\eqref{eqs:table_abe_loss_f2}. In what follows, we do this for the 2333- and 2233-scenarios of interest, in which the $\mathsf{Q}$-probabilities in Tab.~\eqref{eqs:table_abe_loss_f2} are determined by the CHSH-optimal measurements~\eqref{eq:standard_measurements} performed on a shared maximally entangled state~$\ket{\Phi^{+}}$---and are given by \eqnref{eqs:biased_CHSH} depending on the key settings $\xk$, $\yk$.

%%%%%%%%%%%%%%%%%%%%%%%%%
\paragraph{2333-protocol.}
Substituting the $\mathsf{Q}$-probabilities of the 2333-protocol with $\xk, \yk = \{0, 2\}$, as well as the form of the optimal local weight $q^\cL$ in Eq.~\eqref{eqs:local_weight_lossy}, into the tripartite correlation \eqref{eqs:table_abe_loss_f2}, we find $I(A\!:\!B|F)=0$ at
\begin{equation}
\crit{\eta} = \frac{1}{4} \left(2 + \sqrt{2}\right) \approx 85.36\%,
\label{eq:eta_crit_3323}
\end{equation}
which we state in \tabref{tab:critical} of the main text, see the column labelled `two-way'.

As a consequence, $\crit{\eta}$ constitutes a lower bound on the detection efficiency required by any (even) two-way DIQKD protocol based on the 2333-scenario (with $\xk,\yk=\{0,2\}$ key-settings), under the assumption that both parties perform deterministic binning of their non-detection events prior to any preprocessing of their data. In comparison, employing a concrete two-way protocol of advantage distillation has been shown, under the same assumption and additionally restricting to collective attacks, to require detection efficiency $\eta_\text{a.d.} = 93.7\%$~\cite{TLR20}.

%%%%%%%%%%%%%%%%%%%%%%%%%
\paragraph{2233-protocol.}
Performing an analogous calculation for the 2233-protocol, we find that the upper bound depends on the specific choice of key settings. If the parties choose $\xk, \yk = \{0, 0\},\,\{0, 1\}\text{ or }\{1, 0\}$, we find $I(A\!:\!B|F)=0$ for $\p{ABF}$ in Eq.~\eqref{eqs:table_abe_loss_f2} at 
\begin{equation}
\crit{\eta}=3(1-1/\sqrt{2})\approx87.87\%,
\label{eq:eta_crit_3322_corr}
\end{equation}
which we state in \tabref{tab:critical} in the `two-way' column. In comparison, the corresponding upper bound on tolerable detection efficiency in two-way DIQKD protocols (when restricting to collective attacks) obtained for the same lossy correlations supplemented by deterministic binning and advantage distillation is $\eta_\text{a.d.} = 91.7\%$~\cite{TLR20}.

However, if the parties choose $\xk, \yk = \{1, 1\}$ as the key settings, the mapping $p(F|\tilde{E})$ introduced in Eq.~\eqref{eq:pfe_map} is no longer sufficient to make $I(A\!:\!B|F)$ vanish for $\eta<1$ in the nonlocal regime. This can be fixed noting that the map $p(F|\tilde{E})$ should now rather equally mix the outcomes of Eve:~$\tilde{e}=\{(0,0),(1,1),?\}$; and not $\tilde{e}=\{(0,1),(1,0),?\}$, as before. This formally corresponds to permuting the columns of the map $p(F|\tilde{E})$ in Eq.~\eqref{eq:pfe_map} or equivalently adding now the three rows denoting $\tilde{e}=\{(0,0),(1,1),?\}$ in Tab.~\eqref{eqs:table_abe_loss_f1} when computing $\p{ABF}$ in Eq.~\eqref{eqs:table_abe_loss_f2}. In this way, one obtains the desired equivalent of critical detection efficiency \eqref{eq:eta_crit_3322_corr} reading
\begin{equation}
\label{eq:eta_crit_3322_anticorr}
\crit{\eta} = \frac{1}{24} \left(24-3 \sqrt{2}+\sqrt{6 \left(32 \sqrt{2}-45\right)}\right) \approx 87.47 \%,
\end{equation} 
which is slightly lower than for other choices of key settings. This follows from the fact that the choice of $\xk, \yk = \{1, 1\}$ leads to a higher probability of Alice and Bob having different outcomes, which doesn't fall in line with their symmetric, deterministic binning. It turns out, however, that if Bob and Alice bin $\varnothing$ deterministically, but one of them to 0 and the other to 1, the threshold can again be shown to read $\crit{\eta} =3(1-1/\sqrt{2})$, as in Eq.~\eqref{eq:eta_crit_3322_corr}. 

%%%%%%%%%%%%%%%%%%%%%%%%%%%%%%%%%%%%%%%%%%%%%%%%%%%%%%
\subsubsection{Finite visibility}
\label{app:2way_maxent_UBs_V}
In this section, we consider the case of the visibility being finite ($V<1$) instead, for which we now study not only the $2222$- and $2322$-scenarios, but also for completeness the $2422$-scenario introduced in \citeref{SGP+20}. Similarly to the above, we compute non-trivial upper bounds on two-way key rates in the form of \eqnref{eq:r_twoway_UB_mapEF} of the main text, $\rtwoway(A\leftrightarrow B)\le I(A\!:\!B|F)$, by identifying sufficient forms of the conditional mutual information $I(A\!:\!B|F)$. Since in the scenario that we are considering Alice and Bob announce their inputs, the tripartite correlations from which Alice and Bob attempt to extract a secure key can be written as in \eqnref{eq:CC_tripart} of the main text, from which one can then compute the necessary conditional mutual information after choosing a suitable post-processing map, $E\to F$, for Eve. Nonetheless, let us note for completeness that we are primarily reproducing here calculations from our \citeref{Farkas2021}.

\paragraph{$2322$-scenario.} Consider here the CHSH-based DIQKD protocol with an added key setting for Bob. Let us recall that the ideal correlation shared between Alice and Bob is given by \eqnref{eqs:biased_CHSH}, i.e.:
\begin{align}
\label{eq:2322ideal}
Q_{AB}(a,b|x,y) =
\begin{cases}
  \frac{1}{4}\left[1+\frac{(-1)^{a+b+xy}}{\sqrt2}\right], &\text{ if } x,y\in\{0,1\}\\
  \frac{1}{2}\,\delta_{a,b}, &\text{ if }(x,y)=(0,2)\\
  \frac{1}{4}, &\text{ if }(x,y)=(1,2)
\end{cases},
\end{align}
while the observed (noisy) correlation reads $\pABobs(a,b|x,y) = V Q_{AB}(a,b|x,y) + \frac{1-V}{4}$, see Eqs.~\eref{eq:visibility} or \eref{eq:Pab_Qab}. Similarly to \eqnref{eq:pfe_map}, we use the CC attack with the post-processing for Eve given by $p_{F|E}(f|e)=(P_{F|E})_{f,e}$, where
\begin{align}
P_{F|E}= \left(\begin{array}{ccccc}
1 & 0 & 0 & 0 & 0\\
0 & 0 & 0 & 1 & 0\\
0 & 1 & 1 & 0 & 1
\end{array}\right),
\label{eq:PFEpostprocessing}
\end{align}
and the variables $e \in \left \{ (0, 0), (0, 1), (1, 0), (1, 1), ?\right \}$ are transformed onto $f \in \left \{ (0,0), (1,1), \text{``other''} \right \}$.

We can now compute $I(A\!:\!B|F)$ to obtain an upper bound \eref{eq:r_twoway_UB_mapEF} on the two-way key rate, i.e:
\begin{align}\label{eq:2way_2322}
\rtwoway
\leq 
\frac{1+\sqrt{2}}{2} \left\{ \log_2 \left[ \frac{j(V)^{j(V)} k(V)^{k(V)} }{ [j(V)+k(V)]^{j(V)+k(V)}} \right] + j(V) + k(V) \right\},
\end{align}
where $j(V) := (1-V)(\sqrt{2}-1)$ and $k(V) := 2( \sqrt{2} V - 1 )$.
One can verify that the upper bound is zero at $\crit{V}=\frac{1}{17}(7+4\sqrt{2}) \approx 74.45\%$ ($>\frac{1}{\sqrt 2}=V_\text{loc}$), as stated in \tabref{tab:critical} in the `two-way' column.

\begin{figure}[t!]
\centering
\includegraphics[width=0.5\columnwidth]{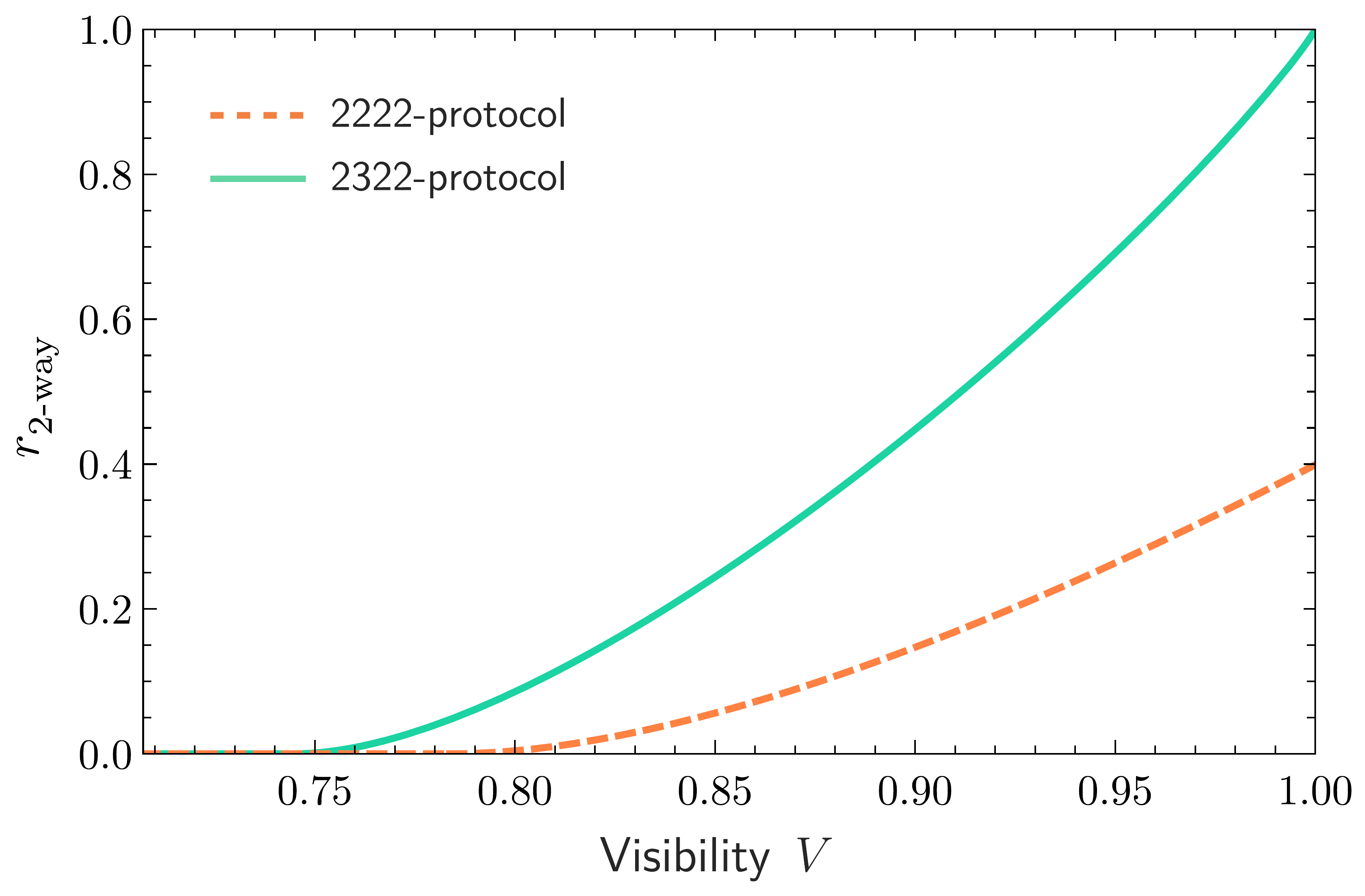}
\caption{Upper bounds on the two-way key rates for the $2222$ and the $2322$-protocols based on the CHSH inequality.}
\label{fig:2322vs2222}
\end{figure}

\paragraph{$2222$-scenario.} Consider now the case where Alice and Bob make two measurements each, that is, $x, y \in \{0, 1\}$ and there is no extra key setting for Bob. The ideal correlation shared between them is then given by the appropriate case in~\eqnref{eq:2322ideal} (or~\eqnref{eqs:biased_CHSH}) and reads
\begin{align}
\label{eq:2222ideal}
Q_{AB}(a,b|x,y) = \frac{1}{4}\left[1+\frac{(-1)^{a+b+xy}}{\sqrt2}\right].
\end{align}
As before, we consider a noisy version of this correlation with visibility $V<1$, and we apply the CC attack with the same post-processing of Eve given by \eqref{eq:PFEpostprocessing}. In this case, the two-way key rate is upper-bounded by
\begin{align}
\rtwoway
\leq
\frac{1+\sqrt{2}}{2} \left\{ \log_2 \left[ \frac{ \tilde{\jmath}(V)^{\tilde{\jmath}(V)} \tilde{k}(V)^{\tilde{k}(V)} }{ [\tilde{\jmath}(V)+\tilde{k}(V)]^{\tilde{\jmath}(V)+\tilde{k}(V)}} \right] + \tilde{\jmath}(V) + \tilde{k}(V) \right\},
\end{align}
where $\tilde{\jmath}(V) := (1-V/\sqrt{2})(\sqrt{2}-1)$ and $\tilde{k}(V) := (1+1/\sqrt{2})( \sqrt{2} V - 1 )$.
One can verify that the bound is zero at $\crit{V}=\frac{3}{7}(2\sqrt{2}-1)\approx 78.36\%$ ($>\frac{1}{\sqrt 2}=V_\text{loc}$), as stated in \tabref{tab:critical} in the `two-way' column.
The difference between the $2222$-protocol and the $2322$-protocol bounds are shown in Fig.~\ref{fig:2322vs2222}, and it is clear that the from the perspective of the honest users, the $2322$-protocol performs better at all visibilities.

\paragraph{$2422$-scenario.} Let us consider as well the CHSH-based DIQKD protocol with two added key settings for Bob, $y\in\{0,1,2,3\}$, as proposed in \citeref{SGP+20}, i.e.:~with one setting $y=2$ chosen again to be correlated with the setting $x=0$ of Alice, but also another extra setting $y=3$ correlated with the setting $x=1$ of Alice instead. The ideal correlation shared between Alice and Bob is then given by
\begin{align}
\label{eq:2422ideal}
Q_{AB}(a,b|x,y) =
\begin{cases}
  \frac{1}{4}\left[1+\frac{(-1)^{a+b+xy}}{\sqrt2}\right], &\text{ if } x,y\in\{0,1\}\\
  \frac{1}{2}\delta_{a,b}, &\text{ if }(x,y)=(0,2) \text{ or } (1,3)\\
  \frac{1}{4}, &\text{ if }(x,y)= (0,3) \text{ or } (1,2)
\end{cases}.
\end{align}
The key setting pairs $(\xk, \yk)$ are therefore either $(0,2)$ or $(1,3)$, and it is clear that both of these choices give rise to the same upper bound as the $2322$-protocol, that is, Eq.~\eqref{eq:2way_2322}.

%%%%%%%%%%%%%%%%%%%%%%%%%%%%%%%%%%%%%%%%%%%%%%%%%%%%%%%%%%%%%%%%%%%%%%%%%%%%%%%%%%%%%%%%%%%%%%%%%%%%%%%%%%%%%%%%
\subsection{One-way protocols involving partially entangled states}
\label{app:1way_partent_UBs}
%%%%%%%%%%%%%%%%%%%%%%%%%%%%%%%%%%%%%%%%%%%%%%%%%%%%%%%%%%%%%%%%%%%%%%%%%%%%%%%%%%%%%%%%%%%%%%%%%%%%%%%%%%%%%%%%
%
%%%%%%%%%%%%%%%%%%%%%%%%%%%%%%%%%%%%%%%%%%%%%%%%%%%%%%
\subsubsection{Finite detection efficiency}
In this section, we utilise the formulae derived in \appref{sec:key_rates_CC} for the EC- and PA-terms in the scenario of finite detection efficiency ($\eta<1$) and the purely lossy correlation~\eqref{eq:lossy_corr2} being shared, in order to determine the CC-based upper bounds on one-way key rates when Alice and Bob now ideally measure the partially entangled state~\eqref{eq:part_ent_state} parametrised by $\theta$, while performing projective measurements that maximise the CHSH violation~\eqref{eq:S_det}---the setting we introduced in \secref{subsec:gen_non-loc_corr} of the main text. 

Given particular preprocessing strategies or optimising over these, we compute then numerically the corresponding thresholds on the tolerable detection efficiency, $\crit{\eta}(\theta)$, that are depicted in \figref{fig:partial_thresholds} of the main text as a function of the state-parameter $\theta$. However, we are also importantly able to determine analytically the lowest possible thresholds applicable in the regime of $\theta\to0$ that is known to exhibit highest robustness to imperfect detection~\cite{E93}---these correspond to the smallest critical values presented in the left-most part of \figref{fig:partial_thresholds}, i.e~the values at which all the corresponding curves start from at $\theta=0$.

Specifically, we treat here two preprocessing strategies of Alice applied on the purely lossy correlation~\eqref{eq:lossy_corr2}, namely, deterministic binning with and without noisy preprocessing. Crucially, for these two choices it is beneficial for Alice and Bob to tune $\theta$, in order to lower the critical detection efficiencies set by the CC attack. In contrast, for random binning of inconclusive outcomes (or their absence in the finite visibility model) it is the maximally-entangled case of $\theta = \pi/2$ discussed in the preceding section that remains optimal.

Moreover, we focus here on the 2333-protocol, as it generally exhibits lower noise thresholds than the 2233-protocol. In such a case, the parties generate the key from the lossy correlation~\eqref{eq:lossy_corr2} using measurement settings $\xk,\yk=\{0,2\}$, in which the $\mathsf{Q}$-probabilities then simply read
\begin{equation}
\label{eq:limit_probs}
\mathsf{Q}_{00}=\cos^2 \frac{\theta}{2},\quad \mathsf{Q}_{11}=\sin^2 \frac{\theta}{2}, \quad \mathsf{Q}_{01}=\mathsf{Q}_{10}=0, \quad \mathsf{Q}_0^{\mathrm{A}} = \mathsf{Q}_0^{\mathrm{B}} = \cos^2 \frac{\theta}{2}, \quad \mathsf{Q}_1^{\mathrm{A}} = \mathsf{Q}_1^{\mathrm{B}} = \sin^2 \frac{\theta}{2}. 
\end{equation}
However, let us emphasise that other measurement settings, which are chosen to maximise the CHSH functional~\eqref{eq:S_det}, determine the maximal local weight $q^\cL$ employed within the CC attack that now also depends on the $\theta$-parameter of the partially entangled state. Although for the important case of $\theta\to0$ we possess an analytic expression for $q^\cL$, see \eqnref{eq:local_weight_partial}, for any other $\theta$ we may still evaluate efficiently the maximal local weight via the linear program \eref{eq:linear_program}, in particular, when computing all the curves that represent thresholds on tolerable detection efficiency in \figref{fig:partial_thresholds} as a function of $\theta$.

%%%%%%%%%%%%%%%%%%%%%%%%%%%%%%%%%%%%%%%%%%%%%%%%%%%%%%%%
\paragraph{Deterministic binning.}
In case Alice applies deterministic binning as her preprocessing strategy, we use again Eqs.~\eref{eq:ec_det} and \eref{eq:pa_noprep} to compute the upper bound on the one-way rate, i.e.~$\UB{\roneway{det}} = H(A'|E)_\text{det} - H(A'|B)_{\text{det}}$. However, in contrast to \eqnref{eq:UB_roneway_det}, we now substitute for the $\mathsf{Q}$-probabilities~\eqref{eq:limit_probs} that apply to the 2333-protocol involving partially entangled states, and for the optimal local weight applicable when $\theta \to 0$, i.e.~$q^\cL$ in Eq.~\eqref{eq:local_weight_partial}, in order to obtain
\begin{equation}
\UB{\roneway{det}}(\eta,\theta)
\;\underset{\theta\to0}{=}\;
\eta (3\eta - 2)\; h\!\left[\cos^2 \frac{\theta}{2}\right] - \eta \sin^2 \frac{\theta}{2} h[\eta] - (1-\eta) h\!\left[\eta \sin^2 \frac{\theta}{2}\right],
\end{equation}
which we may expand further in the limit of $\theta \to 0$, as follows:
\begin{equation}
\UB{\roneway{det}}(\eta,\theta)
\;\underset{\theta\to0}{=}\;
\frac{\eta\theta^{2}\left(-3+4\eta(1+\ln4)-\eta\ln(1-\eta)+\ln(\eta(1-\eta)/64)\right)}{\ln16}+\frac{\eta\theta^{2}}{\ln16}(6-8\eta)\ln\theta + \mathcal{O}\!\left(\theta^4\right).
\end{equation}
Now, it becomes evident that as $\theta\to0$ it is the second term above that dominates over the first term, as the ratio of the former to the latter is proportional to $\ln \theta$ and diverges in that limit. Therefore, it must be made zero by choosing adequately $\eta$ if $r^\uparrow_{\text{det},\theta\to0}$ is to exhibit a root as $\theta\to0$. Hence, this proves that $\lim_{\theta\to0}\{\UB{\roneway{det}}(\eta,\theta)\}$ may be vanishing only at the critical detection efficiency:
\begin{equation}
\crit{\eta}^\text{det}(\theta\to0) = 75 \%,
\end{equation}
which is indeed clearly observed in \figref{fig:partial_thresholds}---see the \emph{solid red curve} at $\theta=0$.

%%%%%%%%%%%%%%%%%%%%%%%%%%%%%%%%%%%%%%%%%%%%%%%%%%%%%%%%
\paragraph{Deterministic binning with noisy preprocessing.}
If Alice decides to apply noisy preprocessing apart from deterministically binning her inconclusive outcomes $\varnothing$, we similarly to \eqnref{eq:UB_roneway_det_np} compute the upper bound on the one-way rate with help of Eqs.~\eref{eq:ec_np} and \eref{eq:pa_np} as $\UB{\roneway{det+n.p}} = H(A'|E)_\text{det+n.p.} - H(A'|B)_{\text{det+n.p.}}$. However, this time after substituting for the $\mathsf{Q}$-probabilities~\eqref{eq:limit_probs} that apply for the 2333-protocol with partially entangled states, and the optimal local weight valid in the $\theta \to 0$ limit, i.e.~$q^\cL$ in Eq.~\eqref{eq:local_weight_partial}, we firstly verify that the critical detection efficiency gets smaller with the bit-flip probability approaching $\pp \to \frac{1}{2}_\pm$. Hence, substituting further for $\pp = \frac{1}{2} \pm \delta$ and expanding in small $\delta$, we get
\begin{equation}
\UB{\roneway{det+n.p}}(\eta,\theta)
\;\underset{\theta\to0}{=}\; 
\frac{4 \eta \sin^2 \frac{\theta}{2}}{\ln 2} \left(-4 + \eta(4+\eta) + (-2 + \eta(2+\eta)) \cos \theta \right) \delta^2 + \mathcal{O}\!\left(\delta^4\right),
\end{equation}
which we expand further in the $\theta \to 0$ limit to obtain
\begin{equation}
\UB{\roneway{det+n.p}}(\eta,\theta)
\;\underset{\theta\to0}{=}\; 
\left(\frac{2\eta(-3+3\eta+\eta^2)\theta^2}{\ln 2} + \mathcal{O}\!\left(\theta^4\right)\right) \delta^2 + \mathcal{O}\!\left(\delta^4\right).
\end{equation}
As a result, we may now explicitly identify that $\UB{\roneway{det+n.p}}$ evaluated for $\theta\to0$ vanishes when $-3+3\eta+\eta^2$ is zero, which exhibits a positive root at
\begin{equation}
\crit{\eta}^\text{det+n.p.}(\theta\to0)= \frac{\sqrt{21} - 3}{2} \approx 79.13\%.
\end{equation}
The above threshold value corresponds importantly to the starting point of the \emph{solid blue curve} at $\theta=0$ in \figref{fig:partial_thresholds}, which, as claimed in the main text, we can now state analytically.

%%%%%%%%%%%%%%%%%%%%%%%%%%%%%%%%%%%%%%%%%%%%%%%%%%%%%%%%
\paragraph{Optimising over all preprocessing maps.}
Finally, in a similar manner to \appref{app:1way_maxent_opt_maps}, we compute \emph{general} upper bounds~\eqref{eq:oneway_rate} that apply to one-way key rates independently of the preprocessing strategy employed when the parties share partially entangled states and observe lossy correlations \eref{eq:lossy_corr2}. In particular, we further perform by heuristic numerical methods the optimisation in \eqnref{eq:oneway_rate} over all stochastic maps $p_{A'|A}$ applied by Alice on her variable $A$, as well as stochastic maps $p_{M|A'}$ resulting in an extra message $M$ sent publicly to Bob.

The upper bounds can be then translated onto universal thresholds on tolerable detection efficiency, $\critmin{\eta}(\theta)$, below which no key can be distilled with one-way communication. These appear in~\figref{fig:partial_thresholds} of main text as \emph{dashed lines} with \emph{diamonds}, \emph{circles}, and \emph{squares} corresponding to the optimization being performed, respectively:~over the mappings $A \to A'$ and $A \to M$, and both of them simultaneously. Let us emphasise that we perform the optimisation over preprocessing strategies for the same CHSH-optimal correlations for which the thresholds with deterministic binning of inconclusive outcomes (with and without noisy preprocessing) in~\figref{fig:partial_thresholds} were derived.

Strikingly, it follows from~\figref{fig:partial_thresholds} that it is not only the special case of $\theta=\pi/2$, previously discussed in~\appref{app:1way_maxent_opt_maps}, but actually independently of the $\theta$-angle parametrising the partially entangled state, when it is sufficient for Alice to utilize only the $A \to M$ mapping and omit the $A \to A'$ preprocessing in order to achieve highest robustness against the CC attack. Moreover, we find that it is again always sufficient to consider $\cS_{A \to M}$ in \eqnref{eq:S_A->M} as the stochastic map $A \to M$, i.e.~the strategy in which Alice effectively signals the occurrence of inconclusive events to Bob. Furthermore, the preprocessing-optimized critical thresholds $\critmin{\eta}$ obtained in the $\theta\to\pi/2$ limit (right-most values of \emph{dashed lines} in \figref{fig:partial_thresholds}) consistently coincide with the values \eref{eq:general_bound_AAp} and \eref{eq:general_bound_AAp_ApM} determined in~\appref{app:1way_maxent_opt_maps} for protocols utilising maximally entangled states.

%%%%%%%%%%%%%%%%%%%%%%%%%%%%%%%%%%%%%%%%%%%%%%%%%%%%%%%%%%%%%%%%%%%%%%%%%%%%%%%%%%%%%%%%%%%%%%%%%%%%%%%%%%%%%%%%
\subsection{One-way protocols involving partially entangled states with postselection}
\label{app:1way_postselect}
%%%%%%%%%%%%%%%%%%%%%%%%%%%%%%%%%%%%%%%%%%%%%%%%%%%%%%%%%%%%%%%%%%%%%%%%%%%%%%%%%%%%%%%%%%%%%%%%%%%%%%%%%%%%%%%%
%
In this last section, we derive an upper bound on the DW rate, \eqnref{eq:DW_rate} of the main text, that now incorporates a postselection (PS) step, being used to certify the security of the protocol considered in~\citeref{xu_device-independent_2021}, i.e.:
\begin{equation}
\rDW^\text{PS}:=p_{\mathcal{V}_{p}}\left[H(A|\mrm{E},\mathcal{V}_{p})-H(A|B,\mathcal{V}_{p})\right]\ge p_{\mathcal{V}_{p}}\left[H_{\mathrm{min}}(A|\mrm{E},\mathcal{V}_{p})-H(A|B,\mathcal{V}_{p})\right],
\end{equation}
where $\mathcal{V}_{p}$ indicates successful postselection occurring with probability $p_{\mathcal{V}_p}$;~$H(A|\mrm{E},\dots)$ denotes the von Neumann entropy conditioned on the information possessed by the most general quantum eavesdropper Eve (hence the roman letter $\mrm{E}$ instead of an italic $E$ that would correspond to a classical random variable), which can in turn be lower-bounded by the min-entropy $H_\mrm{min}(A|\mrm{E},\dots)$. The CC attack, in which Eve holds a classical variable $E$, allows us to directly compute the upper bound on the PA-term, i.e.:
\begin{equation}
H_{\mathrm{min}}(A|\mrm{E},\mathcal{V}_{p})\le H(A|\mrm{E},\mathcal{V}_{p})\le H(A|E,\mathcal{V}_{p}),
\end{equation}
however, we drop for simplicity the conditioning on the postselected subset $\mathcal{V}_{p}$ in what follows. 

In particular, we write the upper bound determined by the CC attack as just $\frac{\rDW^\text{PS}}{p_{\mathcal{V}_p}} \le H(A|E)-H(A|B)$, and determine the analytic form of the corresponding EC- and PA-terms, $H(A|B)$ and $H(A|E)$, respectively. These may then be evaluated explicitly given a particular form of the observed correlation $\pABobs(a,b|x,y)$---here, see \secref{sec:postselection} of the main text, we consider the purely lossy correlation (\ref{eq:lossy_corr},\ref{eq:lossy_corr2}) with $\eta<1$  and $V=1$ in \eqnref{eqs:simplified_lossy_corr}.

%%%%%%%%%%%%%%%%%%%%%%%%%%%%%%%%%%%%%%%%%%%%%%%%%%%%%%%%
\subsubsection{The EC-term $H(A|B)$}
We adopt the notation of~\citeref{xu_device-independent_2021} and consider the 2333-protocol with the honest users, Alice and Bob, ideally sharing a partially entangled two-qubit state:
\begin{equation}
\left|\psi(\theta)\right\rangle_\sub{AB} =\cos\theta\,\left|00\right\rangle +\sin\theta\,\left|11\right\rangle .\label{eq:state}
\end{equation}
Alice uses two measurement settings $x\in\{1,2\}$, while Bob three settings $y\in\{1,2,3\}$, each corresponding to a projective measurement $\{\Pi(\phi),\mathbf{1}-\Pi(\phi)\}$ parametrised by the angle $\phi$, where 
\begin{equation}
\Pi(\phi)=\left(\begin{array}{cc}
\cos\left(\phi/2\right)^{2} & \cos\left(\phi/2\right)\sin\left(\phi/2\right)\\
\cos\left(\phi/2\right)\sin\left(\phi/2\right) & \sin\left(\phi/2\right)^{2}
\end{array}\right).\label{eq:pi_phi_measurement}
\end{equation}
This is equivalent to considering projective measurements $\{\frac{\mathbf{1}\pm\Pi(\phi)}{2}\}$ and $\Pi(\phi)=\cos\phi\,\sigma_{z}+\sin\phi\,\sigma_{x}$. For future reference, the key settings are set to $\xk=1$, $\yk=3$.

The lossy correlation shared by Alice and Bob is defined in Eqs.~\eref{eq:lossy_corr} and \eref{eq:lossy_corr2} for finite detection efficiency $\eta$ (and $\bar{\eta}\eqdef 1-\eta)$ as---dropping the $(x,y)$-dependence and the `obs' superscript for convenience, as in \eqnref{eqs:simplified_lossy_corr}:
\begin{equation}
\pAB(a,b)\quad=\quad\begin{array}{|c|c|c|c|}
\hline a\setminus b & 0 & 1 & \varnothing\\
\hline 0 & \eta^{2}\,\mathsf{Q}_{00} & \eta^{2}\,\mathsf{Q}_{01} & \eta\bar{\eta}\QA{0}\\
\hline 1 & \eta^{2}\,\mathsf{Q}_{10} & \eta^{2}\,\mathsf{Q}_{11} & \eta\bar{\eta}\QA{1}\\
\hline \varnothing & \bar{\eta}\eta\QB{0} & \bar{\eta}\eta\QB{1} & \bar{\eta}^{2}
\\\hline \end{array}\,,
\label{eq:correlation_AB}
\end{equation}
where $\varnothing$ denotes the no-click event, while $\mathsf{Q}_{ab}$ are the ideal probabilities observed in the absence of losses, i.e.:
\begin{equation}
\mathsf{Q}_{ab}=\tr{ A_{a|\xk}\otimes B_{b|\yk}\;\ket{\psi(\theta)}_\sub{AB}\bra{\psi(\theta)}}
\label{eq:ideal_probabilities}
\end{equation}
with 
\begin{equation}
A_{a|\xk}=\delta_{a,0}\Pi(\phi_{\xk})+\delta_{a,1}\left(1-\Pi(\phi_{\xk})\right),\quad B_{b|\yk}=\delta_{b,0}\Pi(\phi_{\yk})+\delta_{b,1}\left(1-\Pi(\phi_{\yk})\right).
\end{equation}
The marginals are defined as $\QA{0}=\mathsf{Q}_{00}+\mathsf{Q}_{01}$, $\QA{1}=\mathsf{Q}_{10}+\mathsf{Q}_{11}$, and similarly for Bob.

After Alice bins deterministically $\varnothing\to1$ (but not Bob~\citep{ML11}), the shared correlation reads 
\begin{equation}
\pAB^{\text{det}}(a,b)\quad=\quad\begin{array}{|c|c|c|c|}
\hline a\setminus b & 0 & 1 & \varnothing\\
\hline 0 & \eta^{2}\,\mathsf{Q}_{00} & \eta^{2}\,\mathsf{Q}_{01} & \eta\bar{\eta}\QA{0}\\
\hline 1 & \eta^{2}\,\mathsf{Q}_{10}+\bar{\eta}\eta\QB{0} & \eta^{2}\,\mathsf{Q}_{11}+\bar{\eta}\eta\QB{1} & \eta\bar{\eta}\QA{1}+\bar{\eta}^{2}
\\\hline \end{array}\,.
\label{eq:det_correlation}
\end{equation}
She then performs the postselection step, i.e.~she keeps the events when she measured `1' with probability $\qq$ and discards them with probability $1-\qq$. Bob does the same but, importantly, he postselects both `1' and $\varnothing$ results, because of $\varnothing$ being equivalent to `1' with respect to postselection (which should be done as if Bob also binned his data~\citep{ML11}). This corresponds to both of them defining a new variable `D' signifying discarded bits. We will ignore these discarded outcomes, but let us for completeness write down the full correlation including the `D' outcome. In order to do so, let us express \eqref{eq:det_correlation} for short as 
\begin{equation}
\pAB^{\text{det}}(a,b)\quad\eqdef \quad\begin{array}{|c|c|c|c|}
\hline a\setminus b & 0 & 1 & \varnothing\\
\hline 0 & q_{00} & q_{01} & q_{0\varnothing}\\
\hline 1 & q_{10} & q_{11} & q_{1\varnothing}
\\\hline \end{array}\,,
\end{equation}
so that the `full' postselected (fPS) correlation then reads 
\begin{equation}
\pAB^{\text{fPS}}(a,b)\quad=\quad\begin{array}{|c|c|c|c|c|}
\hline a\setminus b & 0 & 1 & \varnothing & \mathrm{D}\\
\hline 0 & q_{00} & \qq \,q_{01} & \qq\,q_{0\varnothing} & (1-\qq) \,(q_{01}+q_{0\varnothing})\\
\hline 1 & \qq\,q_{10} & \qq^{2}\,q_{11} & \qq^{2}\,q_{1\varnothing} & \qq(1-\qq)(q_{11}+q_{1\varnothing})\\
\hline \mathrm{D} & (1-\qq)\,q_{10} & \qq(1-\qq)\,q_{11} & \qq(1-\qq)\,q_{1\varnothing} & (1-\qq)^{2}(q_{11}+q_{1\varnothing})
\\\hline \end{array}\,.\label{eq:ps-1}
\end{equation}
Throwing away the discarded bits, i.e.~columns and rows of Tab.~\eqref{eq:ps-1} labelled by `D', we arrive at the desired postselected (PS) correlation: 
\begin{equation}
\pAB^{\text{PS}}(a,b)\quad=\quad\frac{1}{P(\mathcal{V}_{p})}\;\begin{array}{|c|c|c|c|}
\hline a\setminus b & 0 & 1 & \varnothing\\
\hline 0 & q_{00} & \qq\,q_{01} & \qq\,q_{0\varnothing}\\
\hline 1 & \qq\,q_{10} & \qq^{2}\,q_{11} & \qq^{2}\,q_{1\varnothing}
\\\hline \end{array}\,,\label{eq:ps-2}
\end{equation}
where the probability of successful postselection is then given by
\begin{align}
P(\mathcal{V}_{p}) & \eqdef q_{00}+\qq(q_{01}+q_{0\varnothing}+q_{10})+\qq^{2}(q_{11}+q_{1\varnothing})\nonumber\\
 & =\eta^{2}\,\mathsf{Q}_{00}+\qq\left[\eta^{2}\left(\mathsf{Q}_{01}+\mathsf{Q}_{10}\right)+\bar{\eta}\eta\left(\QA{0}+\QB{0}\right)\right]+\qq^{2}\left[\eta^{2}\,\mathsf{Q}_{11}+\eta\bar{\eta}\left(\QA{1}+\QB{1}\right)+\bar{\eta}^{2}\right],
 \label{eq:pVp-1}
\end{align}
with the full expression obtained after substituting explicitly for $\pAB^{\text{det}}(a,b)$ according to \eqnref{eq:det_correlation}.

In general, the postselected marginals of Bob read
\begin{equation}
\pB^\text{PS}(b)=\frac{1}{P(\mathcal{V}_{p})}\begin{cases}
q_{00}+\qq\,q_{10} & \text{if }b=0\\
\qq\,q_{01}+\qq^{2}\,q_{11} & \text{if }b=1\\
\qq\,q_{0\varnothing}+\qq^{2}\,q_{1\varnothing} & \text{if }b=\varnothing
\end{cases},\label{eq:bobs_marginals}
\end{equation}
and the relevant conditional probability distribution is obtained by dividing the columns of \eqref{eq:ps-2} by the corresponding terms in \eqref{eq:bobs_marginals}, i.e.: 
\begin{equation}
\pAB^{\text{PS}}(a|b)\quad=\quad\begin{array}{|c|c|c|c|}
\hline a\setminus b & 0 & 1 & \varnothing\\
\hline 0 & q_{00}/(q_{00}+\qq\,q_{10}) & \qq\,q_{01}/(\qq\,q_{01}+\qq^{2}\,q_{11}) & \qq\,q_{0\varnothing}/(\qq\,q_{0\varnothing}+\qq^{2}\,q_{1\varnothing})\\
\hline 1 & \qq\,q_{10}/(q_{00}+\qq\,q_{10}) & \qq^{2}\,q_{11}/(\qq\,q_{01}+\qq^{2}\,q_{11}) & \qq^{2}\,q_{1\varnothing}/(\qq\,q_{0\varnothing}+\qq^{2}\,q_{1\varnothing})
\\\hline \end{array}\,.\label{eq:ps-cond}
\end{equation}
Hence, we can now evaluate explicitly the conditional entropy $H(A|B)$, i.e.~the EC-term, as:
\begin{equation}
H(A|B)=
\sum_{b}\pB^\text{PS}(b)\,H(A|B=b)=
\pB^\text{PS}(0)\,h\!\left[\frac{q_{00}}{q_{00}+\qq\,q_{10}}\right]
+
\pB^\text{PS}(1)\,h\!\left[\frac{\qq\,q_{01}}{\qq\,q_{01}+\qq^{2}\,q_{11}}\right]
+
\pB^\text{PS}(\varnothing)\,h\!\left[\frac{\qq\,q_{0\varnothing}}{\qq\,q_{0\varnothing}+\qq^{2}\,q_{1\varnothing}}\right],\label{eq:ec-term}
\end{equation}
where $h[x]$ is again the binary entropy function.

%%%%%%%%%%%%%%%%%%%%%%%%%%%%%%%%%%%%%%%%%%%%%%%%%%%%%%%%
\subsubsection{The PA-term $H(A|E)$}
Let us recall again that within the CC attack here considered Eve distributes a noiseless correlation in each non-local round that occur with probability $q^{\cNL}\eqdef 1-q^{\cL}$. In contrast, whenever she distributes a local correlation with probability $q^{\cL}$ (the local weight), she perfectly knows the outcome of both Alice and Bob. Hence, denoting generally by $\mathsf{P}_{ab}^{\cL}$ the resulting correlation shared by Alice and Bob within the local rounds after they bin their $\varnothing$-outcomes onto `1', we may write the overall tripartite correlation specifying the CC attack as
\begin{equation}
\p{ABE}(a,b,e)=q^{\cNL}\begin{bmatrix}\mathsf{Q}_{00} & \mathsf{Q}_{01}\\
\mathsf{Q}_{10} & \mathsf{Q}_{11}
\end{bmatrix}\delta_{e,?}+q^{\cL}\begin{bmatrix}\mathsf{P}_{00}^{\cL}\delta_{e,00} & \mathsf{P}_{01}^{\cL}\delta_{e,01}\\
\mathsf{P}_{10}^{\cL}\delta_{e,10} & \mathsf{P}_{11}^{\cL}\delta_{e,11}
\end{bmatrix},\label{eq:pABE}
\end{equation}
where the random variable of Eve, $E$, consists of two bits (one for Alice and one for Bob) and an extra outcome `?' representing her lack of knowledge. Note that, as it will be clear from the calculation below, without loss of generality we could have also ignored the second bit of Eve, because in the following we are interested only in the correlations between her and Alice---assumed to be the party performing PA within the one-way paradigm of key distribution. Moreover, recall that, after tracing out Eve, Alice and and Bob must recover their correct shared correlation that corresponds the case of finite detection efficiency $\eta$ and binning $\varnothing$-outcomes onto `1'---obtained by adding the last column onto the second one and similarly then for the rows in \eqnref{eq:correlation_AB}. Hence, we also have a constraint that:
\begin{equation}
\pAB(a,b)=q^{\cNL}\begin{bmatrix}\mathsf{Q}_{00} & \mathsf{Q}_{01}\\
\mathsf{Q}_{10} & \mathsf{Q}_{11}
\end{bmatrix}+q^{\cL}\begin{bmatrix}\mathsf{P}_{00}^{\cL} & \mathsf{P}_{01}^{\cL}\\
\mathsf{P}_{10}^{\cL} & \mathsf{P}_{11}^{\cL}
\end{bmatrix}
\;\equiv\;
\begin{bmatrix}\eta^{2}\mathsf{Q}_{00} & \eta^{2}\mathsf{Q}_{01}+\eta\bar{\eta}\QA{0}\\
\eta^{2}\mathsf{Q}_{10}+\bar{\eta}\eta\QB{0} & \eta^{2}\mathsf{Q}_{11}+\bar{\eta}\eta\left(\QB{1}+\QA{1}\right)+\bar{\eta}^{2}
\end{bmatrix},\label{eq:pAB}
\end{equation}
which specifies unambiguously the correlation $\mathsf{P}_{ab}^{\cL}$ distributed by Eve in the local rounds.

Now, we consider the tripartite correlation \eqref{eq:pABE} after both Alice and Bob postselect their `1' outcomes with probability $\qq$, as in Eq.~\eqref{eq:ps-2}, i.e.:
\begin{equation}
p_{\mathrm{ABE}}^{\text{PS}}(a,b,e)=\frac{1}{P(\mathcal{V}_{p})}\left(q^{\cNL}\begin{bmatrix}\mathsf{Q}_{00} & \qq\,\mathsf{Q}_{01}\\
\qq\,\mathsf{Q}_{10} & \qq^{2}\,\mathsf{Q}_{11}
\end{bmatrix}\delta_{e,?}+q^{\cL}\begin{bmatrix}\mathsf{P}_{00}^{\cL}\delta_{e,00} & \qq\,\mathsf{P}_{01}^{\cL}\delta_{e,01}\\
\qq\,\mathsf{P}_{10}^{\cL}\delta_{e,10} & \qq^{2}\,\mathsf{P}_{11}^{\cL}\delta_{e,11}
\end{bmatrix}\right),\label{eq:pABE_PS}
\end{equation}
where the probability of successful postselection, $P(\mathcal{V}_{p})$, can be naturally decomposed as 
\begin{equation}
P(\mathcal{V}_{p})=P(\mathcal{V}_{p},\cNL)+P(\mathcal{V}_{p},\cL)=P(\cNL)\,P(\mathcal{V}_{p}|\cNL)+P(\cL)\,P(\mathcal{V}_{p}|\cL),
\end{equation}
with $P(\cL)=q^{\cL}$, $P(\cNL)=q^{\cNL}$ by the definition of local weight, and
\begin{align}
P(\mathcal{V}_{p}|\cNL) & =\mathsf{Q}_{00}+\qq\,(\mathsf{Q}_{01}+\mathsf{Q}_{10})+\qq^{2}\,\mathsf{Q}_{11},\\
P(\mathcal{V}_{p}|\cL) & =\mathsf{P}_{00}^{\cL}+\qq\,(\mathsf{P}_{01}^{\cL}+\mathsf{P}_{10}^{\cL})+\qq^{2}\,\mathsf{P}_{11}^{\cL},
\end{align}
being the probabilities of successful postselection given that, respectively, the non-local or the local box is distributed. In particular, we have that
\begin{align}
P(\mathcal{V}_{p}) & =q^{\cNL}\left(\mathsf{Q}_{00}+\qq\,(\mathsf{Q}_{01}+\mathsf{Q}_{10})+\qq^{2}\,\mathsf{Q}_{11}\right)+q^{\cL}\left(\mathsf{P}_{00}^{\cL}+\qq\,(\mathsf{P}_{01}^{\cL}+\mathsf{P}_{10}^{\cL})+\qq^{2}\,\mathsf{P}_{11}^{\cL}\right)\nonumber\\
 & =\eta^{2}\mathsf{Q}_{00}+\qq\,\left[\eta^{2}\left(\mathsf{Q}_{01}+\mathsf{Q}_{10}\right)+\bar{\eta}\eta\left(\QA{0}+\QB{0}\right)\right]+\qq^{2}\,\left[\eta^{2}\mathsf{Q}_{11}+\bar{\eta}\eta\left(\QA{1}+\QB{1}\right)+\bar{\eta}^{2}\right],\label{eq:pVp}
\end{align}
where we consistently reproduce the expression \eqref{eq:pVp-1} after imposing the constraint \eqref{eq:pAB} on $\mathsf{P}_{ab}^{\cL}$. 

Tracing out Bob from \eqref{eq:pABE_PS}, we obtain
\begin{equation}
\p{AE}^{\text{PS}}(a,e)
\quad=\quad 
\frac{1}{P(\mathcal{V}_{p})} \; \begin{array}{|c|c|c|c|c|c|}
\hline a\setminus e & 00 & 01 & 10 & 11 & ?\\
\hline 0 & q^{\cL}\mathsf{P}_{00}^{\cL} & q^{\cL}\qq\,\mathsf{P}_{01}^{\cL} & 0 & 0 & q^{\cNL}\left(\mathsf{Q}_{00}+\qq\,\mathsf{Q}_{01}\right)\\
\hline 1 & 0 & 0 & q^{\cL}\qq\,\mathsf{P}_{10}^{\cL} & q^{\cL}\qq^{2}\,\mathsf{P}_{11}^{\cL} & q^{\cNL}\left(\qq\,\mathsf{Q}_{10}+\qq^{2}\,\mathsf{Q}_{11}\right)
\\\hline \end{array}\,,\label{eq:pAE_PS}
\end{equation}
where, as expected, Eve is perfectly correlated with Alice in the local rounds, so that $\forall e\ne?:\;H(A|E=e)=0$.

As a result, the desired entropy of Alice's outcomes conditioned on Eve is fully determined by the case when Eve distributes a non-local correlation, which occurs with probability 
\begin{align}
p_{E}(?) & =P(\cNL|\mathcal{V}_{p})=\frac{P(\cNL,\mathcal{V}_{p})}{P(\mathcal{V}_{p})}=q^{\cNL}\frac{P(\mathcal{V}_{p}|\cNL)}{P(\mathcal{V}_{p})},
\end{align}
so that the PA-term reads
\begin{align}
H(A|E)  =p_{E}(?)\;H(A|E=?)=q^{\cNL}\frac{P(\mathcal{V}_{p}|\cNL)}{P(\mathcal{V}_{p})}\;h\!\left[\frac{\mathsf{Q}_{00}+\qq\,\mathsf{Q}_{01}}{P(\mathcal{V}_{p}|\cNL)}\right],
\end{align}
being defined completely by the last column of the correlation \eqref{eq:pAE_PS}.

%%%%%%%%%%%%%%%%%%%%%%%%%%%%%%%%%%%%%%%%%%%%%%%%%%%%%%%%%%%%%%%%%%%%%%%%%%%%%%%%%%%%%%%%%%%%%%%%%%%%%%%%%%%%%%%%%%%%%%%%%%%%%%%%%

\end{document}